\documentclass[12pt]{article}
\usepackage{amsmath}
\usepackage{graphicx,psfrag,epsf,epstopdf}
\usepackage{enumerate}
\usepackage{natbib}
\usepackage{amsfonts}

\newcommand{\blind}{0}

\newcommand{\real}{\ensuremath{\mathbb{R}}}

\newcommand{\ltwo}{\ensuremath{\mathbb{L}^2}}

\graphicspath{{figuresnew/}}

\begin{document}

\def\spacingset#1{\renewcommand{\baselinestretch}%
{#1}\small\normalsize} \spacingset{1}

%%%%%%%%%%%%%%%%%%%%%%%%%%%%%%%%%%%%%%%%%%%%%%%%%%%%%%%%%%%%%%%%%%%%%%%%%%%%%%

\if0\blind
{
  \title{\bf A Geometric Approach to Visualization of Variability in Functional Data}
  \author{Weiyi Xie$^{1}$, Sebastian Kurtek$^{1}$, Karthik Bharath$^{2}$ and Ying Sun$^{3}$\\
    $^{1}$Department of Statistics, The Ohio State University\\
    $^{2}$School of Mathematical Sciences, University of Nottingham\\
    $^{3}$Division of Computer, Electrical and Mathematical Sciences\\ and Engineering, King Abdullah University of Science and Technology}
  \maketitle
} \fi

\if1\blind
{
  \bigskip
  \bigskip
  \bigskip
  \begin{center}
    {\LARGE\bf A Geometric Approach to Visualization of Variability in Functional Data}
\end{center}
  \medskip
} \fi

\bigskip
\begin{abstract}
We propose a new method for the construction and visualization of boxplot-type displays for functional data. We use a recent functional data analysis framework, based on a representation of functions called square-root slope functions, to decompose observed variation in functional data into three main components: amplitude, phase, and vertical translation. We then construct separate displays for each component, using the geometry and metric of each representation space, based on a novel definition of the median, the two quartiles, and extreme observations. The outlyingness of functional data is a very complex concept. Thus, we propose to identify outliers based on any of the three main components after decomposition. We provide a variety of visualization tools for the proposed boxplot-type displays including surface plots. We evaluate the proposed method using extensive simulations and then focus our attention on three real data applications including exploratory data analysis of sea surface temperature functions, electrocardiogram functions and growth curves.
\end{abstract}

\noindent%
{\it Keywords:} amplitude and phase variabilities; Fisher--Rao metric; functional outlier detection; square-root slope function
\vfill

\newpage
\spacingset{1.45} % DON'T change the spacing!
\section{Introduction}
\label{sec:intro}

Functional data analysis refers to a set of tools, including alignment, comparison, and statistical modeling, developed to study complex, modern data objects that are represented as 1D functions, shapes of curves and surfaces, and images. Thus, analysis of functional data has potential for application to a number of fields including medical imaging, biology, computer vision, environmetrics, biometrics, and bioinformatics. Thanks to recent progress in the functional data and shape analysis communities (\cite{kneip-ramsay:2008,muller-biometrika:2008}; \cite*{EFA:10}; \cite{marron2015,Cheng,younes-elastic-distance,Beg:2005}; \cite*{srivastava_etal_PAMI:10}; \cite{kurtekmi,eccv12:jermyn}), statistical methods for analyzing such datasets are now fairly well established. However, not enough consideration has been given to visualization of functional data, which is an important part of data exploration. Effective visualization techniques for such challenging datasets are also indispensable for communicating results of analysis to experts across a variety of fields. Here, our focus is to add to the limited selection of literature on functional data graphics by developing boxplot-type displays for one-dimensional functional data. The key premise of the proposed method is that the `boxplot' is constructed on the infinite-dimensional function space, thus truly capturing the full structure of the given data.

%kneip-gasser-annals:92,ramsay-li-RSSB:98,gervini-gasser-RSSB:04,muller-JASA:2004,ramsay-silverman-2005,james:07

Boxplot displays for univariate Euclidean data were pioneered by \cite{tukey1977}, and have proved to be very effective for exploratory data analysis. In recent years, there have been some efforts to extend these methods to functional datasets, a direction which is gaining interest. The method of \cite{citeulike:7579189} focused on generating functional bagplots and functional highest-density-region boxplots using only the first two functional principal component (PC) scores. The approach of \cite{Fraiman:2012:QFI:2170034.2170095} is also projection-based, where principal quantile directions are defined (similar in nature to principal directions of variability in PC analysis). Although useful in some settings, these methods are not truly functional because the resulting data for which the displays are generated is lower dimensional (bivariate or multivariate). In fact, this drawback applies to any method that first uses a basis expansion to represent the functional data and subsequently generates displays via the basis coefficients. As a result, variability not captured by the few chosen basis elements is lost from the display.

There has been some success in extending the concept of data depth (\cite{mahalanobis1936generalized,tukey_74_mathematics,oja,citeulike:2056627,fraiman,vardi_00_multivariate}) from the multivariate setting to the function setting (\cite{Fraiman1,Febrero,febrero2008outlier}). \cite{citeulike:10107686} were the first to generate boxplot displays for given functions rather than for their multivariate summaries. In their work, they defined ordering of functions using the notion of a functional data depth measure called band depth (\cite{lopez}). However, their method has some shortcomings. First, because some aspects of the boxplot are constructed in a pointwise manner (i.e., the $50\%$ central region and the minimum/maximum envelopes), the full functional interpretation of the display is lost and the structure of the underlying function space is ignored. Second, boxplots generated using band depth as a measure are not directly applicable to functional data observed under hidden temporal warping variability, as they are not effective at capturing such variability of the sample functions in the display. An extension of the original functional boxplots to more general functional data including shapes and images was presented in \cite{HDM}.

Another set of approaches relied on multivariate functional depth measures to construct displays and for outlier detection (\cite{doi:10.1080/03610926.2012.746368,doi:10.1080/01621459.2013.856795,LP2014,Hubert2015}). However, these methods inherit some of the previously mentioned drawbacks. \cite{doi:10.1080/01621459.2013.856795} noted the importance of accounting for temporal variability in functional data when generating displays and in outlier detection. Their solution was to form bivariate functions using the amplitude and phase components and perform subsequent analysis based on this representation. The main issue in this approach is that the amplitude and phase variabilities are estimated using a completely unrelated procedure to that used for boxplot construction. It seems more appropriate to use the same framework (functional representation spaces, metrics, etc.) for both tasks. Additionally, at the initial exploratory stage, it may be beneficial to consider the two sources of variability separately in the construction of visualization techniques.

To date, efforts to visualize functional data in the presence of temporal variability have been sparse. The problem of extracting the underlying warping variability from a given set of functions has been referred to by many names in the literature including registration, alignment, and separation of amplitude and phase. The focus of the current paper is not on the problem of amplitude and phase separation in functional data. Instead, we propose a functional boxplot-type display that can accommodate these two types of variability. In particular, we argue that most functional datasets contain a minimum of three main components: translation, amplitude, and phase. The translation component refers to the action of a translation group on some function space, where the same constant is added to all values of the function. The amplitude and phase components are much more difficult to formally define. The definitions used in this work are those by \cite*{EFA:10}, who define the phase component of a function as its warping under the group of diffeomorphisms and the amplitude component as the values of the function after the warping has been accounted for. The complexity of many functional datasets demands that translation, amplitude, and phase components be separated prior to further statistical analysis; this can also be very helpful for visualization, where a different display can be generated for each source of variability. In some settings, the translation component of functional data is dominant compared to the other two, and can thus be effectively extracted and visualized through principal component analysis (PCA). In this paper, we focus on a more general case where this may not be true.

\begin{figure}[!t]
\begin{center}
    \includegraphics[width=6in]{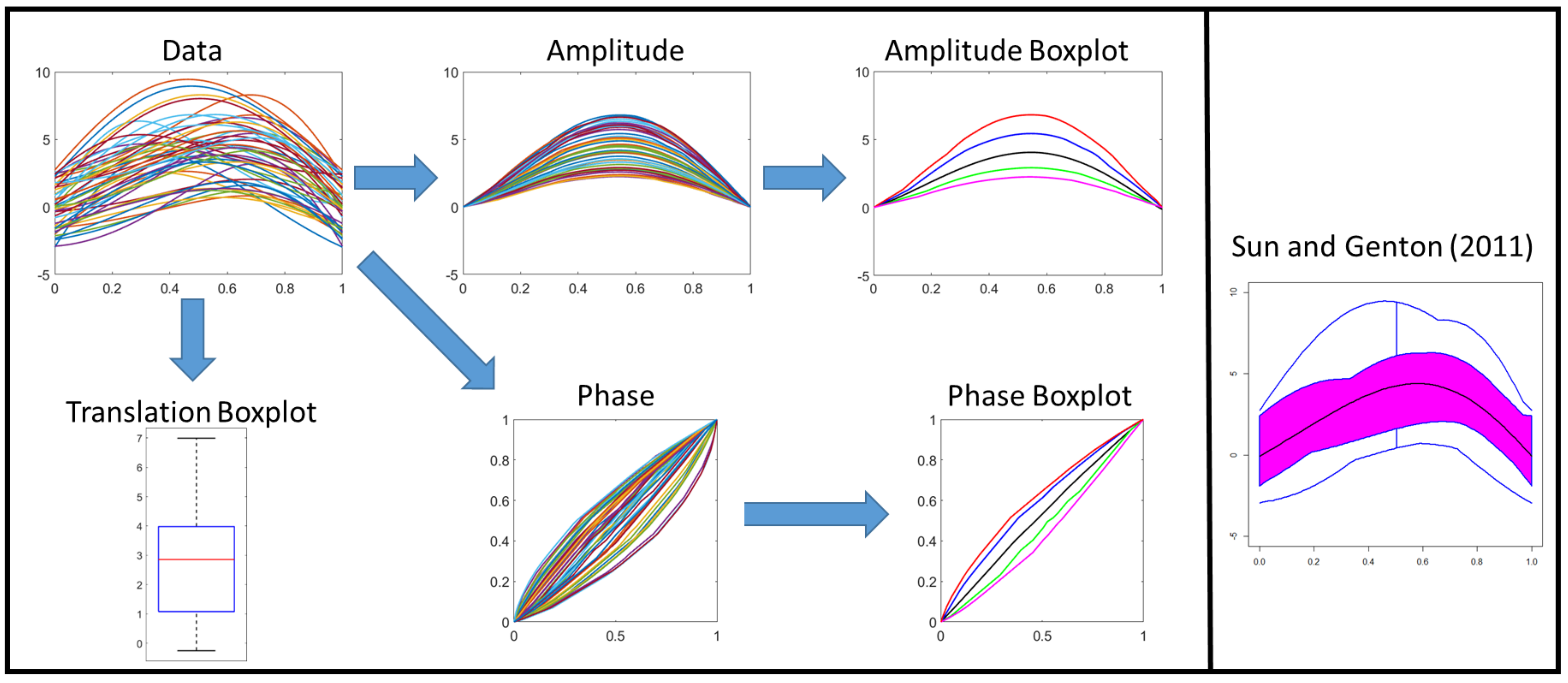}
    \caption{\small Visualization of functional data in the presence of temporal variability. The proposed method (left) first decomposes the data into the translation, amplitude and phase components, and then generates a boxplot-type display for each one separately (black=median, blue and green=quartiles, red and magenta=extremes). This allows for effective visualization of variability in each component. In comparison, the method of \cite{citeulike:10107686} (right) generates a single boxplot display where the three sources of variability are difficult to distinguish.}\label{fig:motdata}
    \end{center}
\vspace{-7mm}
\end{figure}

The definition and characterization of functional outliers is a complicated process. Our proposal to first separate the functions into more fundamental components and then detect outliers based on these components provides a convenient simplification. Furthermore, it allows us to view function outlyingness with respect to each individual component as well as jointly using multiple components. We provide a motivating toy example for the proposed method in Figure \ref{fig:motdata}. The data here was generated as follows. First, we generated multiple amplitude components by rescaling a template function (the start and end points of the template were both zero). The phase variation was applied next using a one-parameter family of warping functions (similar to those in Section \ref{sec:simu}). Finally, we randomly translated the functions. Next, we visualize the functions using the proposed approach. The extracted translation, amplitude and phase components closely resemble the corresponding original simulated components. Furthermore, the amplitude and phase boxplots reflect the natural variation in the simulated data, i.e., the amplitude boxplot shows scale changes of the template while the phase boxplot reflects variability in warping functions similar to those observed in the simulated one-parameter family. We contrast these results with those generated by the popular method of \cite{citeulike:10107686} where the varaibilities are all displayed in one boxplot, making interpretation much more difficult in this context.

This work defines a principled method for the construction of boxplot-type displays for functions in the presence of phase or warping variability. In fact, the construction holds for any functional data on an ordered domain. First, the given functional data is decomposed into translation, amplitude, and phase components using the methods described in \cite*{EFA:10}. Then, the proposed approach relies on the Riemannian geometry and metrics of the three respective representation spaces for ordering, and subsequent visualization. The geometric tools are important for constructing reliable visualization tools as they provide a principled background for summarization of the three sources of variability. Furthermore, these methods are based on a warping-invariant distance, which allows one to naturally extract the temporal variability from functional data. Note that our approach is fully functional and respects the geometry of the underlying function space (i.e., the displays are intrinsic data summaries). This is in contrast to standard methods that perform analysis using cross-sectional (pointwise) summaries, which are known to be deficient when functional data contains temporal variability (\cite*{EFA:10}; \cite{RePEc:eee:csdana:v:61:y:2013:i:c:p:50-66}).

We study the ability of the proposed method to detect outliers in several simulated datasets, and we compare our method to the state-of-the-art functional boxplots by \cite{citeulike:10107686}. Before proceeding, we would like to clarify our terminology of boxplot-type displays. As will be clear in later sections, the display generated via the proposed method is a boxplot on the infinite-dimensional function space of amplitude or phase. The two-dimensional displays of these boxplots do not necessarily resemble a box, which follows intuition as it is generally difficult to display infinite-dimensional objects in 2D. Thus, we additionally provide a 3D surface plot that can be better at displaying the functional boxplot. To summarize, the main contributions of this paper are the following:
\begin{itemize}
\item We separate variability of functional data into the translation, amplitude and phase components using methods of \cite*{EFA:10} for individual visualization of functions in the presence of temporal variability.
\item We use the Riemannian geometry of the amplitude and phase representation spaces to define boxplot-type displays for these components. In particular, we use warping-invariant Riemannian metrics for function ordering and the computation of the five number summary. This is the first work in literature that provides tools for computing the amplitude and phase medians, which are more robust to outliers than their mean counterparts.
\item We provide an approach for outlier detection based on the three components.
\end{itemize}

\subsection{Motivating Applications}

We provide visualizations and the results of outlier detection on three real, functional datasets, which are of importance in environmetrics, biometrics, and medicine: sea surface temperature, Berkeley growth data, and electrocardiogram signals. In particular, we study sea surface temperature in relation to El Ni\~{n}o and La Ni\~{n}a events, which have a strong impact on weather patterns. Given that many global economies are very sensitive to climate fluctuations, especially those heavily dependent on agriculture, the study of such environmental signals in relation to climate changes is pertinent. It is also valuable to note that, at the time of this writing, we are experiencing one of the strongest El Ni\~{n}o events since the beginning of record collection. The phase variability in this data comes from interannual temperature variation (e.g., some years experience longer winters than others). Both the amplitude (temperature) and phase (timing of seasonal patterns) components are important features of sea surface temperature functions (see Figure \ref{fig:elnino}).

Growth curves are of great importance in many applications including plant science and biometrics. Here, we study development patterns of boys and girls through the Berkeley growth data (\cite{BGD}), which can reveal important characteristics of complex growth processes including the number and size of growth spurts across gender or other important covariates such as socioeconomic status. Phase variability again plays an important role in this data (\cite{Cheng}); it tells us about the time at which various interesting traits occur (see Figure \ref{fig:bgd}).

Finally, we showcase the proposed visualization techniques on PQRST complexes extracted from electrocardiogram signals. PQRST refers to the five waves in each complex: the first positive peak is the P wave, the first negative peak is the Q wave, the second positive peak is the R wave, the second negative peak is the S wave, and the third positive peak is the T wave (see Figure \ref{fig:ecg} for an example). Cardiologists routinely use such data to diagnose various heart disorders including congenital heart disease and myocardial infarction. In this application, phase variability directly corresponds to the length of a person's heartbeat, and the occurrence and duration of the P, Q, R, S and T waves.

All of these datasets share a common feature: they are all observed under natural phase variability. Thus, our goal in this paper is to propose a method, which can be used to discover new data patterns by separately visualizing the amplitude, phase, and translation components of the functions.

The rest of this paper is organized as follows. Section \ref{sec:meth} describes the details of the boxplot construction for the amplitude and phase variabilities in functional data. Section \ref{sec:results} provides multiple simulation studies as well as analysis results for the three real datasets described above. We close with a summary and some directions for future work in Section \ref{sec:conc}. The Supplementary Matrial includes a description of the invariance and equivariance properties of the proposed boxplot displays, a brief discussion of the connections to univariate Euclidean boxplots, and additional simulation studies and real data examples.

\section{Construction of Functional Boxplot-Type Displays}
\label{sec:meth}

In this section, we describe the construction of boxplot-type displays for translation, amplitude, and phase components of a set of functions. We begin by reviewing relevant topics from \cite*{EFA:10} including function representations, metrics, and algorithms for amplitude-phase separation. Given these tools, we proceed to describe our novel method for visualization of various functional data components using a unified geometric approach.

%For additional details, we refer the reader to \cite*{EFA:10}.
%While the proposed approach is valid for any method capable of separating amplitude from phase in a function, we use the elegant Riemannian framework based on the extended Fisher--Rao metric proposed by \cite*{EFA:10}.

\subsection{Elastic Functional Data Analysis}

We focus our efforts on the visualization of real-valued functions on the interval $[0,1]$, where we assume that functions are absolutely continuous. Thus, the original function space can be defined as $\mathcal{F} = \{f: [0,1] \rightarrow \mathbb{R} | f \text{ is absolutely continuous}\}$. All methods described herein are valid for functional data on any closed subinterval of the real line. Because we are interested in visualization of elastic functions (i.e., functions with warping variability), we define $\Gamma$ as the set of all warping functions (orientation-preserving diffeomorphisms) of the unit interval $[0,1]$: $\Gamma = \big\{ \gamma: [0,1] \rightarrow [0,1] | \gamma(0)=0, \gamma(1) = 1, 0<\dot{\gamma}<\infty\big\}$, where $\dot{\gamma}$ is the time derivative of $\gamma$. $\Gamma$ is a Lie group with composition as the natural group action; for a function $f \in \mathcal{F}$ and a warping function $\gamma \in \Gamma$, their composition $f \circ \gamma$ denotes a warping of $f$ using $\gamma$. Warping of a function constitutes its phase variability and should thus be visualized and analyzed separately from the function's amplitude. We use $\Gamma$ to model the warping functions nonparametrically. While more complicated than the parametric case, such generality is needed to produce satisfactory results in many applications including the ones presented in the current paper.

To circumvent some of the theoretical issues related to the warping problem, such as lack of isometry under the $\ltwo$ metric (see \cite*{EFA:10}), we define a mapping $Q:{\cal F}\to\ltwo([0,1],\real)$ as (for a function $f\in\cal{F}$) $Q(f)=q=\text{sign}(\dot{f})\sqrt{|\dot{f}|}$, where $|\cdot|$ denotes the absolute value and $\dot{f}$ is the time derivative of $f$; henceforth, we simply use $\ltwo$ to represent $\ltwo([0,1],\real)$. This new representation, $q$, of a function $f$ is called it's \textit{square-root slope function} (SRSF). The mapping $f \leftrightarrow (q, f(0))$, which maps between $\mathcal{F}$ and $\mathbb{L}^2 \times \mathbb{R}$, is a bijection, which can be obtained precisely as $f(t) = f(0) + \int_0^t q(s) |q(s)| ds$. Therefore, for each function $f \in \mathcal{F}$, there exists a $q \in \mathbb{L}^2$, such that $q$ is the SRSF of $f$ (i.e., if $f$ is absolutely continuous, then its SRSF $q$ is square-integrable) (\cite*{EFA:10}; \cite{KLASSEN}). Also, note that the SRSF representation is invariant to function translations.

When a function $f$ is warped by $\gamma$, the SRSF of $f \circ \gamma$ is given by $(q, \gamma) = (q \circ \gamma) \sqrt{\dot{\gamma}}$. One of the most important properties of SRSFs, and their usefulness in separation of amplitude and phase variabilities of elastic functions, is the invariance of the $\ltwo$ distance under warping. That is, for any two functions $f_1, f_2 \in \mathcal{F}$ represented using their SRSFs $Q(f_1)=q_1$ and $Q(f_2)=q_2$ and any warping function $\gamma \in \Gamma$, we have $\|q_1 - q_2\| = \|(q_1 , \gamma) - (q_2, \gamma)\|$; \textit{this is different from standard approaches where the $\ltwo$ metric is used directly on the function space ${\cal F}$: in that case, $\|f_1-f_2\|\neq\|f_1\circ\gamma-f_2\circ\gamma\|$}. \cite*{EFA:10} showed that this property is essential for defining a proper framework to separate amplitude and phase variabilities in functional data. Their methods are attractive for our study because we aim to provide visualization techniques for each of these components individually. Furthermore, the $\ltwo$ distance is a proper metric (i.e., it satisfies symmetry, positive definiteness, and triangle inequality) on the SRSF amplitude space $\ltwo/\Gamma=\{[q]|q\in\ltwo\}$, where $[q]=\{(q,\gamma)|\gamma\in\Gamma\}$. Equivalently, the pullback of the $\ltwo$ metric to $\cal{F}$, which results in the extended Fisher--Rao metric (\cite*{EFA:10}), is a proper metric on the original function amplitude space ${\cal F}/\Gamma=\{[f]|f\in{\cal F}\}$, where $[f]=\{f\circ\gamma|\gamma\in\Gamma\}$ denote equivalence classes under the action of the warping group; thus, each equivalence class contains all possible warpings of a function $f$ (representation of function amplitude), and this quotient space contains all such equivalence classes.

While the above methodological descriptions justify our use of the elastic functional data analysis framework of \cite*{EFA:10}, we omit their full algorithm for separation of amplitude and phase variabilities for brevity. The tools provided by this framework allow us to extract different components of given functions and analyze them separately in their respective representation spaces: (1) amplitude in $\ltwo/\Gamma$ or ${\cal F}/\Gamma$, (2) warping or phase in $\Gamma$, and (3) translation in $\real$. We provide a summary of the relevant representation spaces and corresponding metrics in Table \ref{tab:repsp}. The theoretical foundations of this framework are studied in detail in \cite{KLASSEN} and establish uniqueness of the warping solution in $\Gamma$.

\begin{table}[!t]
\begin{center}
\begin{small}
\begin{tabular}{|c|c|c|c|c|c|}
    \hline
    Source of Variability&Translation&\multicolumn{2}{|c|}{Amplitude}&\multicolumn{2}{|c|}{Phase}\\
    \hline
	Space& $\real$ & ${\cal F}/\Gamma$ & $\ltwo/\Gamma$ & $\Gamma$ & $\Psi$\\
    \hline
	Metric& Euclidean & Extended Fisher--Rao & $\ltwo$ & Fisher--Rao & $\ltwo$\\
	\hline
\end{tabular}
\caption{\small Representation spaces and metrics for different sources of variability in functional data.}\label{tab:repsp}
\end{small}
\end{center}
\vspace{-7mm}
\end{table}

\begin{figure}[!t]
\begin{center}
    \begin{tabular}{|c|c|c|c|}
    \hline
    (a)&(b)&(c)&(d)\\
    \hline
    \includegraphics[width=1.3in]{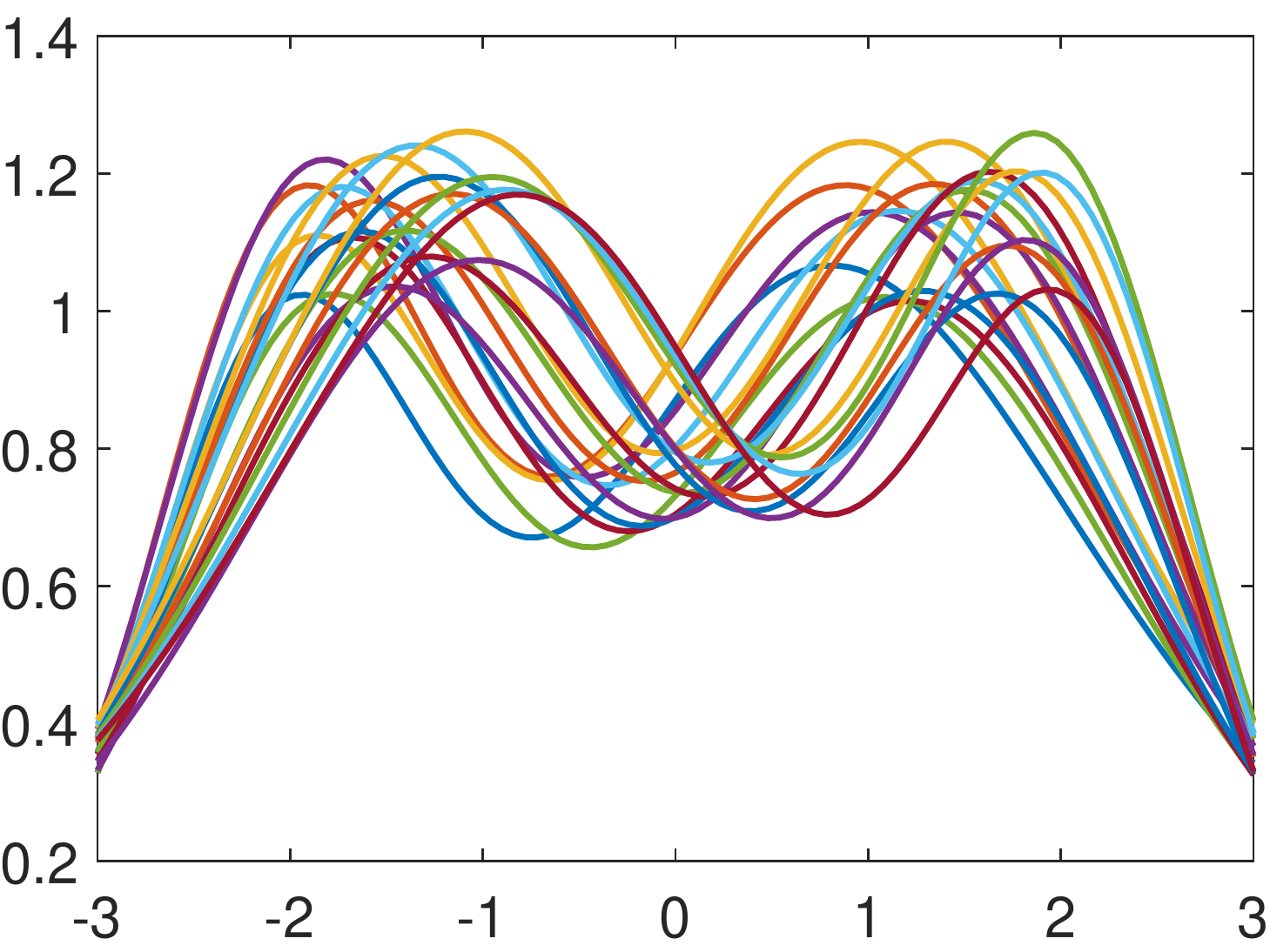}&\includegraphics[width=1.3in]{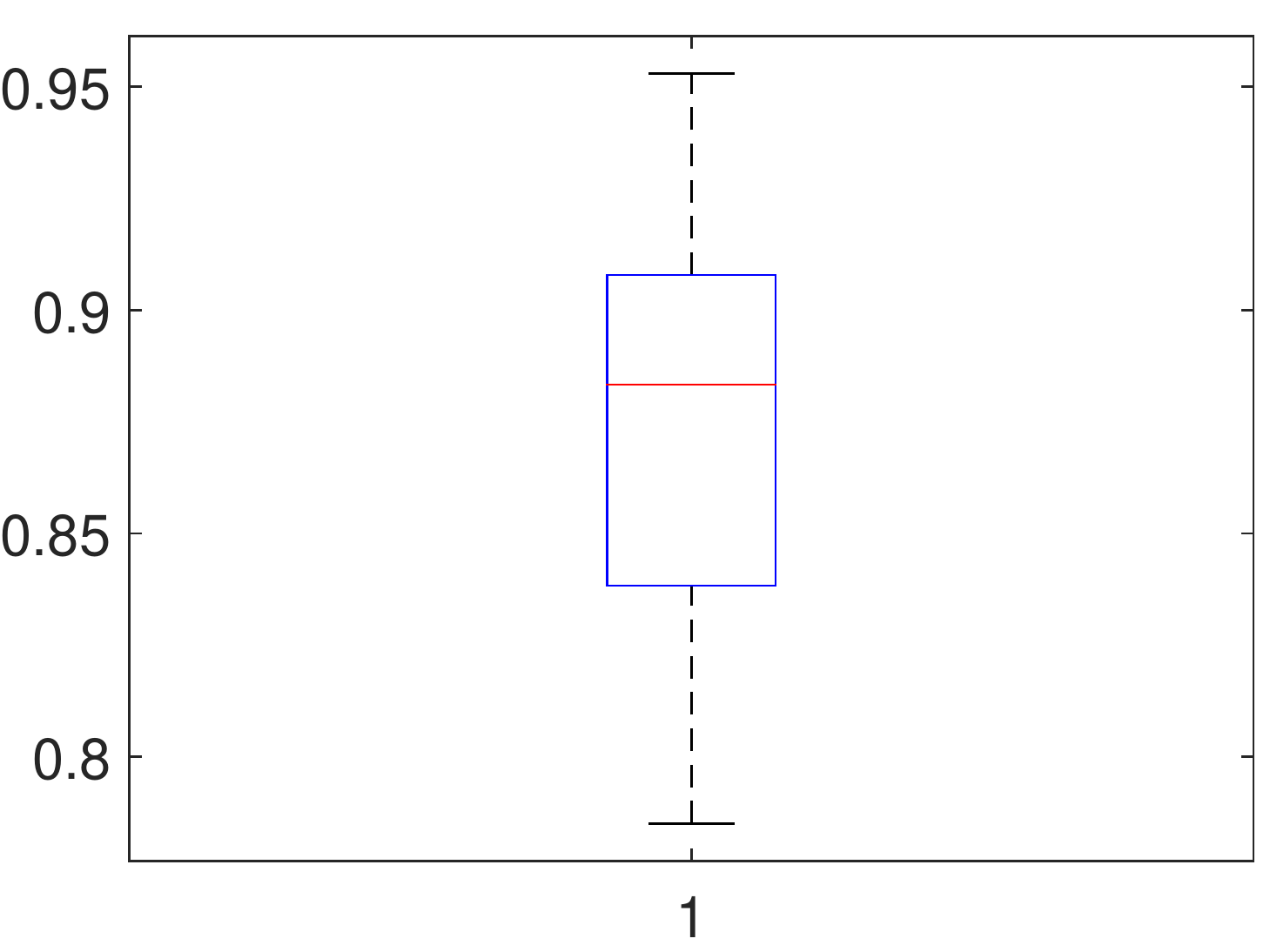}&\includegraphics[width=1.3in]{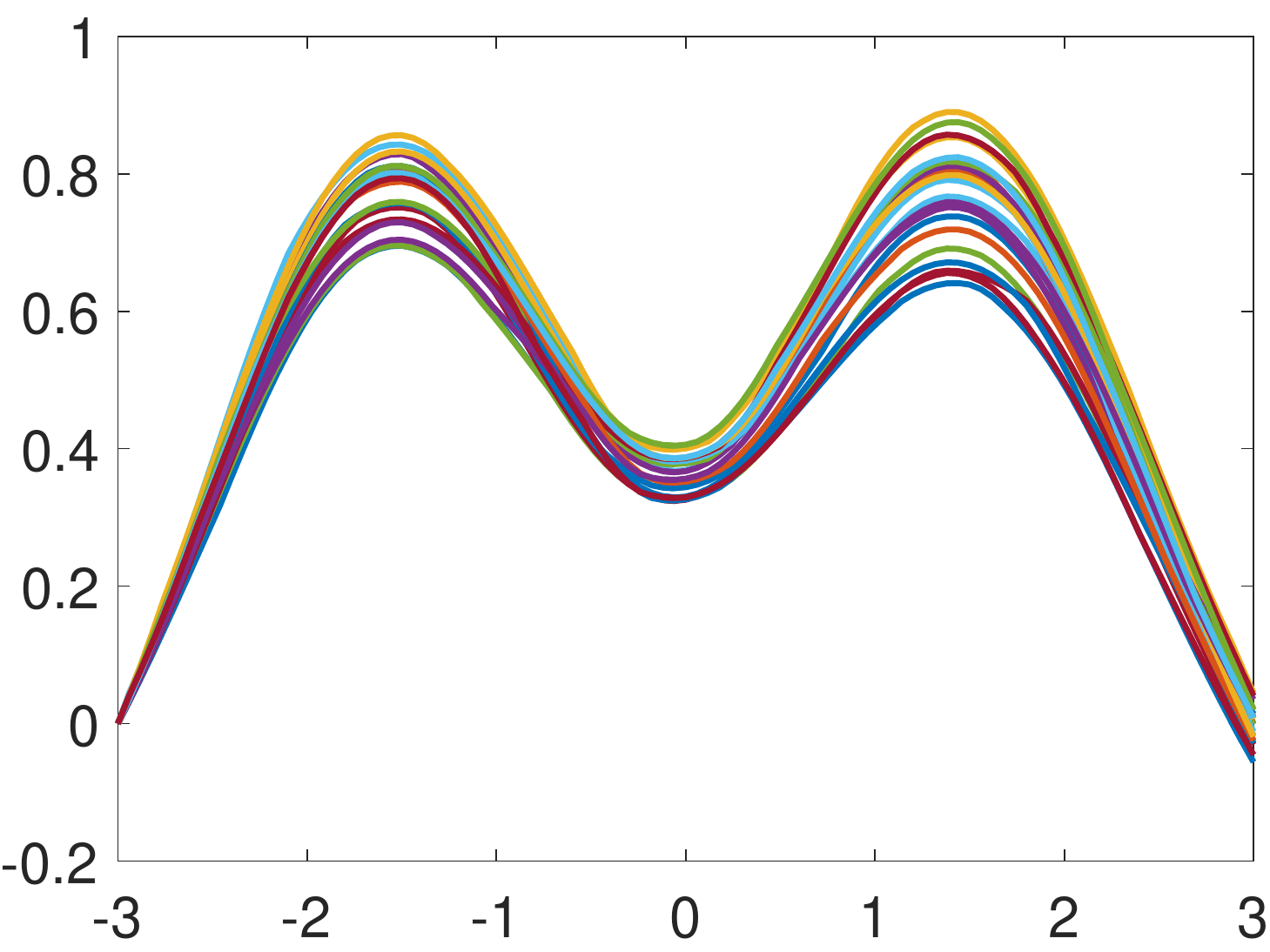}&\includegraphics[width=1.3in]{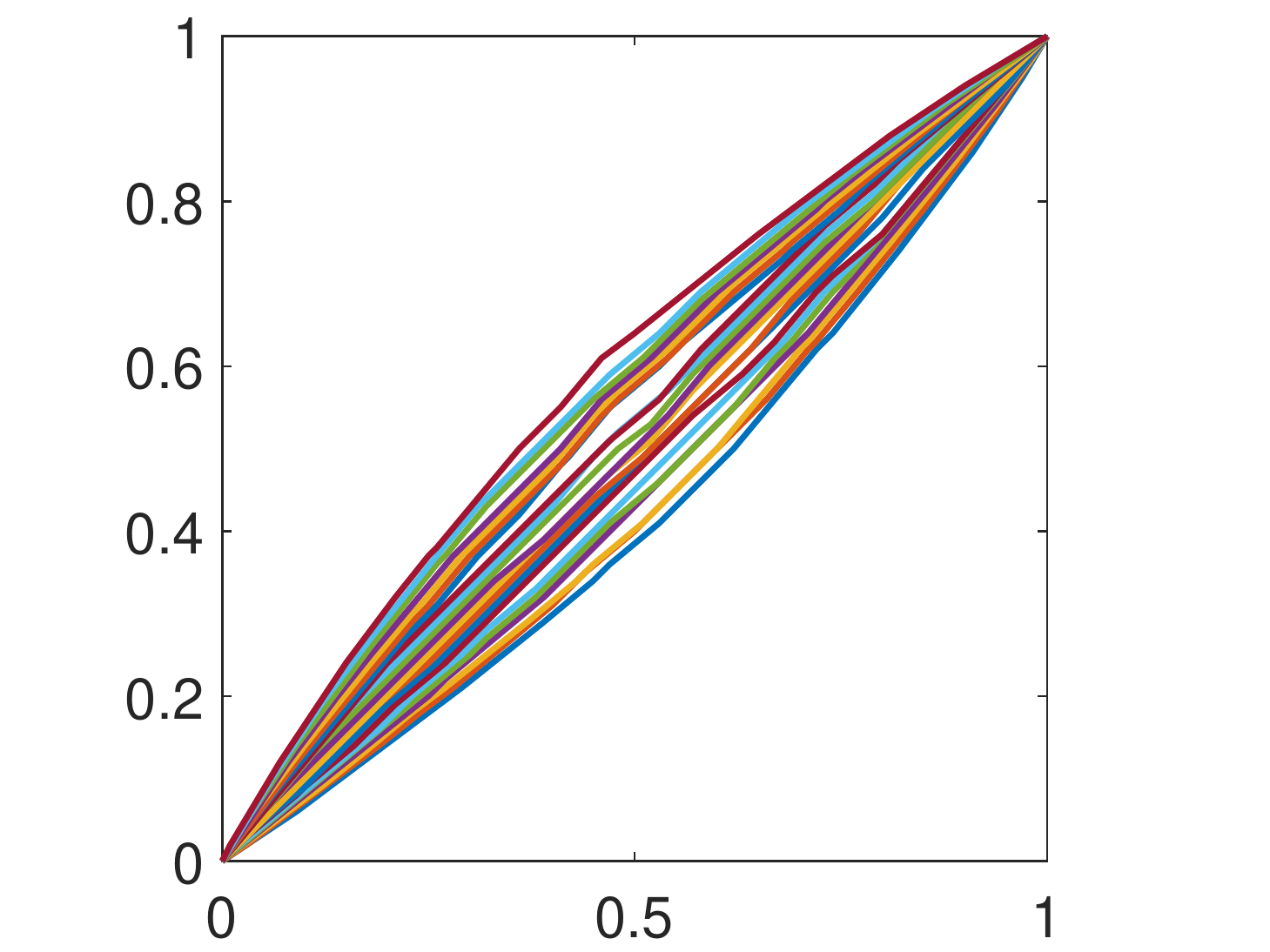}\\
    \hline
    \end{tabular}
    \caption{\small Separation of translation, amplitude, and phase variabilities in elastic functional data. (a) Original functions. (b) Translation. (c) Amplitude. (d) Phase.}\label{fig:data}
    \end{center}
\vspace{-7mm}
\end{figure}

In this paper, we approach the problem of functional data visualization by first extracting the translation component. Then, we map all given functions to their corresponding SRSFs and use the framework of \cite*{EFA:10} to extract the amplitude and phase components of the functions. Once each component has been separated out, we construct individual metric-based, geometric boxplot-type displays to more accurately visualize various sources of variability in the functional data. As an example, consider the simulated functions given in Figure \ref{fig:data}(a). These functions clearly differ in three main aspects: (1) vertical translation (Figure \ref{fig:data}(b)), (2) relative heights of peaks and valleys (Figure \ref{fig:data}(c)), and (3) relative positions of peaks and valleys (Figure \ref{fig:data}(d)). The proposed method will construct a boxplot for each source of variability separately. Such a construction is trivial for the translation component, which in our framework is defined as the average height of the function; we use standard boxplots for this case. Instead of using the average height of the functions one could also use the median or any other measure of translation. But, the construction of boxplots on the functional spaces of amplitude, ${\cal F}/\Gamma$, and phase, $\Gamma$, is much more challenging and requires geometric tools from those representation spaces.

\subsection{Construction of the Amplitude Boxplot}
\label{sec:ampbox}

The construction of a functional boxplot for the amplitude component requires the computation of the amplitude median, two quartiles, and two extremes. We use the geometry of the functional representation space to define these boxplot features. Additionally, we provide a recipe for amplitude outlier identification. Given a set of functions $\{f_1,\dots,f_n\}$, we begin by finding the amplitude geometric median defined as $[\bar{f}] = \underset{[f] \in \mathcal{F}/\Gamma}{\text{argmin}} \displaystyle \sum_{i = 1}^n D_a ([f_i],[f])$, where $D_a([f_1],[f_2]) = \underset{\gamma \in \Gamma}{\text{inf}} \|q_1 - (q_2, \gamma)\|$ is the amplitude distance (\cite{SCI:Fle2009a}). The solution to this optimization problem can be found using a standard gradient descent algorithm. There are three main advantages to using the proposed geometric median over the previously mentioned band depth methods. First, the definition of the geometric median in one-dimensional Euclidean space with the absolute value as the metric reduces to the standard definition of a median on that space (i.e., a point that splits the data into two equal-sized portions). Second, this definition allows one to automatically separate the amplitude and phase variabilities under a unified framework, which is not possible under the band depth setup. Finally, the functional band depth is based on a probability measure on the function space, which is known to have degeneracy issues (\cite{kuelbs,chack}). In contrast, our notion of a median only requires the definition of a proper metric on the function space.

The definition of the geometric median is very similar to the definition of the Frechet or Karcher mean (\cite{dryden-mardia_book:98,le-mean-shape}); however, it is more robust to outliers as shown in \cite{SCI:Fle2009a}. The amplitude median is actually an entire equivalence class of functions $[\bar{f}]$; we select one element of this equivalence class as a representative, $\bar{f}\in[\bar{f}]$, using the orbit centering method (\cite*{EFA:10}). Next, we compute $\bar{q}$, the SRSF of $\bar{f}$, and align all original functions $\{f_1, f_2, \dots, f_n\}$ to the amplitude geometric median $\bar{f}$ using $D_a$. This operation results in three pieces of information: (1) amplitude distances of all functions from the amplitude median $\{D_a^1,\dots,D_a^n\}$, (2) aligned functions or amplitude components $\{\tilde{f}_1, \dots, \tilde{f}_n\}$ and their corresponding SRSFs $\{\tilde{q}_1, \dots, \tilde{q}_n\}$, and (3) optimal warping or phase functions $\{\gamma_1, \dots, \gamma_n\}$.

We use the computed amplitude distances to order the corresponding amplitude components $\{\tilde{q}_1,\dots,\tilde{q}_n\}$ according to their proximity in $\ltwo/\Gamma$ to the amplitude geometric median. Then, we extract the $50\%$ of amplitude functions that are closest to $\bar{q}$, resulting in the ordered amplitude functions $\{\tilde{q}_{(1)},\dots,\tilde{q}_{(\lfloor n/2\rfloor)}\}$; these functions define the $50\%$ central amplitude region of $\ltwo/\Gamma$. To define the two amplitude quartiles, $\tilde{q}_{Q_1}$ ($\tilde{f}_{Q_1}$) and $\tilde{q}_{Q_3}$ ($\tilde{f}_{Q_3}$), we solve the following optimization problem over the $50\%$ central amplitude region:
\begin{small}
\begin{eqnarray}
\nonumber(\tilde{q}_{Q_1},\tilde{q}_{Q_3})&=&\underset{\tilde{q}_1,\tilde{q}_2\in\{\tilde{q}_{(1)},\dots,\tilde{q}_{(\lfloor n/2\rfloor)}\}}{\text{argmax}} (1-\lambda)\left(\frac{\|\tilde{q}_1 - \bar{q}\|}{\underset{i}{\text{max}} \|\tilde{q}_{(i)} - \bar{q}\|} + \frac{\|\tilde{q}_2 - \bar{q}\|}{\underset{i}{\text{max}} \|\tilde{q}_{(i)} - \bar{q}\|}\right)\\ &-& \lambda \left(\left\langle\frac{\tilde{q}_1 - \bar{q}}{\|\tilde{q}_1 - \bar{q}\|}, \frac{\tilde{q}_2 - \bar{q}}{\|\tilde{q}_2 - \bar{q}\|}\right\rangle + 1\right),\label{eq:quartiles}
\end{eqnarray}
\end{small}where $\|\cdot\|$ and $\langle\cdot,\cdot\rangle$ denote the $\ltwo$ norm and inner product, respectively. The intuition behind this approach is as follows. The first term in the expression measures the cumulative distance between the two quartiles normalized using the maximum distance from the amplitude median to any of the amplitude functions within the $50\%$ central region; we want this quantity to be as large as possible to ensure that the two quartiles are far away from the geometric median. The maximum value of this first term is $2$. The second term in this expression measures the inner product between the unit vectors pointing from the amplitude median to each of the amplitude functions in the $50\%$ central region; we want this term to be as small as possible (i.e., $-1$) to maximize the angle between the two chosen vectors. In this way, we prefer the two vectors which point in opposite directions from the amplitude median. The inner product between two unit vectors has a minimum value of $-1$ and a maximum value of $1$, so we add $1$ to put the second term on the same scale as the first term. The tuning parameter $\lambda \in [0,1]$ controls the weight of the two terms; we use $\lambda=0.5$ to achieve a balance between the two terms. Note that the chosen directions for the first and third quartiles are quite different from the projection-based principal quantile direction defined in \cite{Fraiman:2012:QFI:2170034.2170095}.

Given the two quartiles $\tilde{q}_{Q_1}$ and $\tilde{q}_{Q_3}$, we define the amplitude interquartile range (IQR) as the sum of the amplitude distances from each quartile to the geometric median: $IQR_a = \|\tilde{q}_{Q_1}-\bar{q}\| + \|\tilde{q}_{Q_3}-\bar{q}\|$. Then, the two amplitude outlier cutoffs can be defined as $\tilde{q}_{W_1} =\tilde{q}_{Q_1} + k_a \times IQR_a \times \frac{\tilde{q}_{Q_1}-\bar{q}}{\|\tilde{q}_{Q_1}-\bar{q}\|}$ and $\tilde{q}_{W_3} = \widetilde{q}_{Q_3} + k_a \times IQR_a \times \frac{\tilde{q}_{Q_3}-\bar{q}}{\|\tilde{q}_{Q_3}-\bar{q}\|}$. This definition is similar to the standard boxplot definition where $k_a=1.5$. In the case of amplitude functions, the choice of $k_a$ is not as obvious and we will study the behavior of outlier detection with respect to this constant in later sections. An amplitude outlier is then defined as any amplitude function $\tilde{f}$ whose $\tilde{q}$ is further away from the geometric median $\bar{q}$ than the larger of the two amplitude outlier cutoffs; that is, $\tilde{f}$ is identified as an amplitude outlier if $\|\tilde{q}-\bar{q}\|> \max\{\|\tilde{q}_{W_1}-\bar{q}\|,\|\tilde{q}_{W_3}-\bar{q}\|\}$. If one would like to be less conservative, the minimum can be used instead of the maximum. Thus, the proposed outlier detection procedure considers an entire region with a specific radius governed by the distance of the amplitude outlier cutoffs from the amplitude median. This is in stark contrast to projection-based approaches where one is able to identify outliers in specific directions only. Finally, the two extreme amplitude functions are defined as those $\tilde{f}$ whose $\tilde{q}$ are closest to each of the two amplitude outlier cutoffs $\tilde{q}_{W_1}$ and $\tilde{q}_{W_3}$, under the requirement that they lie outside of the $50\%$ central amplitude region and have not been flagged as amplitude outliers.

\begin{figure}[!t]
\begin{center}
    \begin{tabular}{|c|c|c|c|}
    \hline
    (a)&(b)&(c)&(d)\\
    \hline
    \includegraphics[width=1.3in]{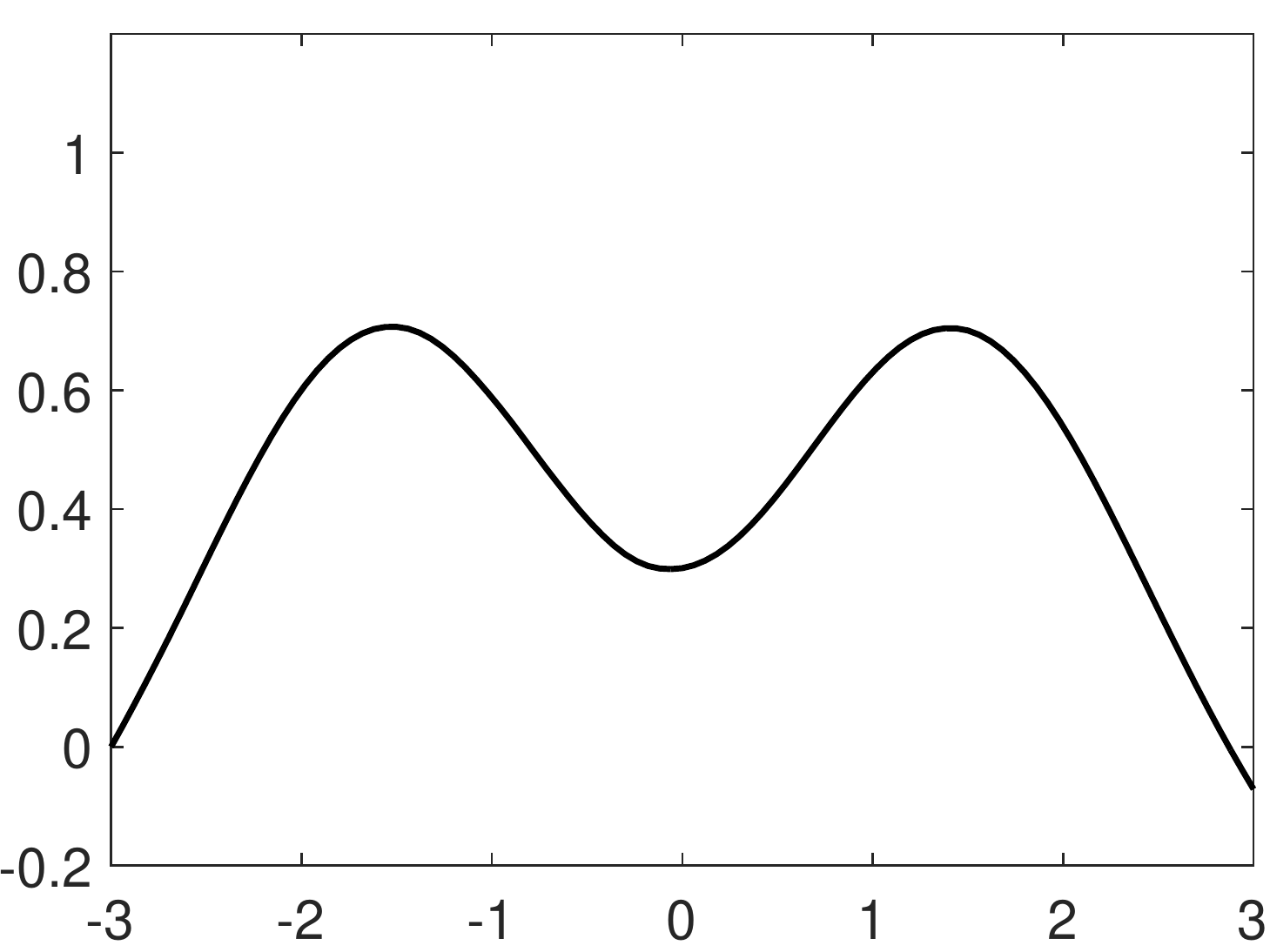}&\includegraphics[width=1.3in]{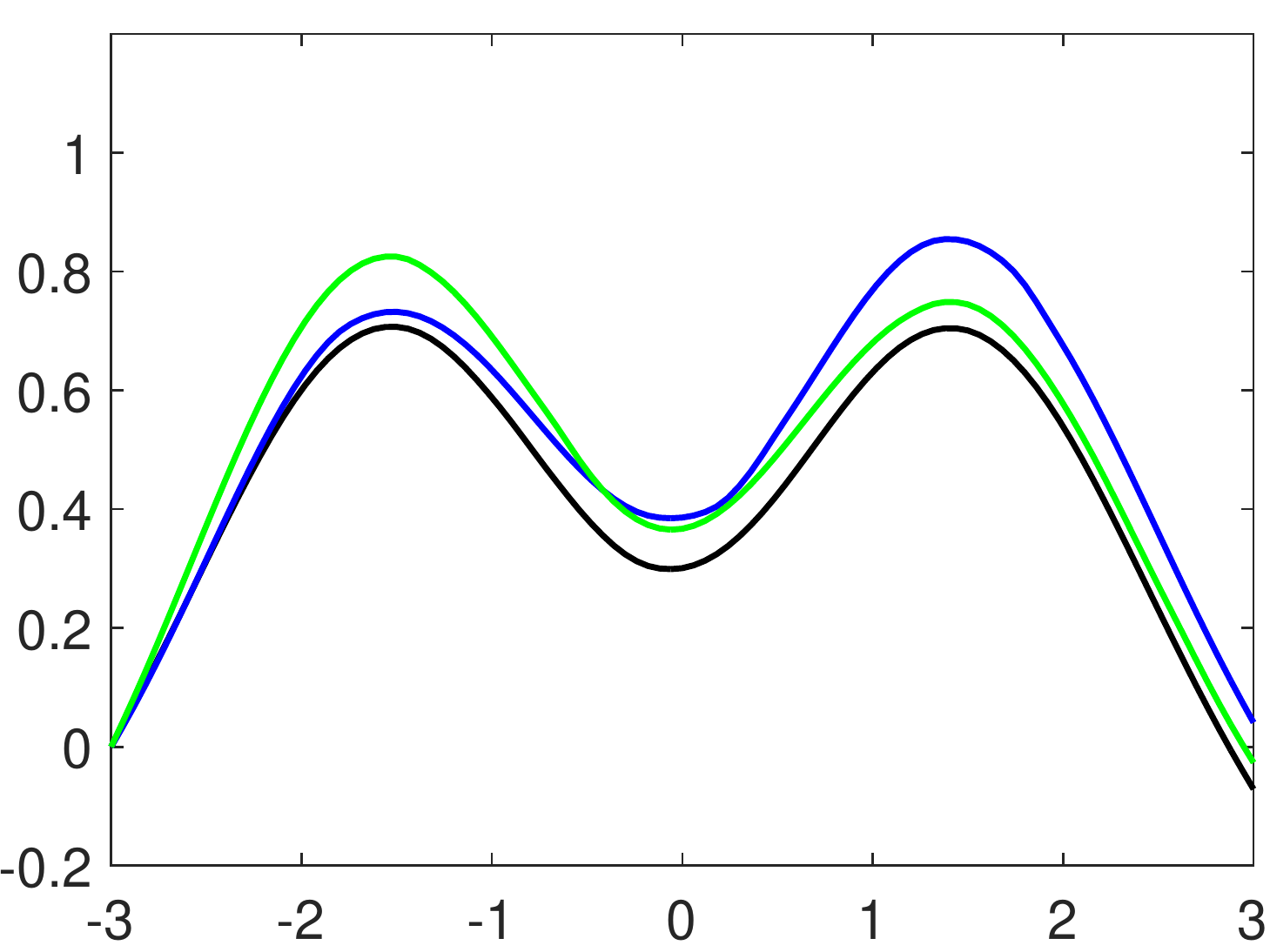}&\includegraphics[width=1.3in]{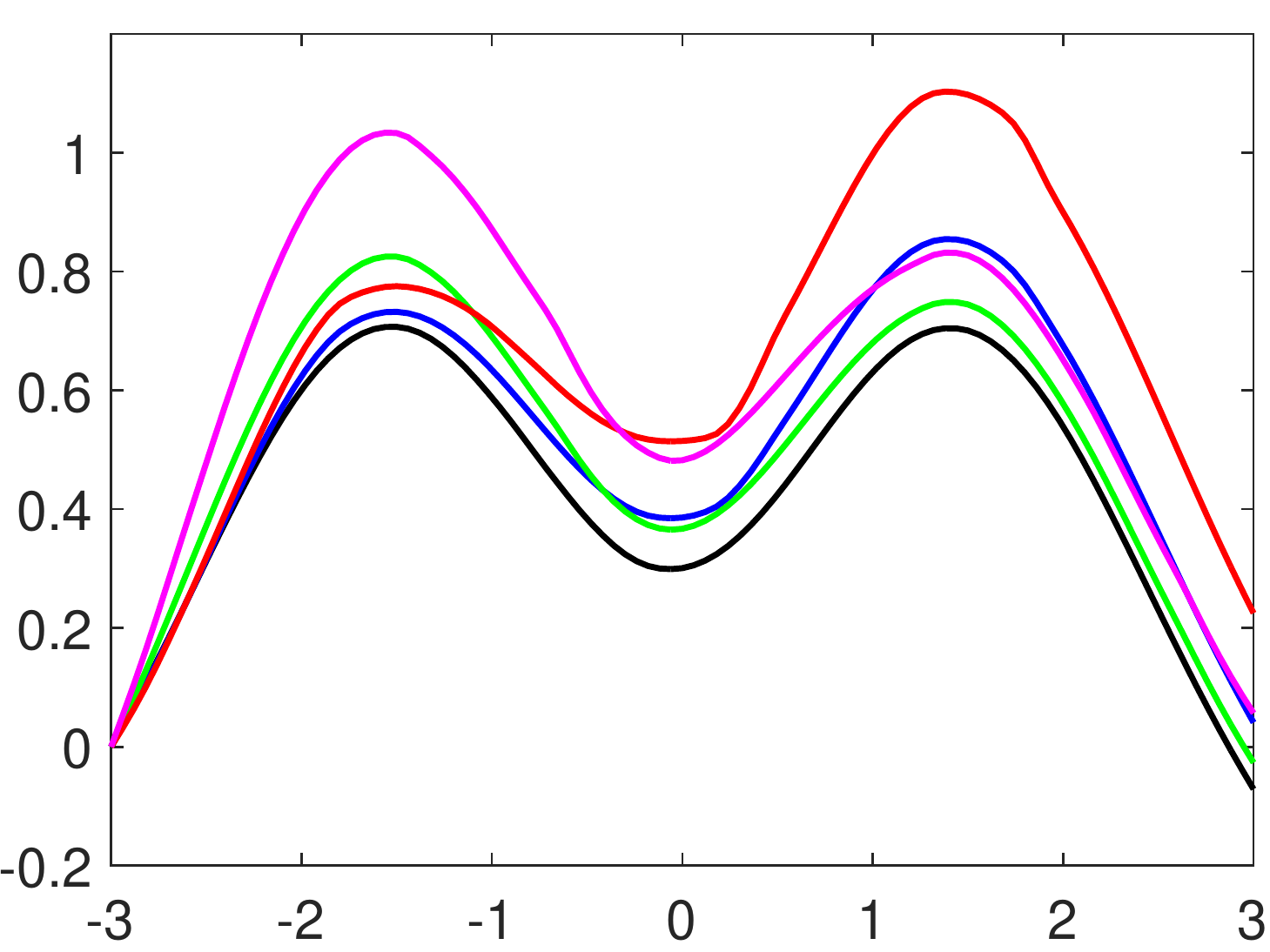}&\includegraphics[width=1.3in]{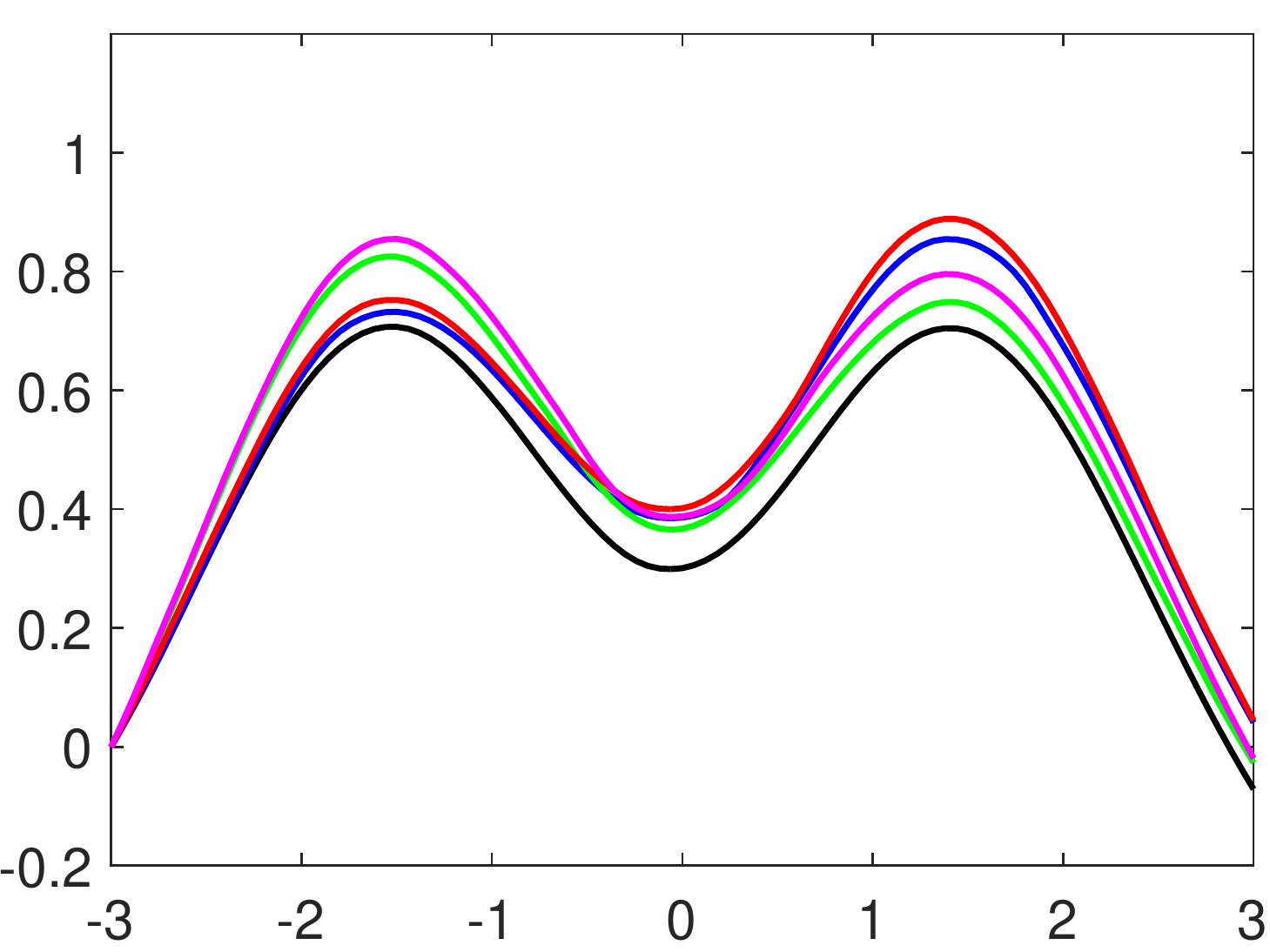}\\
    \hline
    \end{tabular}
    \caption{\small Construction of the amplitude boxplot for the data given in Figure \ref{fig:data}. (a) Amplitude median. (b) Amplitude median with quartiles $\tilde{f}_{Q_1}$ (blue) and $\tilde{f}_{Q_3}$ (green). (c) Amplitude median with quartiles $\tilde{f}_{Q_1}$ and $\tilde{f}_{Q_3}$, and outlier cutoffs $\tilde{f}_{W_1}$ (red) and $\tilde{f}_{W_3}$ (magenta). (d) Full amplitude boxplot (same as (c) except outlier cutoffs have been replaced by extreme amplitude functions).}\label{fig:ampboxplot}
    \end{center}
\vspace{-7mm}
\end{figure}

Figure \ref{fig:ampboxplot} provides an example of constructing an amplitude boxplot for the functional data shown in Figure \ref{fig:data}. In panel (a), we show the amplitude median in black, which is a nice representative of the original data. Then, we compute the two amplitude quartiles using Equation \ref{eq:quartiles} displayed in panel (b). The amplitude boxplot shown is invariant to function translations, and thus, one should only be concerned with `shape' differences of these summaries. In fact, the two quartiles capture a very intuitive source of variation in the given data. The median has two peaks of approximately equal size. The first quartile (blue) has a similarly-sized first peak but a higher second peak, while the third quartile has a similarly-sized second peak but a higher first peak. The two outlier cutoffs shown in panel (c) in red and magenta signify amplitudes which are used for outlier detection. As expected, no amplitude outliers were found in this simulated example. Finally, in panel (d), we display the full amplitude boxplot with the two extremes shown in red and magenta.

\subsection{Construction of the Phase Boxplot}
\label{sec:phbox}

The construction of the amplitude boxplot was aided by the linear geometry of the SRSF functional space. Thus, we were able to add and subtract amplitude functions without leaving the relevant representation space. But, the representation space of phase variability is $\Gamma$, which is a nonlinear manifold with a nontrivial geometry. Luckily, a simple transformation of the phase functions, similar to the SRSF, is able to greatly simplify the Riemannian geometry of $\Gamma$.

For a phase function $\gamma\in\Gamma$, define its \textit{square-root transform} (SRT) representation as $\psi = \sqrt{\dot\gamma}$, where $\dot\gamma$ denotes the time derivative of $\gamma$. The phase function $\gamma$ can be recovered from $\psi$ using $\gamma(t) = \int_0^t \psi^2 (s) ds$. This representation of phase variability has two very important properties. First, notice that $\psi(t)>0$, for all $t$, and $\|\psi\| = 1$. Therefore, all SRTs lie on the positive orthant of the unit Hilbert sphere denoted by $\Psi$, which has a well known geometry. Second, a complicated Fisher--Rao Riemannian metric on $\Gamma$ simplifies to the standard $\ltwo$ metric on $\Psi$ (\cite{srivastava-etal-Fisher-Rao-CVPR:2007}) (see Table \ref{tab:repsp}). These two results imply that standard tools from Riemannian geometry are available analytically and can thus be used to construct functional boxplots on $\Gamma$. We begin by defining all required tools and then show how they can be applied to the present problem.

Define the distance between two warping functions, $\gamma_1$ and $\gamma_2$, as the arc-length between their corresponding SRTs, $\psi_1$ and $\psi_2$: $D_p(\gamma_1,\gamma_2) = \cos^{-1} (\langle\psi_1, \psi_2\rangle)$, where $\langle\cdot,\cdot\rangle$ is the $\ltwo$ inner product. The tangent space at any point $\psi\in\Psi$ is defined as $T_\psi(\Psi)=\{v:[0,1]\to\real|\langle v,\psi\rangle=0\}$; it is a linear space. Thus, our general approach for constructing phase boxplots will be to (1) compute the phase median, (2) map all phase functions to the tangent space defined at the phase median, (3) construct the phase boxplot in the tangent space, and (4) map the boxplot to the original representation space $\Psi$ (and $\Gamma$). To do this we need additional geometric tools called the exponential and inverse-exponential maps. The exponential map at a point $\psi_1 \in \Psi$, denoted by $\exp_{\psi_1}: T_{\psi_1}(\Psi) \mapsto \Psi$, is defined as (for $v\in T_{\psi_1}(\Psi)$) $\exp_{\psi_1}(v) = \cos(\|v\|)\psi_1+ \sin(\|v\|)\frac{v}{\|v\|}$; $\exp_{\psi_1}$ maps points from the tangent space at $\psi_1$ to the representation space. For $\psi_1,\ \psi_2\in \Psi$, the inverse-exponential map, denoted by $\exp^{-1}_{\psi_1}: \Psi \mapsto T_{\psi_1}(\Psi)$, is given by $\exp^{-1}_{\psi_1}(\psi_2) = \frac{\theta}{\sin(\theta)}\left(\psi_2 - \cos(\theta) \psi_1\right)$ where $\theta=\cos^{-1} (\langle\psi_1, \psi_2\rangle)$; $\exp^{-1}_{\psi_1}$ maps points from the representation space to the tangent space at $\psi_1$.

We begin by finding the geometric phase median in the SRT representation space: $\bar{\psi} = \underset{\psi \in \Psi}{\text{argmin}} \displaystyle \sum_{i = 1}^n D_p (\gamma_i, \gamma) = \underset{\psi \in \Psi}{\text{argmin}} \displaystyle \sum_{i = 1}^n \cos^{-1} (\langle\psi_i, \psi\rangle)$. The phase median $\bar{\psi}$ can be converted to $\bar{\gamma}$ using the inverse mapping defined earlier. To find the two phase quartiles, we map all $\psi_i$s to the tangent space at the phase median using the inverse-exponential map: $v_i = \exp^{-1}_{\bar{\psi}}(\psi_i)$. Next, we order the phase functions $\{\psi_1,\dots, \psi_n\}$ based on the phase distance $D_p^i = \cos^{-1} (\langle\bar{\psi}, \psi_i\rangle)\approx \|v_i\|,\ i=1,\dots,n$. We extract the $50\%$ of phase functions that are closest to $\bar{\psi}$ resulting in the ordered phase functions $\{\psi_{(1)},\dots,\psi_{(\lfloor n/2\rfloor)}\}$ and corresponding tangent vectors $\{v_{(1)},\dots,v_{(\lfloor n/2\rfloor)}\}$; these functions define the $50\%$ central phase region of $\Psi$. To define the two phase quartiles, $\psi_{Q_1}$ ($\gamma_{Q_1}$) and $\psi_{Q_3}$ ($\gamma_{Q_3}$), we solve the following optimization problem over the $50\%$ central phase region:
\begin{small}
\begin{eqnarray}
\nonumber (\psi_{Q_1},\psi_{Q_3})&=&\underset{\psi_1,\psi_2\in\{\psi_{(1)},\dots,\psi_{\lfloor n/2\rfloor}\}}{\text{argmax}} (1-\lambda)\left(\frac{\|v_1\|}{\underset{i}{\text{max}} \|v_{(i)}\|} + \frac{\|v_2\|}{\underset{i}{\text{max}} \|v_{(i)}\|}\right)\\ &-& \lambda \left(\left\langle\frac{v_1}{\|v_1\|}, \frac{v_2}{\|v_2\|}\right\rangle + 1\right),\label{eq:quartilesph}
\end{eqnarray}
\end{small}and identify the corresponding warping functions, $\gamma_{Q_1}$ and $\gamma_{Q_3}$, as the two phase quartiles. The interpretation of this optimization problem is the same as we explained for amplitude.

\begin{figure}[!t]
\begin{center}
    \begin{tabular}{|c|c|c|c|}
    \hline
    (a)&(b)&(c)&(d)\\
    \hline
    \includegraphics[width=1.3in]{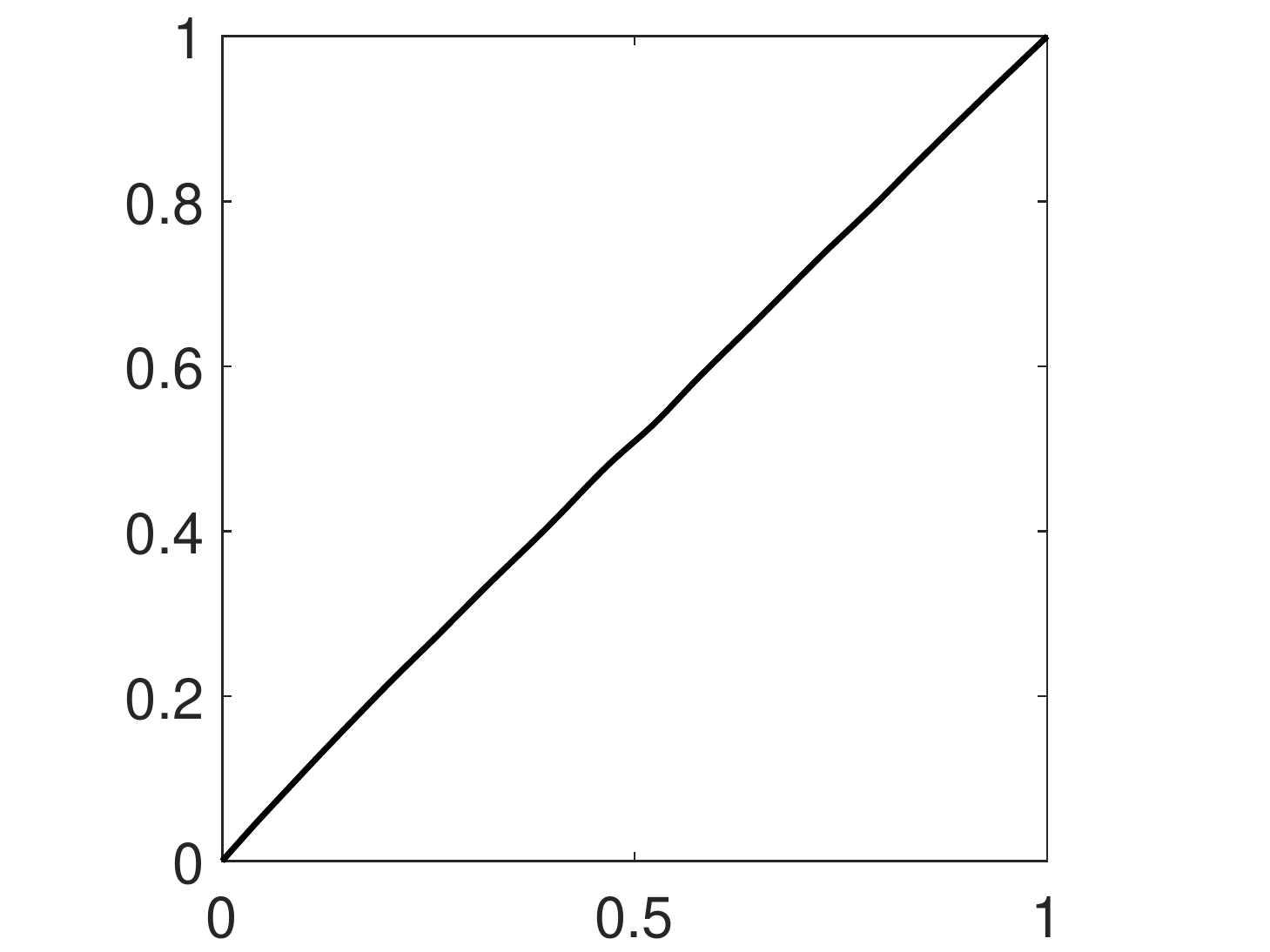}&\includegraphics[width=1.3in]{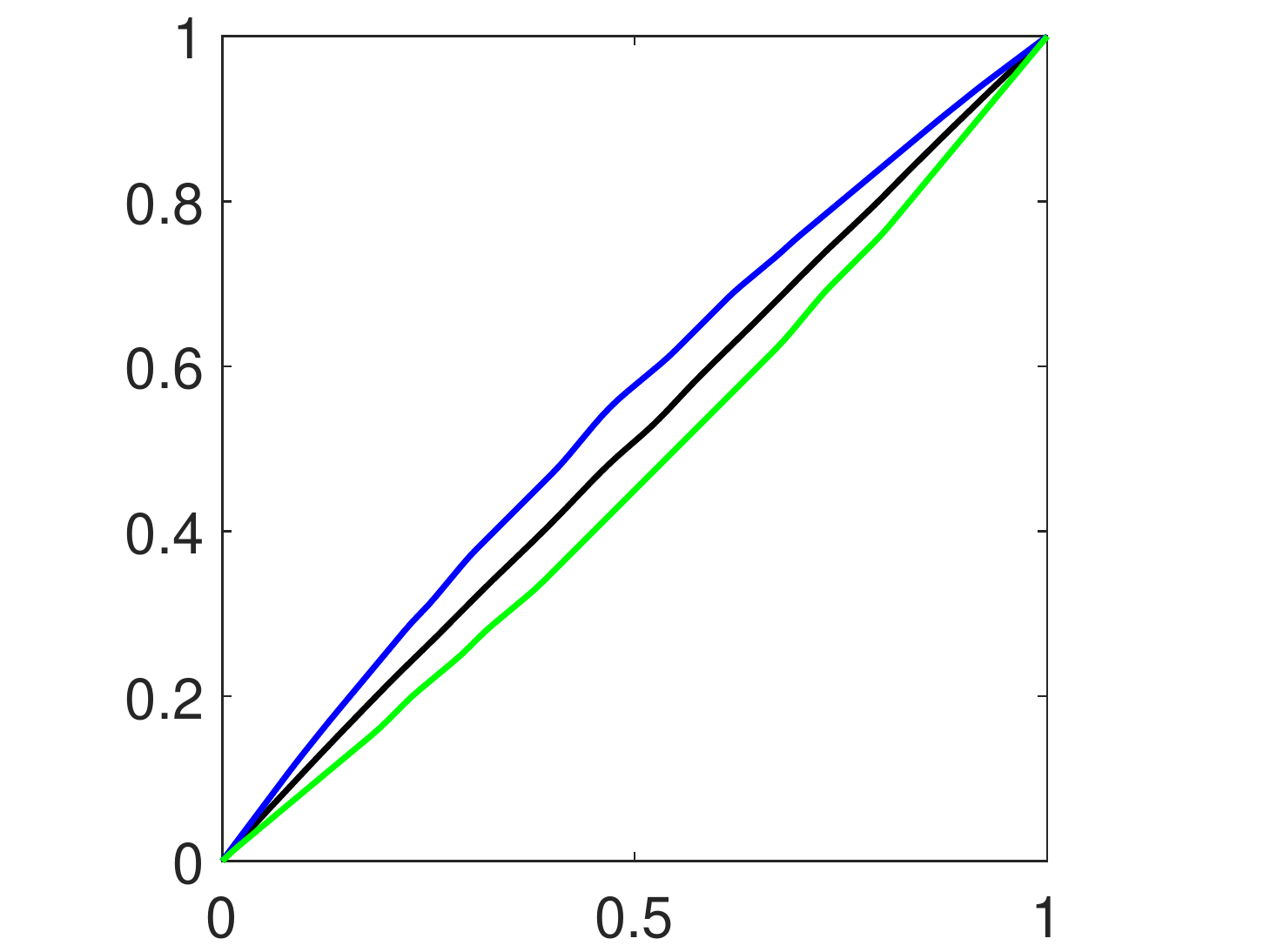}&\includegraphics[width=1.3in]{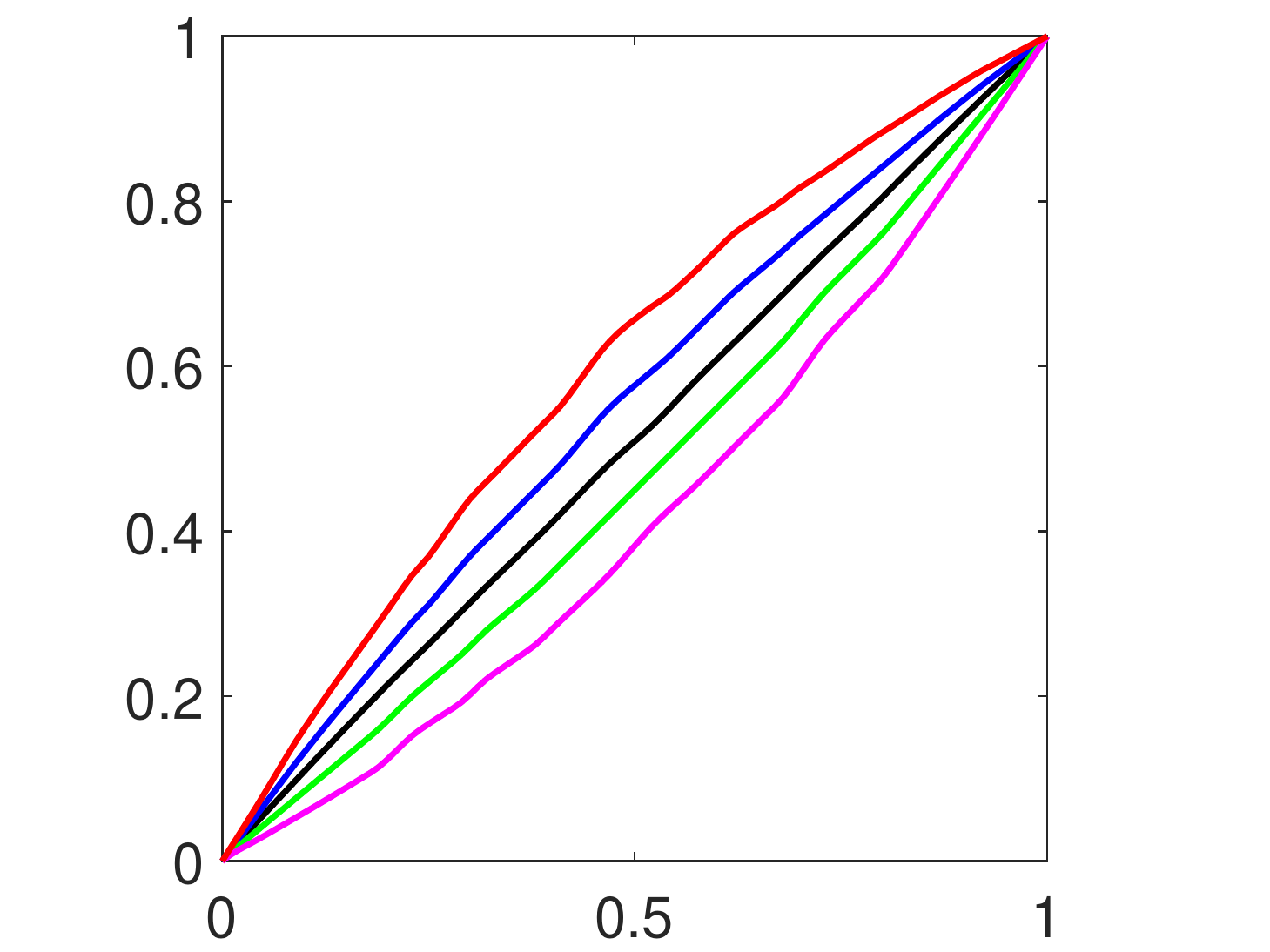}&\includegraphics[width=1.3in]{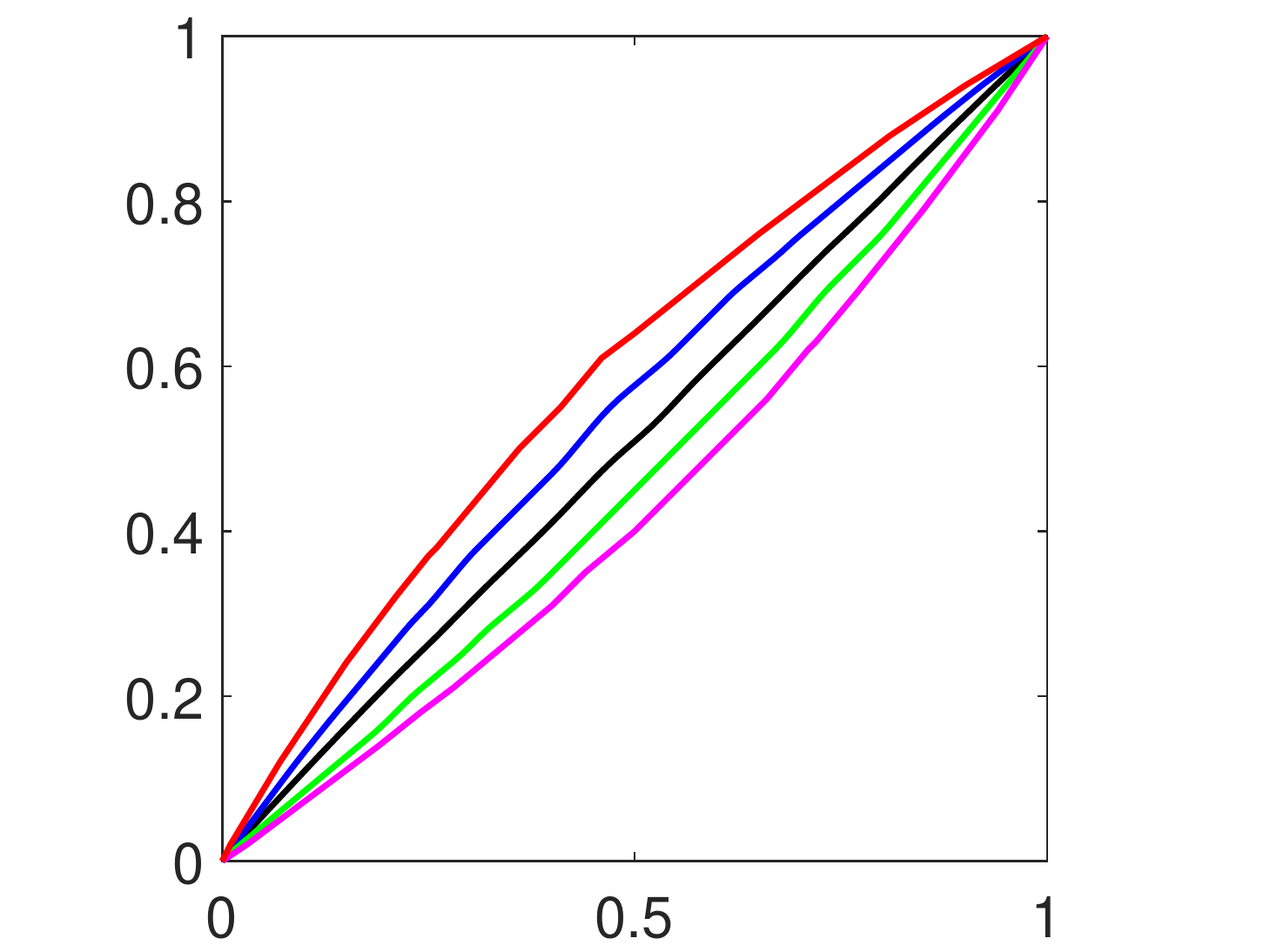}\\
    \hline
    \end{tabular}
    \caption{\small Construction of the phase boxplot for the data given in Figure \ref{fig:data}. (a) Phase median. (b) Phase median with quartiles $\gamma_{Q_1}$ (blue) and $\gamma_{Q_3}$ (green). (c) Phase median with quartiles $\gamma_{Q_1}$ and $\gamma_{Q_3}$, and outlier cutoffs $\gamma_{W_1}$ (red) and $\gamma_{W_3}$ (magenta). (d) Full phase boxplot (same as (c) except outlier cutoffs have been replaced by extreme phase functions).}\label{fig:phaseboxplot}
    \end{center}
\vspace{-7mm}
\end{figure}

Given the two quartiles, we compute their respective tangent space representations (i.e., $v_{Q_1}=\exp^{-1}_{\bar{\psi}}(\psi_{Q_1})$ and $v_{Q_3}=\exp^{-1}_{\bar{\psi}}(\psi_{Q_3})$). Then, we define the phase IQR as $IQR_p = \|v_{Q_1}\| + \|v_{Q_3}\|$, and the two phase outlier cutoffs as $\psi_{W_1} = \exp_{\bar{\psi}}(v_{Q_1} + k_p \times IQR_p \times \frac{v_{Q_1}}{\|v_{Q_1}\|})$ and $\psi_{W_3} = \exp_{\bar{\psi}}(v_{Q_3} + k_p \times IQR_p \times \frac{v_{Q_3}}{\|v_{Q_3}\|})$. As in the case of the amplitude boxplot, the choice of $k_p$ is not trivial and will be discussed in more detail in the experimental results section of the paper. Phase outliers are defined as any phase function $\gamma$ whose $\psi$ has a larger phase distance to the geometric median $\bar{\psi}$ than the larger of the two phase outlier cutoffs; that is, $\gamma$ is identified as an outlier if $D_p(\bar{\gamma},\gamma)> \max\{D_p(\bar{\gamma},\gamma_{W_1}),D_p(\bar{\gamma},\gamma_{W_3})\}$. Finally, the two extreme phase functions are defined as those whose SRT representations are closest to each of the two phase outlier cutoffs $\psi_{W_1}$ and $\psi_{W_3}$, under the requirement that they lie outside of the $50\%$ central phase region and have not been flagged as phase outliers.

Figure \ref{fig:phaseboxplot} provides an example of constructing a phase boxplot for the functional data shown in Figure \ref{fig:data}. In panel (a), we show the phase median in black, which is very similar to the identity element of the warping group ($\gamma_{id}(t)=t$). This is expected as it is due to the orbit-centering step during computation of the amplitude median. Then, we compute the two phase quartiles using Equation \ref{eq:quartilesph} displayed in panel (b). The two quartiles capture a very intuitive source of phase variation in the given data. The median essentially represents a constant speed traversal of the median amplitude function. The first quartile (blue) represents a phase function that goes faster over the entire time interval; the third quartile (green) represents the opposite effect. This type of phase variability corresponds to the peaks in the original functions occuring either later or earlier than the median time. The two outlier cutoffs are shown in panel (c) in red and magenta; in panel (d), we display the full phase boxplot with the two extremes shown in red and magenta. Phase displays throughout this paper normalize the domain of the warping functions to $[0,1]$ for improved display. The invariance/equivariance properties of the amplitude and phase boxplot displays are given in the Supplementary Material.

\section{Simulations and Applications to Real Data}
\label{sec:results}

In this section, we present the results of multiple simulations, where we compare the proposed outlier detection method to the state-of-the-art method by \cite{citeulike:10107686}. We feel that this is the most closely related approach to our method, and thus, use similar simulation settings in the current paper. We also study optimal choices of the outlier cutoff constants $k_a$ and $k_p$. As mentioned in Sections \ref{sec:ampbox} and \ref{sec:phbox}, the choice of these constants for outlier detection is not trivial. Thus, we have compiled a set of simulation results to determine approximate scales for $k_a$ and $k_p$, which provide a definition of mild, regular, and severe amplitude and phase outliers. Another approach would be to derive the empirical distributions of $k_a$ and $k_p$ to match the outlier probabilities in the univariate case, which can then be used to find appropriate values for various applications. Such a study is beyond the scope of this paper and we leave it for future work. Finally, we provide visualization and outlier detection results on three real datasets: annual sea surface temperature, Berkeley growth data, and PQRST complexes extracted from electrocardiogram signals. In each example, we emphasize the effectiveness of the proposed approach to visualize each component of variability in the given data: translation, amplitude, and phase.

\subsection{Simulations}
\label{sec:simu}

Simulations 1-4 are designed similarly to Model 6 in \cite{citeulike:10107686}. The data is generated as follows for each of the four simulations:\\
\vspace{-3mm}{\bf Simulation 1 Data:} We generate 100 functions of the following form:
\begin{small}
$$f_i(t) =
\begin{cases}
a_{1i} \sin(2\pi t) + a_{2i} \cos(2\pi t) \text{ with probability 0.9}\\
b_{1i} \sin(2\pi t) + b_{2i} \cos(2\pi t) \text{ with probability 0.1}
\end{cases},\ i=1,\dots,100,$$\end{small}where $t \in [0,1]$, $a_{1i}$ and $a_{2i}$ follow a uniform distribution on $(0,0.05)$, $U(0,0.05)$, and $b_{1i}$ and $b_{2i}$ follow a $U(0.1,0.15)$. We introduce approximately $10\%$ of severe amplitude outliers into the dataset. This dataset is the same as Model 6 in \cite{citeulike:10107686}; while their approach does not differentiate between amplitude and phase components of functional data, the proposed outlier detection approach considers the two components separately.\\
{\bf Simulation 2 Data:} Simulation 2 is very similar to Simulation 1, except that we add additional phase variability to each of the simulated functions. Thus, we first generate 100 functions of the following form:
\begin{small}
$$f_i (t) =
\begin{cases}
a_{1i} \sin(2\pi t) + a_{2i} \cos(2\pi t) \text{ with probability 0.9}\\
b_{1i} \sin(2\pi t) + b_{2i} \cos(2\pi t) \text{ with probability 0.1}
\end{cases},\ i=1,\dots,100,$$\end{small}where $t \in [0,1]$, $a_{1i}$ and $a_{2i}$ follow a $U(0,0.05)$, and $b_{1i}$ and $b_{2i}$ follow a $U(0.1,0.15)$. Thus, we still introduce $10\%$ of severe amplitude outliers into the dataset. But, in this simulation, we apply an additional random warping of the form $\gamma_i(t) = t + \alpha_i t (t - 1),\ i=1,\dots,100,\ t\in [0,1]$, with $\alpha_i$ coming from a $U(-1, 1)$, to each of the functions $f_i$. This allows us to test the robustness of the proposed method for detection of amplitude outliers under additional phase variability.\\
{\bf Simulation 3 Data:} Here, we focus our attention on the detection of phase outliers and the related behavior of $k_p$.  To perform our study, we first simulate a random function $f(t) = b_1 \sin(2\pi t) + b_2 \cos(2\pi t)$, where $t\in [0,1]$, and $b_1,\ b_2$ are randomly chosen from a $U(0.1,0.15)$. Then, we simulate 100 random warping functions of the following form:
\begin{small}
$$\gamma_i (t) =
\begin{cases}
t + \alpha_{1i} t (t - 1), \text{ with probability 0.9} \\
t + \alpha_{2i} t (t - 1), \text{ with probability 0.1}
\end{cases},\ i=1,\dots,100,$$\end{small}where $t \in [0,1]$, $a_{1i}$ follows a $U(-0.6, 0.6)$, and $a_{2i}$ follows a $U(0.9, 1)$. We apply these warping functions to $f$ to result in the simulated dataset: $f_i(t)=f(\gamma_i(t)),\ i=1,\dots,100$. Thus, we have introduced approximately $10\%$ of phase outliers in this data.\\
{\bf Simulation 4 Data:} In the final simulation study, we want to emphasize the benefits of the proposed method for outlier detection. In particular, we show that in the presence of significant warping variability, the method of \cite{citeulike:10107686} is prone to detecting false outliers. Thus, we simulate a dataset that does not contain any amplitude or phase outliers in the following manner. We simulate 100 functions as $\tilde{f}_i(t) = b_{1i} \sin(2\pi t) + b_{2i} \cos(2\pi t),\ i=1,\dots,100$, where $t\in[0,1]$, and $b_{1i}$ and $b_{2i}$ follow a $U(0.1,0.11)$. Then, to each function $\tilde{f}_i$, we apply a random warping function $\gamma_i(t) = t + \alpha_i t (t - 1),\ i=1,\dots,100$, where $t\in[0,1]$ and $\alpha_i$ follow a $U(-1, 1)$ (i.e., $f_i(t)=\tilde{f}_i(\gamma_i(t))$).

\begin{figure}[!t]
\begin{small}
\begin{center}
    \begin{tabular}{|c|c|c|c|}
    \hline
    Simulation&(a)&(b)&(c)\\
    \hline
    $1$&\includegraphics[width=1.3in]{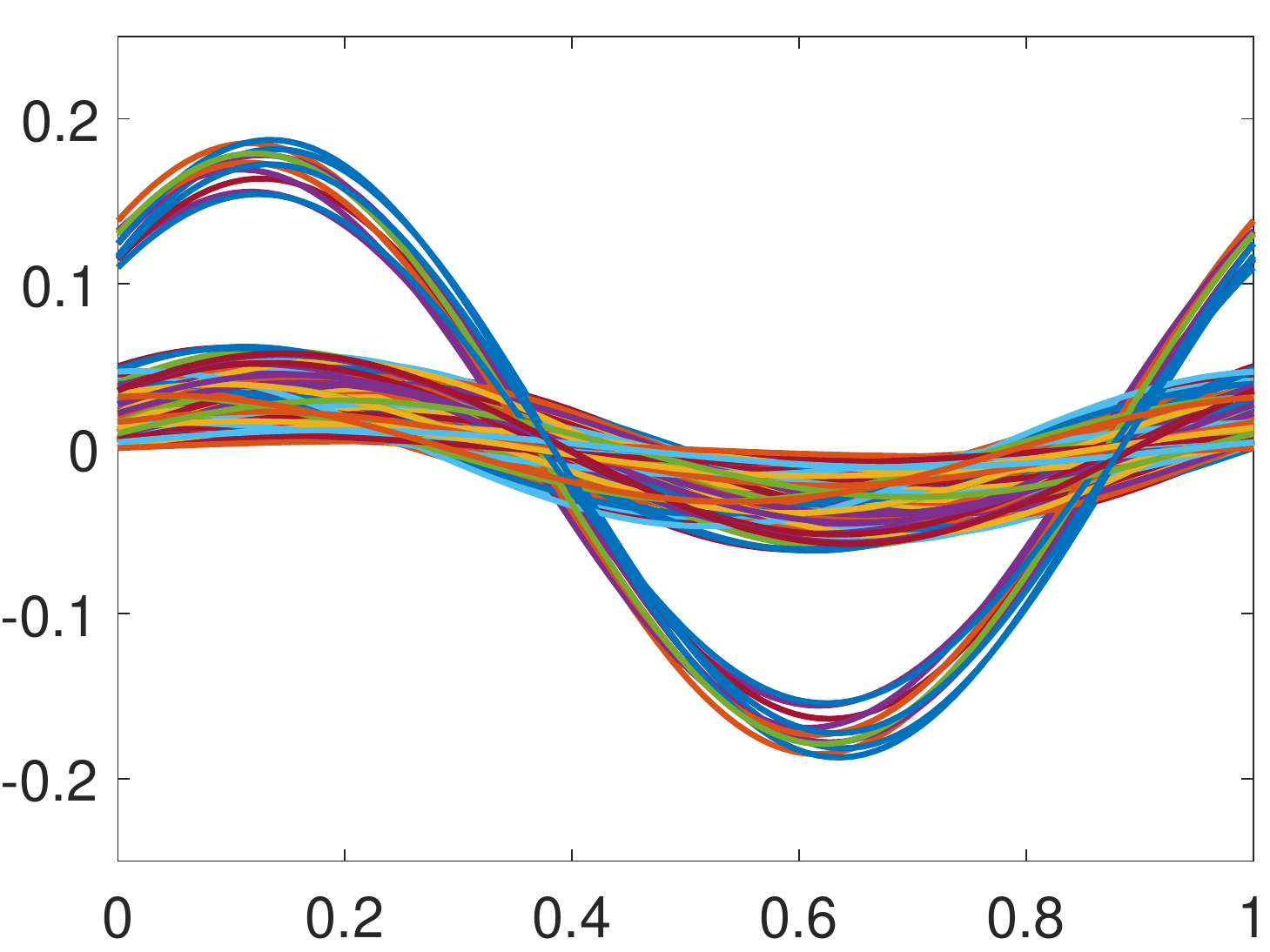}&\includegraphics[width=1.3in]{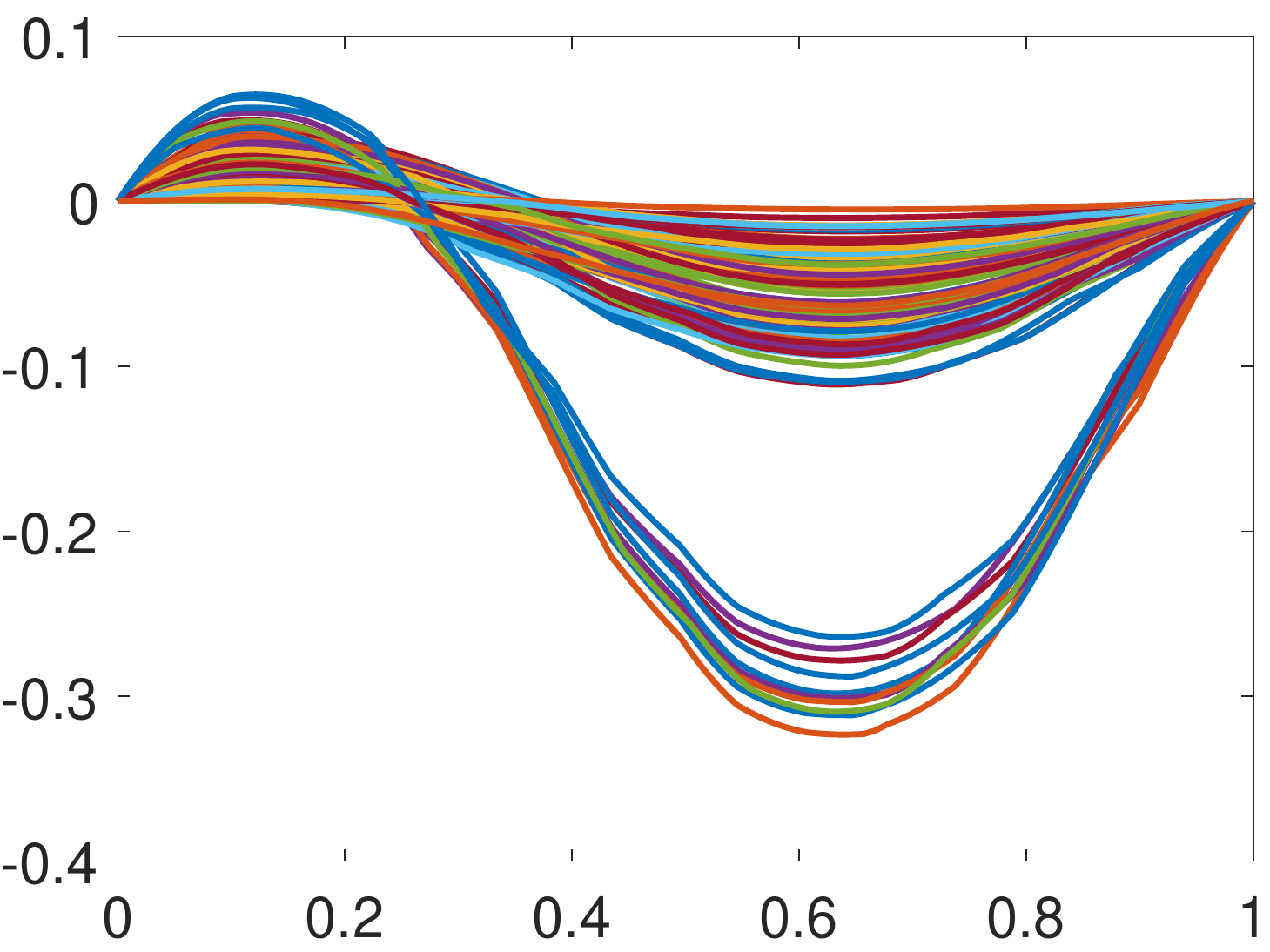}&\includegraphics[width=1in]{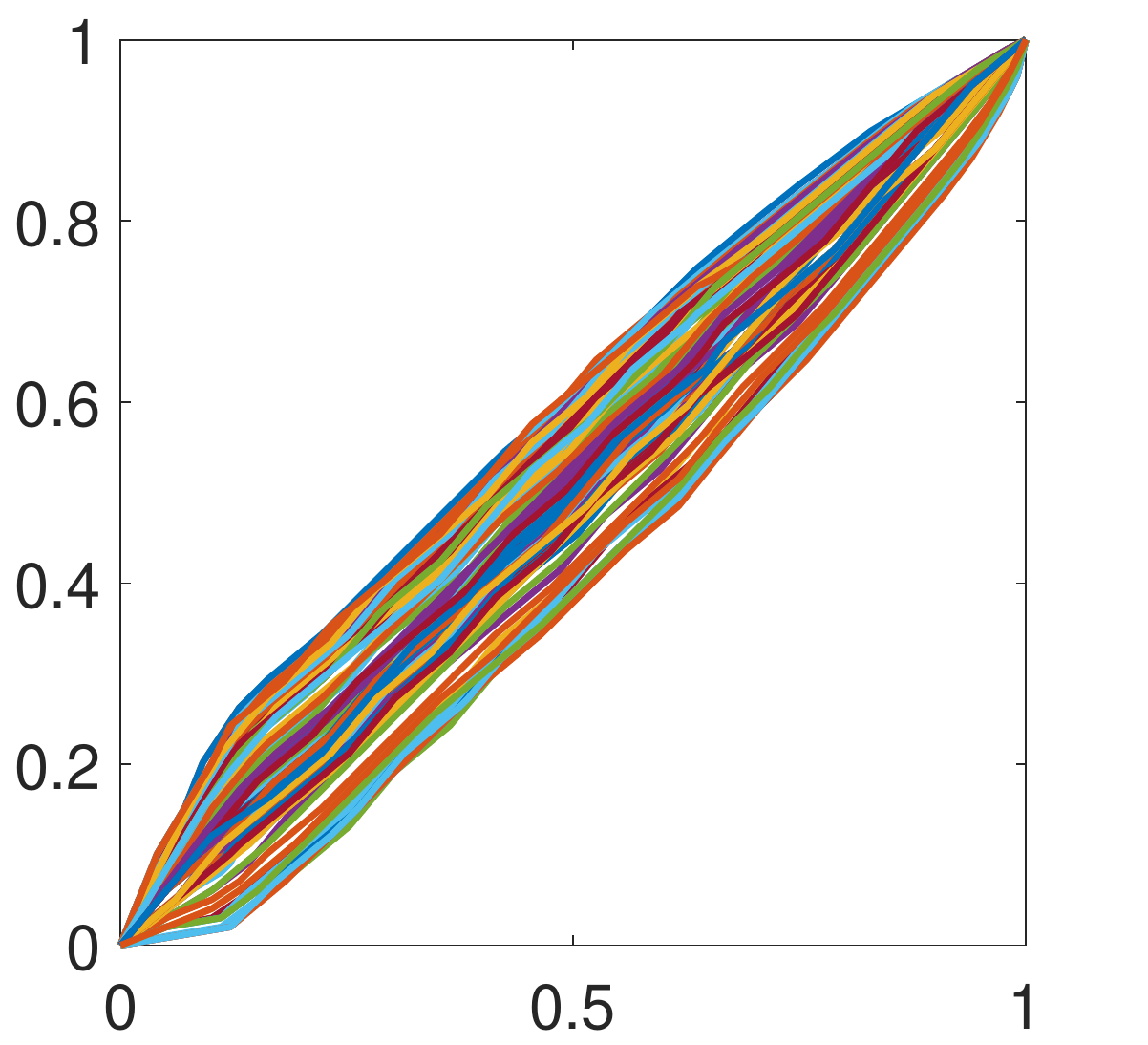}\\
    \hline
    $2$&\includegraphics[width=1.3in]{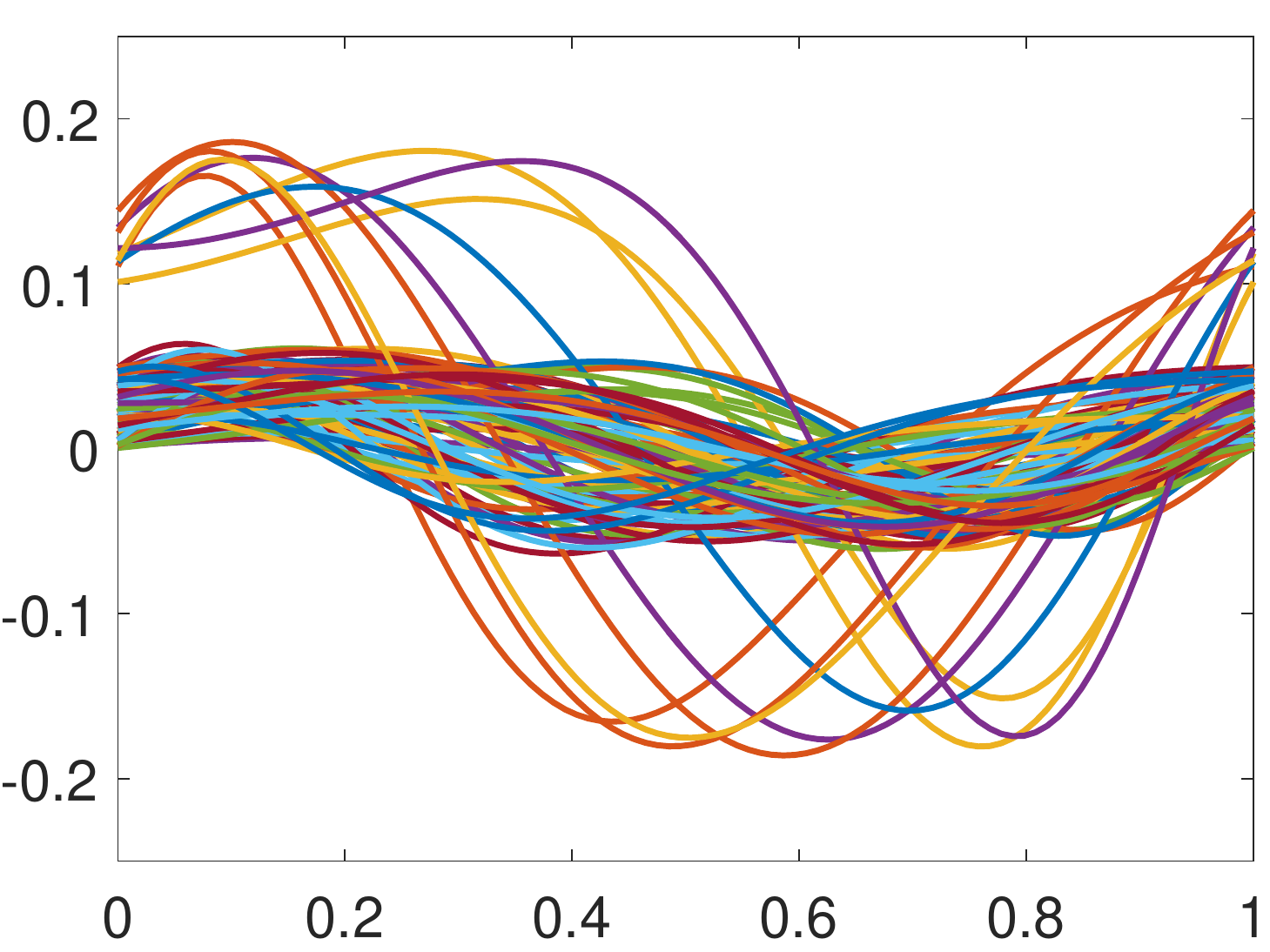}&\includegraphics[width=1.3in]{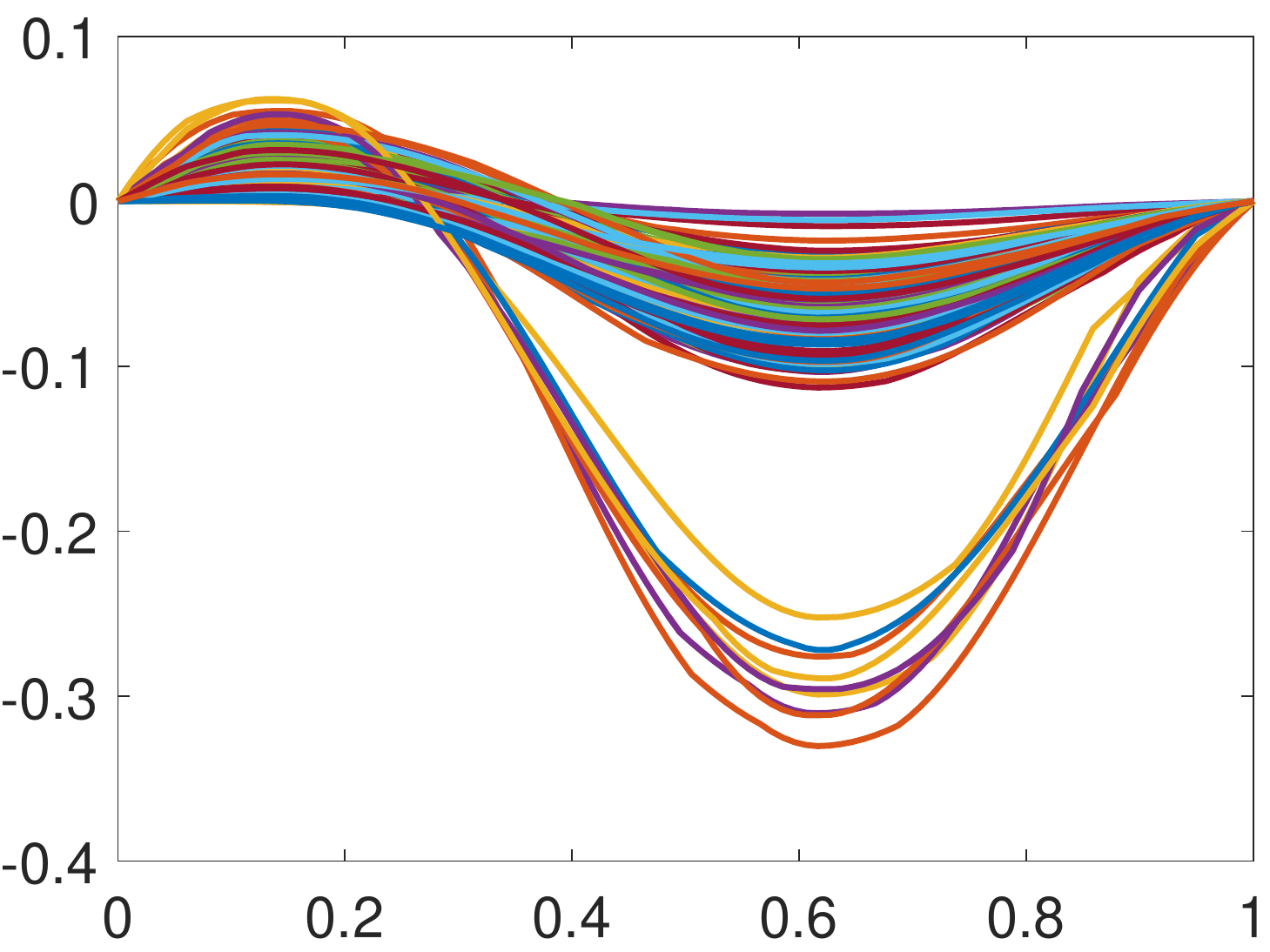}&\includegraphics[width=1in]{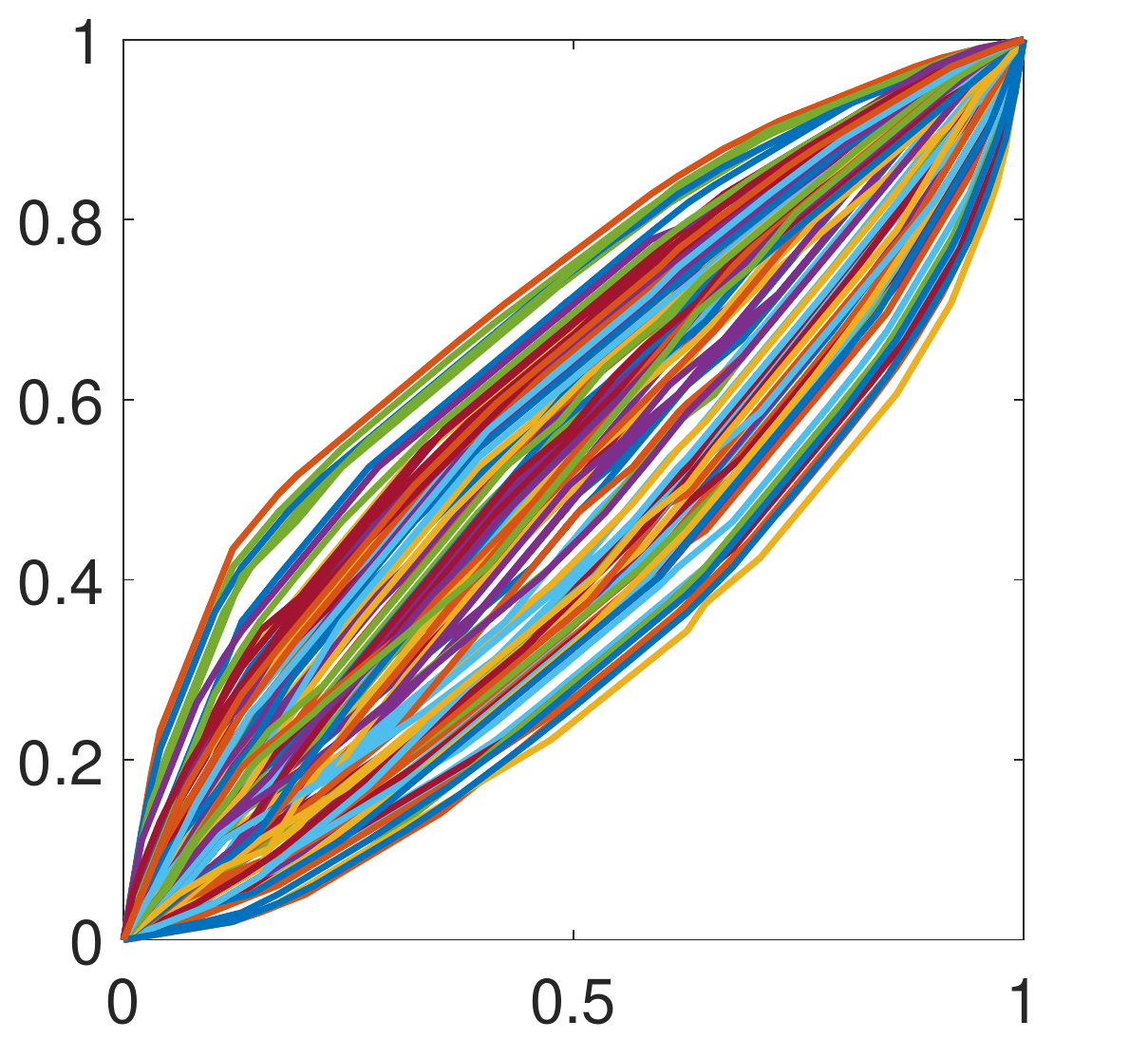}\\
    \hline
    $3$&\includegraphics[width=1.3in]{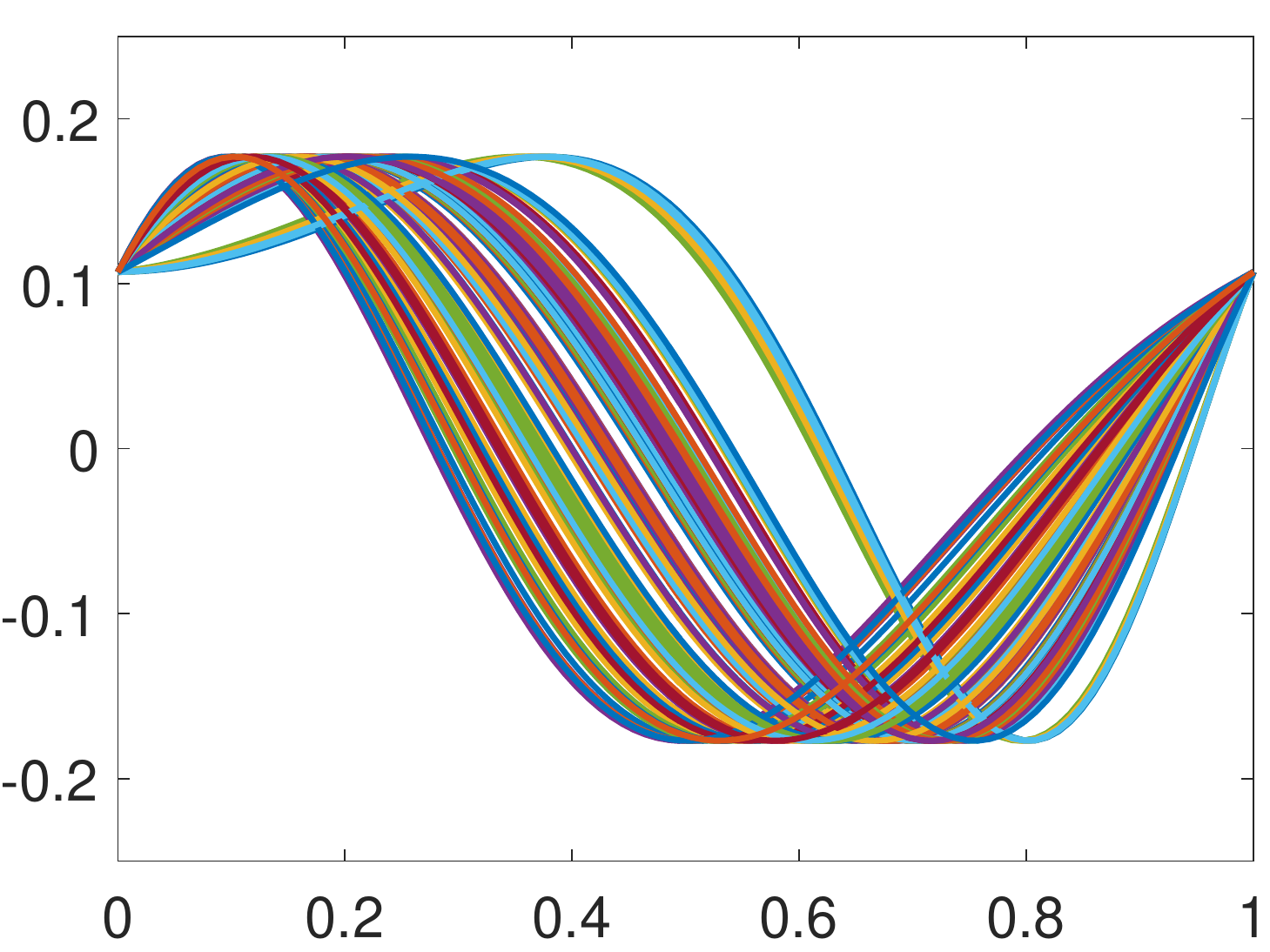}&\includegraphics[width=1.3in]{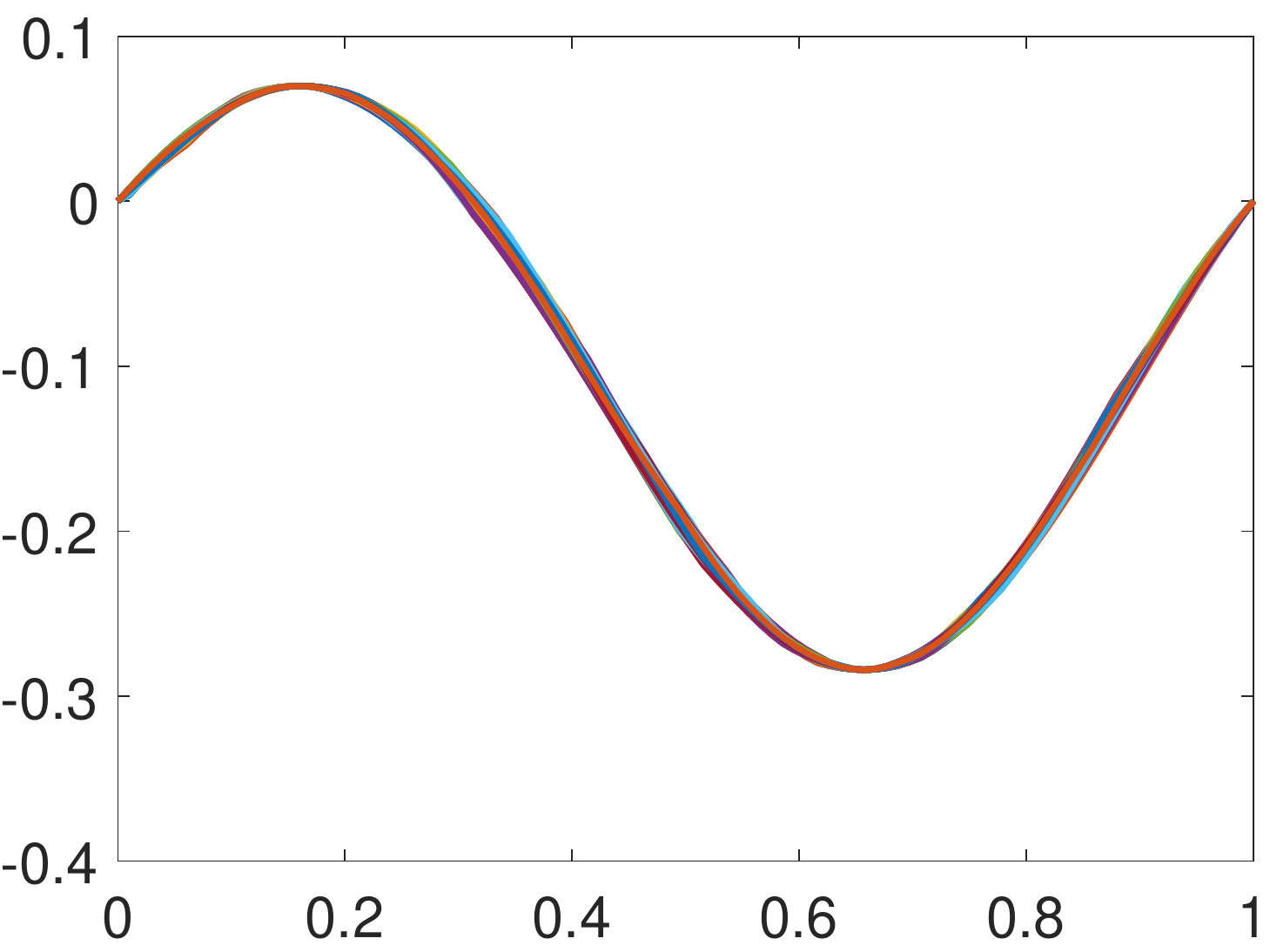}&\includegraphics[width=1in]{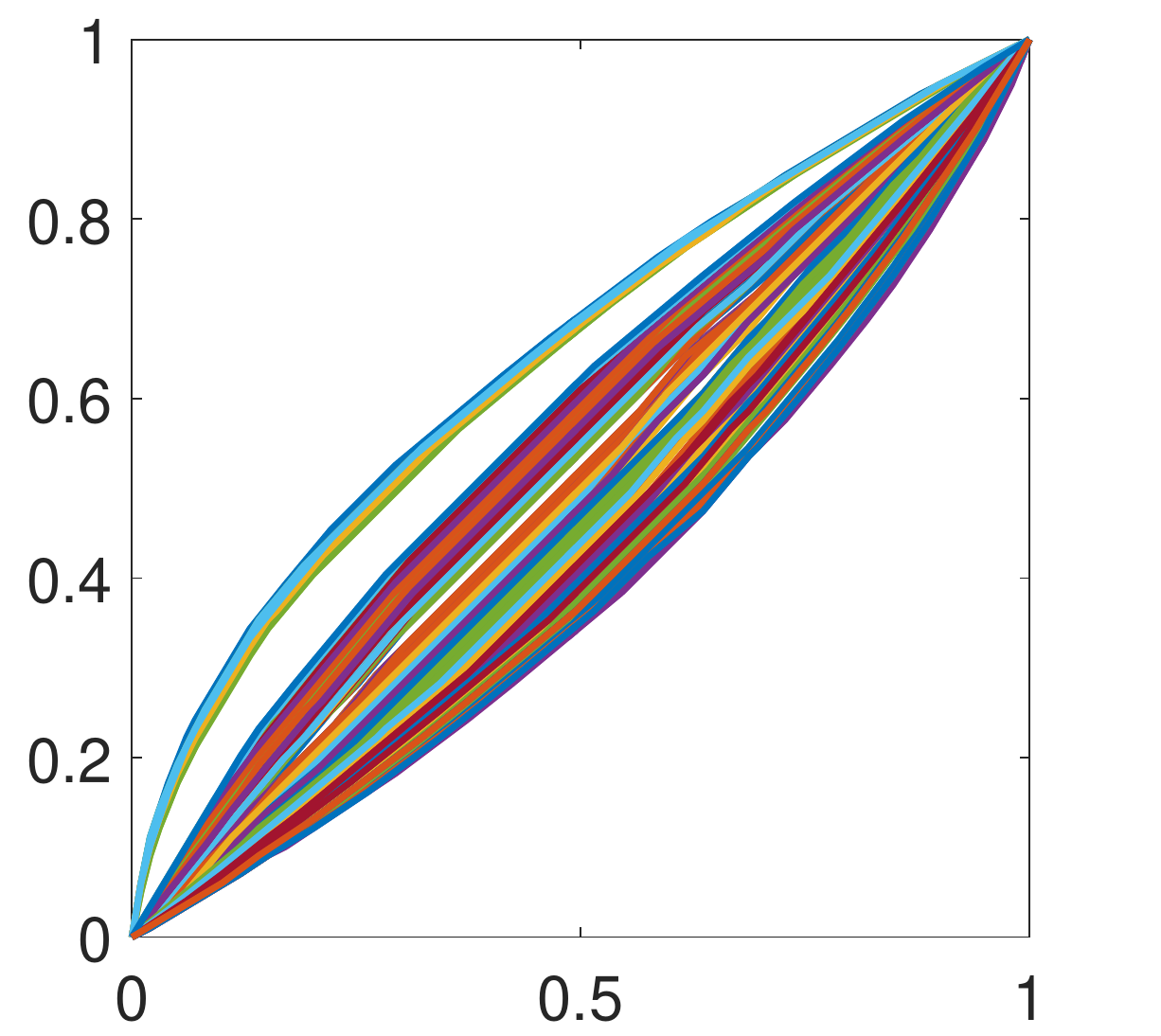}\\
    \hline
    $4$&\includegraphics[width=1.3in]{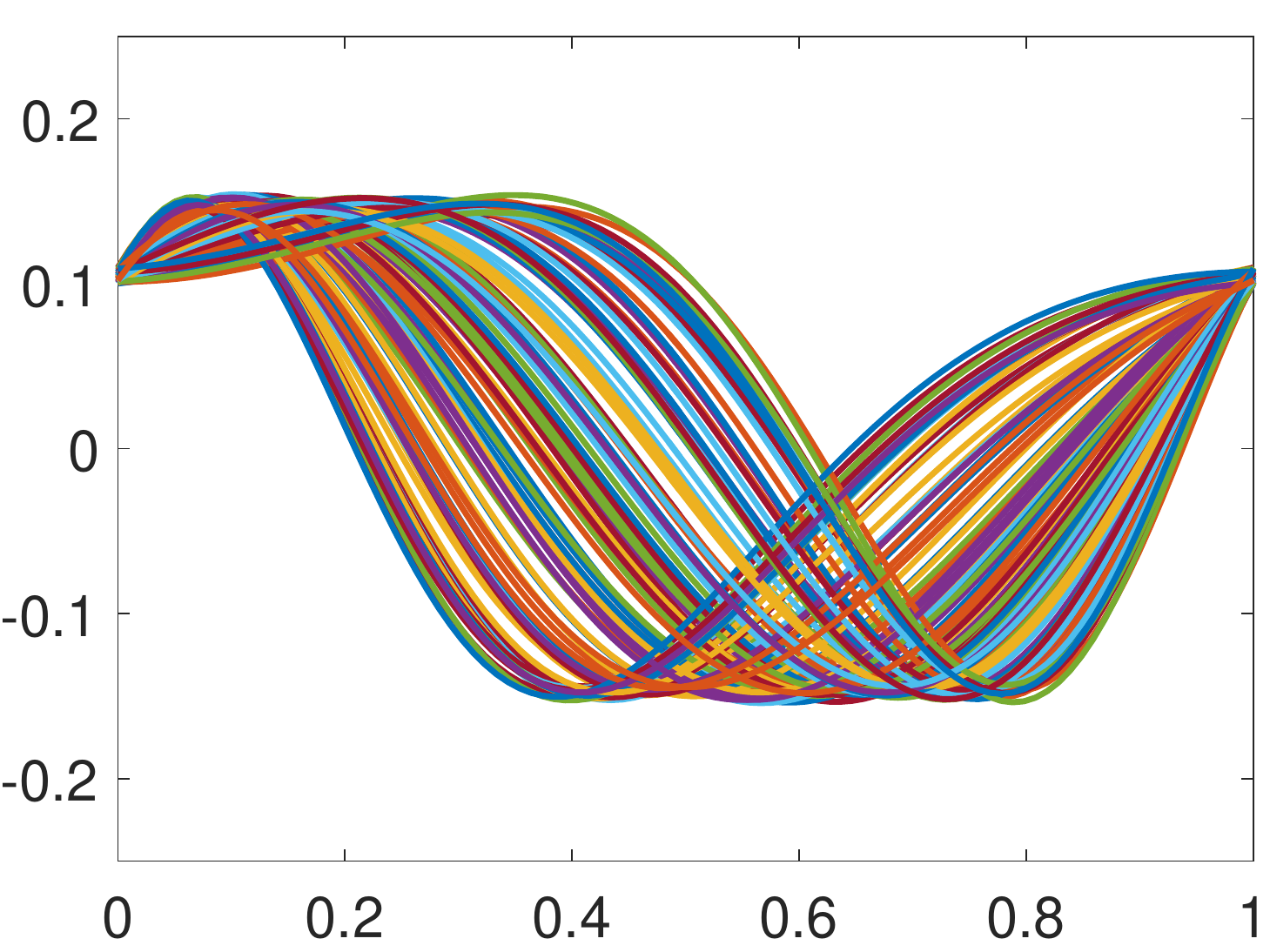}&\includegraphics[width=1.3in]{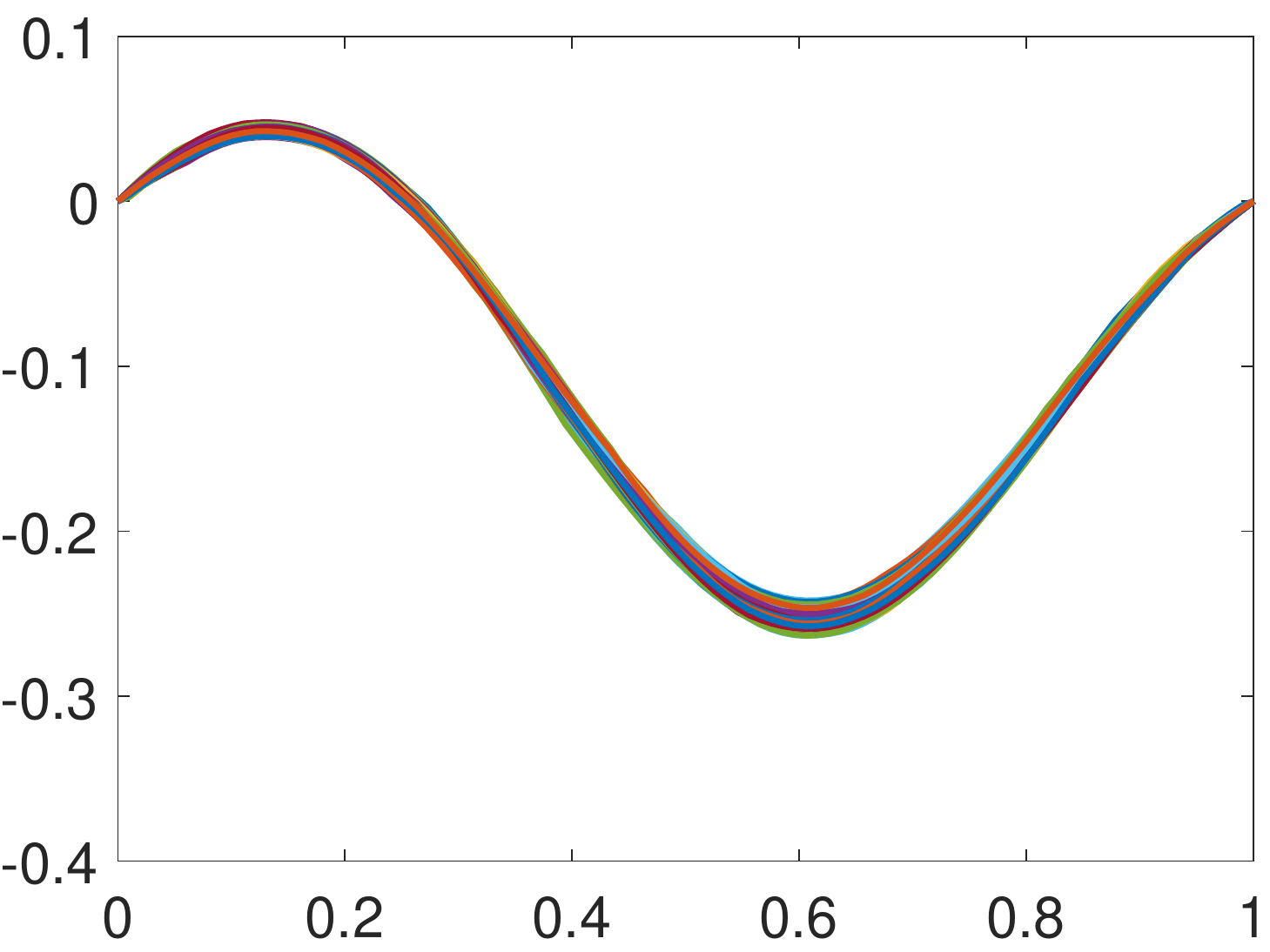}&\includegraphics[width=1in]{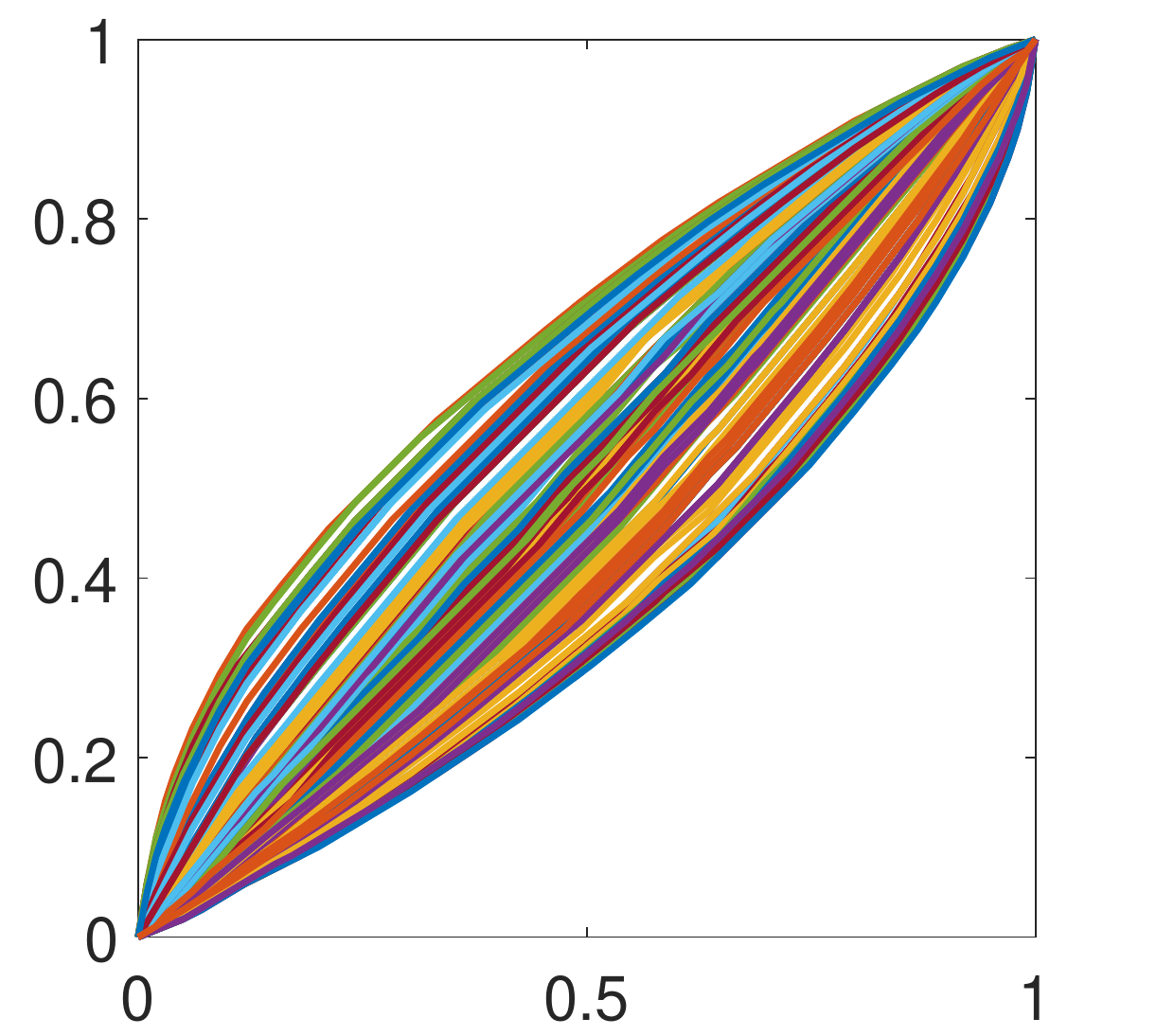}\\
    \hline
    \end{tabular}
    \caption{\small Datasets for Simulations 1-4. (a) Original functions. (b) Amplitude. (c) Phase.}\label{fig:simu2}
    \end{center}
    \end{small}
    \vspace{-5mm}
\end{figure}

We display all of the simulated datasets in Figure \ref{fig:simu2} with the original data in (a), the amplitude in (b), and the phase in (c). As in \cite{citeulike:10107686}, we are interested in the distribution of two quantities: $p_c$, the percentage of correctly detected outliers (number of correctly detected outliers divided by the total number of outliers) and $p_f$, the percentage of falsely detected outliers (number of falsely detected outliers divided by the total number of non-outliers). Thus, for each simulation, we generate 100 replicates and report the estimated average values $\hat{p}_c$ and $\hat{p}_f$ and their standard deviations. For additional simulation studies, designed in a similar manner to Models 1-5 of \cite{citeulike:10107686}, please see the Supplementary Material.

\noindent{\bf Simulation 1 Results:} Note that there are no phase outliers present in this dataset. Thus, we focus on amplitude outlier detection. The results in the top portion of Table \ref{tab:simu2} suggest choosing $k_a \approx 1.3$ for detection of severe outliers. This is when the true detection rate is maximized while the false detection rate is very low. Nonetheless, the proposed method is fairly stable with respect to the choice of $k_a$, with values ranging from $1.0$ to $1.5$ providing very good detection rates. The method of \cite{citeulike:10107686} performs very well on the same dataset with an average true detection rate of $100\%$ with standard deviation $0\%$ and an average false detection rate of $0\%$ with standard deviation $0\%$. The proposed method (with $k_a=1.3$) outperforms the functional bagplot and functional highest-density-region boxplot methods of \cite{citeulike:7579189} (these results were reported in \cite{citeulike:10107686}).

\begin{table}[!t]
\begin{center}
\begin{small}
\begin{tabular}{|c||c|c|c|c|c|c|c|c|}
	\hline
	$k_a$ & 0.6 & 0.8 & 1.0 & 1.2& {\bf 1.3} & 1.5 & 1.7\\
	\hline
    \hline
	$\hat{p}_c$ & 100 & 100 & 100 & 100 & {\bf 100} & 98.95  & 89.56 \\
    &(0) &  (0) &  (0) &  (0) & {\bf (0)} &  (3.75) &  (13.78)\\
	\hline
	$\hat{p}_f$ & 3.69 & 1.60 & 0.63 & 0.15 &{\bf 0.07} & 0.02 & 0.01\\
    & (1.92) & (1.24) & (0.77) & (0.48) &{\bf (0.31)} & (0.15) & (0.11)\\
	\hline
	\hline
	$k_a$ & 0.6 & 0.8 & 1.0 & 1.2& {\bf 1.3} & 1.5 & 1.7\\
	\hline
	\hline
	$\hat{p}_c$ & 100 & 100 & 100 & 100 & {\bf 100} & 98.69 & 89.79 \\
    & (0) & (0) & (0) & (0) & {\bf (0)} & (4.16) & (15.79) \\
	\hline
	$\hat{p}_f$ & 3.48 & 1.59 & 0.64 & 0.21 & {\bf 0.11} & 0.03  & 0 \\
    & (1.87) & (1.15) & (0.88) & (0.52) & {\bf (0.36)} & (0.19) & (0) \\	
    \hline
\end{tabular}
\end{small}
\caption{\small Average true positive and false positive outlier detection rates (with standard deviations in parentheses) for the data in Simulation 1 (top) and Simulation 2 (bottom). Best performance is highlighted in bold.} \label{tab:simu2}
\end{center}
\vspace{-5mm}
\end{table}

\noindent{\bf Simulation 2 Results:} The bottom portion of Table \ref{tab:simu2} provides the results of this simulation. Again, there are no phase outliers present in this dataset, and thus, we focus on the amplitude component only as in the previous simulation. The true and false detection rates provide the best tradeoff when the value of $k_a=1.3$. It is important to note that despite adding significant phase variability to the simulated functions, the performance of the proposed amplitude outlier detection method has not deteriorated; in fact, in some cases it improved slightly. As a comparison, using the method of \cite{citeulike:10107686} on the same dataset, the average value of $\hat{p}_c$ was $100\%$ with standard deviation $0\%$ and the average value of $\hat{p}_f$ was $0$ with standard deviation $0$. This means that their method is also fairly robust to the additional phase variability, when the amplitude outliers are severe. The main advantage of the proposed method is that it can also study phase outliers, which is not possible with the method of \cite{citeulike:10107686}. Combining our observations in Simulations 1 and 2, we advise a general setting of the value of $k_a=1.3$ for detection of severe amplitude outliers. In the remainder of this paper, we use the following scale for $k_a$: mild amplitude outliers are detected with $k_a \in [0.6, 0.8)$, regular amplitude outliers are detected with $k_a \in [0.8, 1.3)$, and severe amplitude outliers are detected with $k_a \in [1.3, \infty)$. This multiscale approach to outlier detection allows for better exploration of complex functional datasets.

%\begin{table}[!t]
%\begin{center}
%\begin{small}
%\begin{tabular}{|c||c|c|c|c|c|c|c|}
%	\hline
%	$k_a$ & 0.6 & 0.8 & 1.0 & 1.2& {\bf 1.3} & 1.5 & 1.7\\
%	\hline
%	\hline
%	$\hat{p}_c$ & 100 & 100 & 100 & 100 & {\bf 99.86} & 96.99 & 86.18 \\
%    & (0) & (0) & (0) & (0) & {\bf (1.42)} & (7.26) & (18.03) \\
%	\hline
%	$\hat{p}_f$ & 3.19 & 1.26 & 0.52 & 0.11 & {\bf 0.04} & 0.01  & 0 \\
%    & (1.92) & (1.22) & (0.77) & (0.37) & {\bf (0.20)} & (0.11) & (0) \\	
%    \hline
%\end{tabular}
%\end{small}
%\caption{\small Average true positive and false positive outlier detection rates (with standard deviations in parentheses) for the data in Simulation 2. Best performance is highlighted in bold.} \label{tab:simu3}
%\end{center}
%\vspace{-5mm}
%\end{table}

\noindent{\bf Simulation 3 Results:} As expected, there is essentially no variability present in the amplitude component and all variability in the given data is captured in the phase component (Figure \ref{fig:simu2}). The outlier detection results are presented in the top portion of Table \ref{tab:simu4}. The phase outlier detection results are best when $k_p\approx 0.7$. Therefore, as a guideline, we define mild phase outliers using $k_p \in [0.5, 0.7)$, regular phase outliers using $k_p \in [0.7, 0.9)$, and severe phase outliers using $k_p \in [0.9, \infty)$. We further emphasize that there are no methods in the current literature that can detect phase outliers in functional data. We tried to compare our results to the method of \cite{citeulike:10107686} on this dataset, but it failed to produce a credible result, which we suspect is due to the definition of the ranking of the curves based on band depth.

\begin{table}[!t]
\begin{center}
\begin{small}
\begin{tabular}{|c||c|c|c|c|c|}
	\hline
	$k_p$ & 0.5 & 0.6 & {\bf 0.7} & 0.8 & 0.9\\
	\hline
	\hline
	$\hat{p}_c$ & 100 (0) & 100 (0) & {\bf 99.83 (1.67)} & 95.43 (12.00) & 78.91 (32.99) \\
	\hline
	$\hat{p}_f$ & 2.75 (3.63)& 0.71 (1.81) & {\bf 0.18 (1.09)} & 0 (0) & 0 (0) \\
	\hline
	\hline
	$k_a$ or $k_p$ & 0.6 & 0.7 & 0.8 & 0.9 & 1.0\\
	\hline
	Amplitude $\hat{p}_{f}$ & 0 (0) & 0 (0) & 0 (0) & 0 (0) & 0 (0) \\
	\hline
	Phase $\hat{p}_{f}$ & 4.56 (4.34) & 1.79 (2.57) & 0.52 (1.32) & 0.12 (0.57) & 0.01 (0.10) \\
	\hline
\end{tabular}
\end{small}
\caption{\small Average true positive and false positive outlier detection rates (with standard deviations in parentheses) for the data in Simulation 3 (top) and Simulation 4 (bottom). Best performance is highlighted in bold.} \label{tab:simu4}
\end{center}
\vspace{-5mm}
\end{table}

\noindent{\bf Simulation 4 Results:} In this simulation, we are only interested in the false detection rates reported in the bottom portion of Table \ref{tab:simu4}. These results confirm that the proposed method is very effective at separating the amplitude and phase variabilities, and as a result makes very few false positive detections in the presence of large amplitude and phase variability. In comparison, the method by \cite{citeulike:10107686} achieved a $6.48\%$ false detection rate with a standard deviation of $5.99\%$ on the same dataset. Thus, in the presence of significant phase variability, our method outperforms their state-of-the-art method.

\subsection{Real Data Study 1: Annual Sea Surface Temperature}

Approximately $71\%$ of the Earth's surface is covered by water. The temperature of the sea surface controls the air-water interaction to/from the atmosphere. In particular, sea surface temperature coordinates air-sea exchange of CO$_2$, which in turn impacts the sequestration of CO$_2$ in the ocean. The ocean absorbs most of the heat caused by increasing atmospheric greenhouse gas levels, which causes ocean temperatures to rise. Therefore, sea surface temperature reflects the overall trend in the climate system and can be regarded as a fundamental measure of global climate change. Moreover, the El Ni\~{n}o phenomenon can also be identified based on sea surface temperature. El Ni\~{n}o is associated with a band of warm ocean water that develops in the central and east-central equatorial Pacific (between approximately the International Date Line and 120$^{\circ}$W), including off the Pacific coast of South America. On the other hand,  La Ni\~{n}a events are associated with abnormally cold ocean water. El Ni\~{n}o Southern Oscillations (ENSO) refer to the cycle of warm and cold sea surface temperatures of the tropical central and eastern Pacific Ocean.

The sea surface temperature (SST) data for the Ni\~{n}o 1+2 region is provided on the Climate Prediction Center website\footnote{http://www.cpc.ncep.noaa.gov/data/indices/ersst3b.nino.mth.81-10.ascii}. In our study, we use annual SST data from 1950 to 2014. We interpolate the 12 monthly temperatures using splines to construct a single SST function for each year (this step is not necessary for our framework). Thus, the dataset contains a total of 65 annual SST functions. It is important to note that this dataset contains natural phase variability. That is, hot and cold months do not always occur at the same time; depending on the year, these events may occur either earlier or later than the typical (median) time. Thus, it is desirable in this application to separate translation, amplitude, and phase variabilities and visualize each of them separately.

\begin{figure}[!t]
\begin{small}
\begin{center}
    \begin{tabular}{|c|c|c|c|}
    \hline
    (a)&(b)&(c)&(d)\\
    \hline
    \includegraphics[width=1.3in]{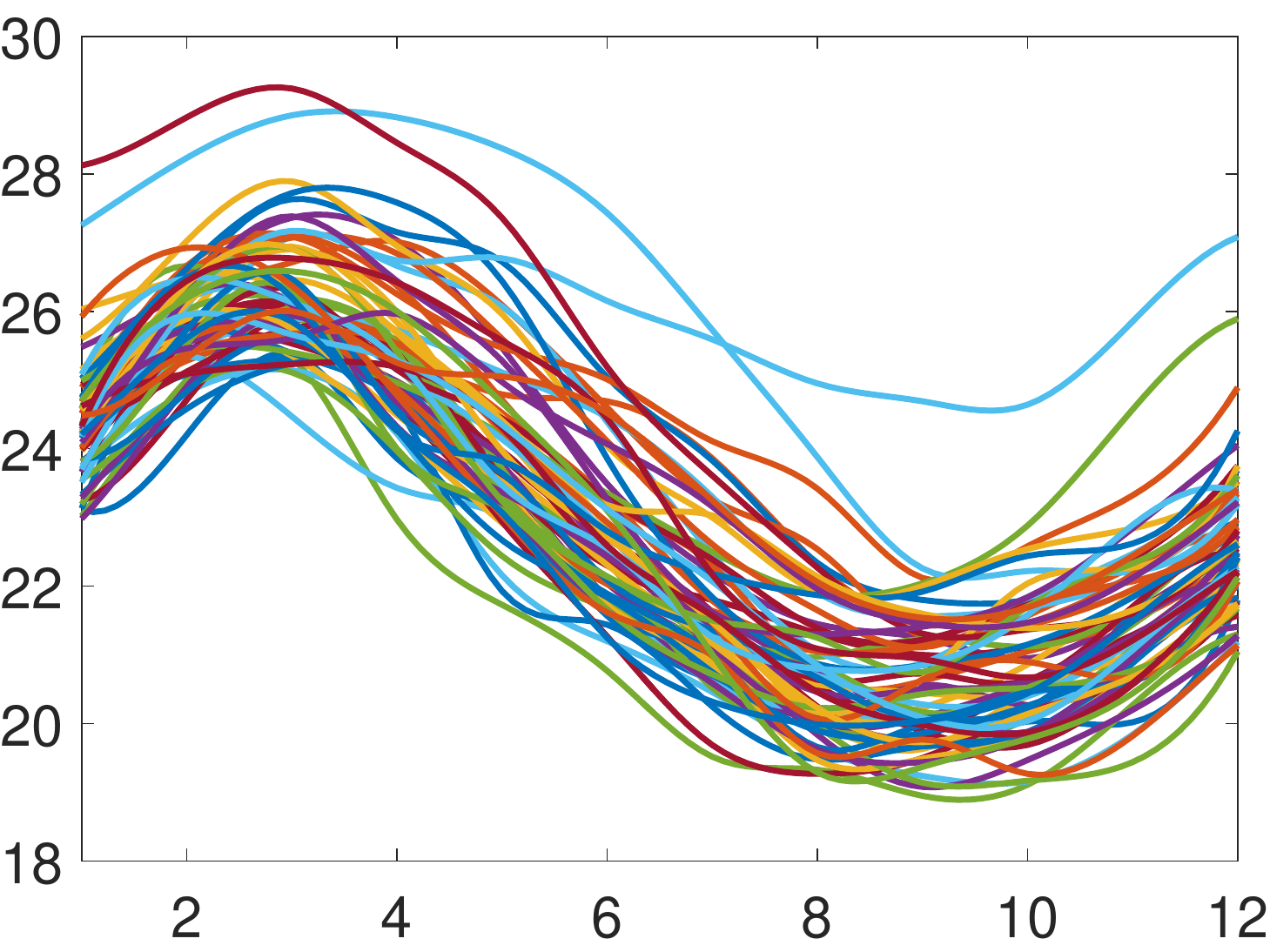}&\includegraphics[width=1.3in]{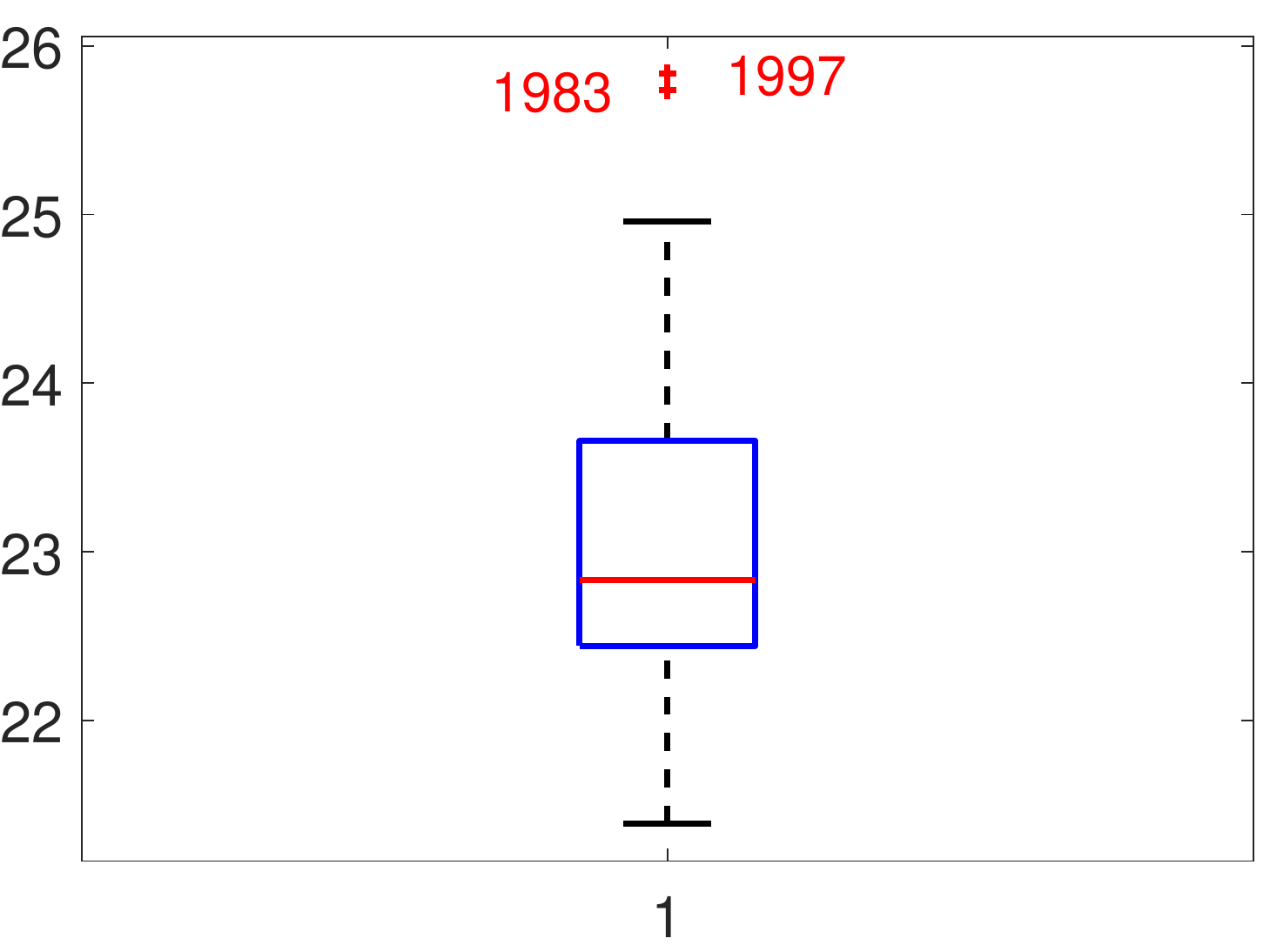}&\includegraphics[width=1.3in]{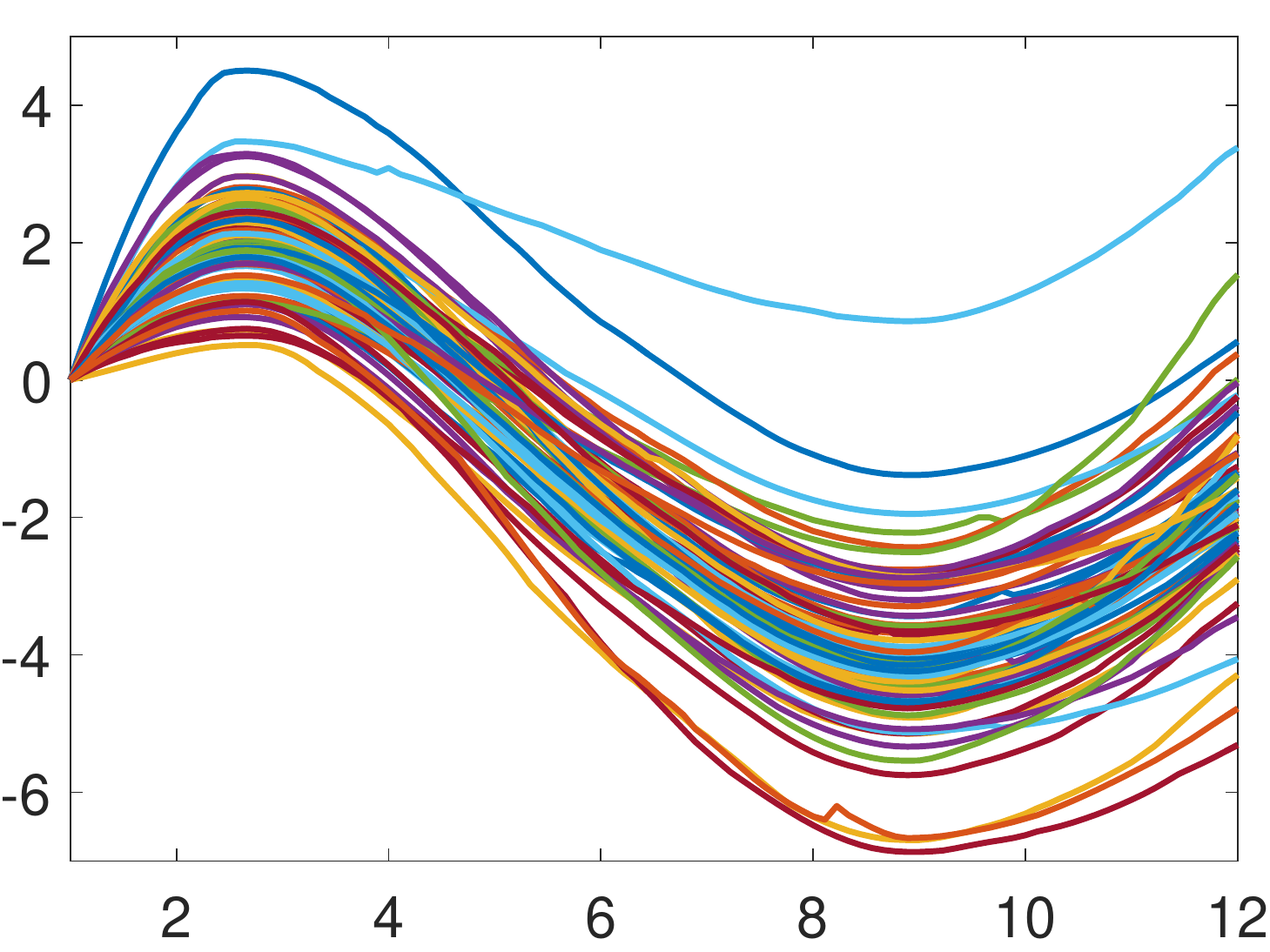}&\includegraphics[width=1.3in]{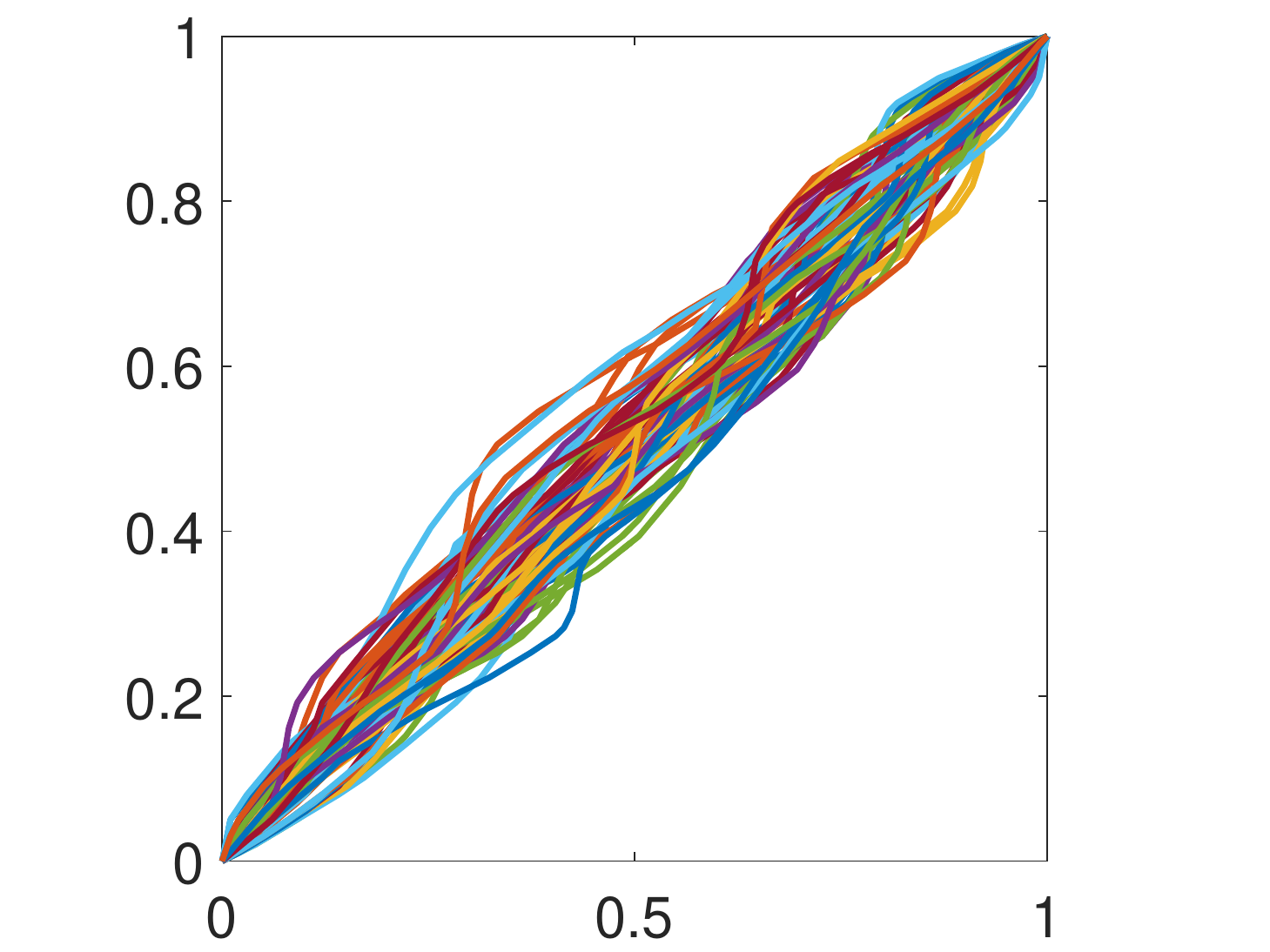}\\
    \hline
    \end{tabular}
    \caption{\small Separation of translation, amplitude, and phase variabilities in the sea surface temperature data. (a) Original functions. (b) Translation. (c) Amplitude. (d) Phase.}\label{fig:elnino}
    \end{center}
    \end{small}
    \vspace{-5mm}
\end{figure}

\begin{figure}[!t]
\begin{small}
\begin{center}
    \begin{tabular}{|cc|cc|}
    \hline
    (a)&(b)&(c)&(d)\\
    \hline
    \includegraphics[width=1.3in]{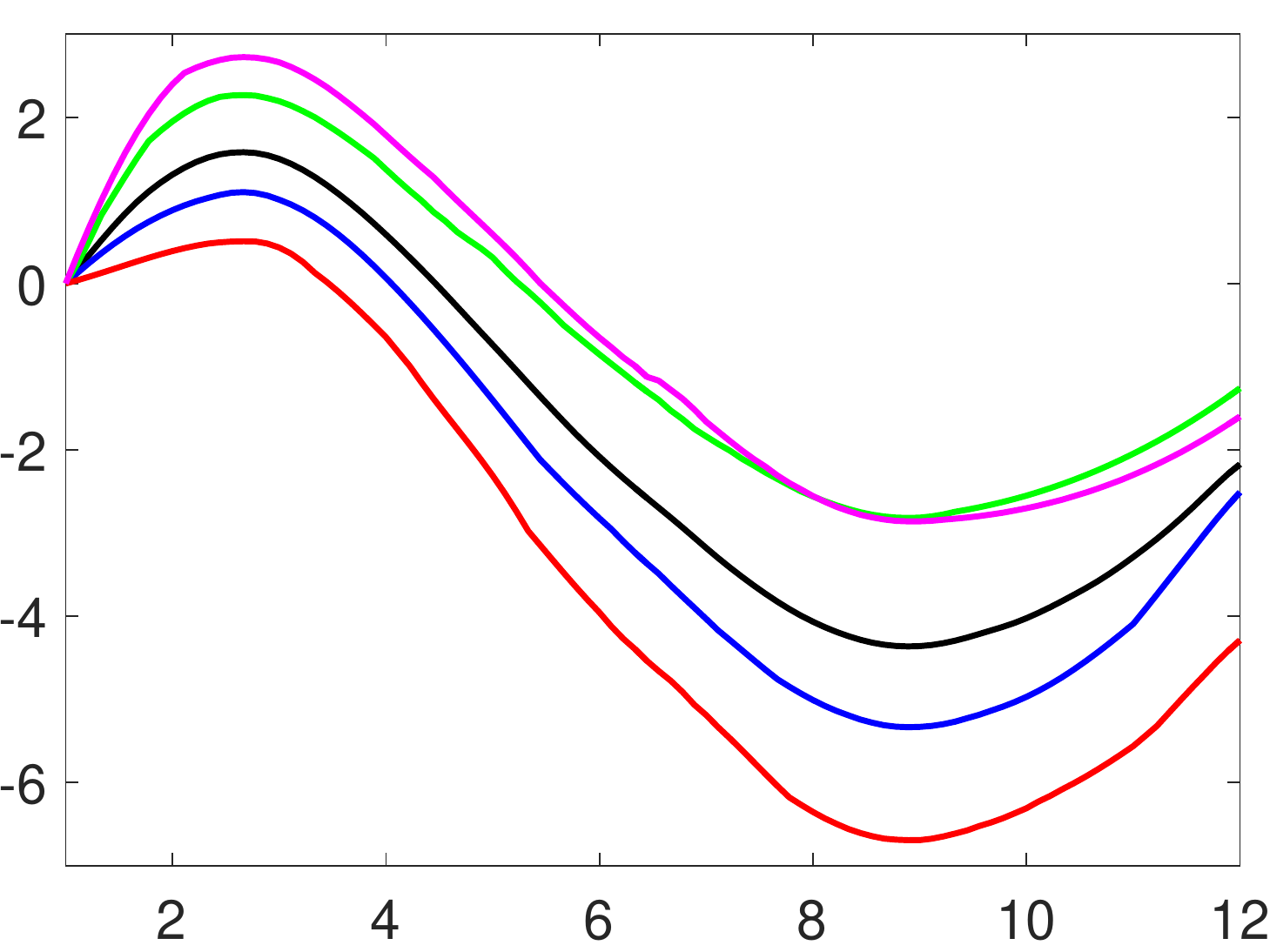}&\includegraphics[width=1.3in]{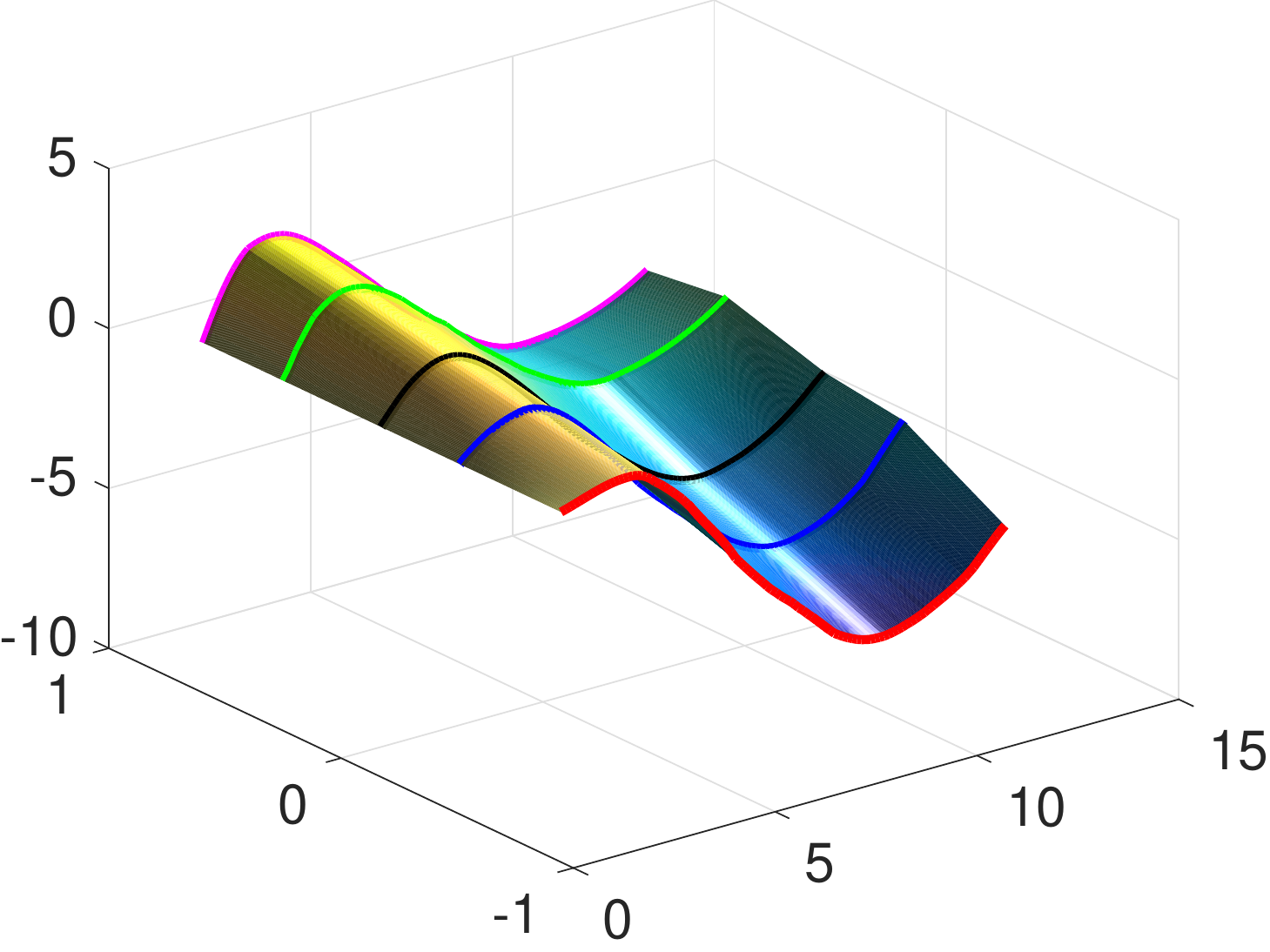}&\includegraphics[width=1.3in]{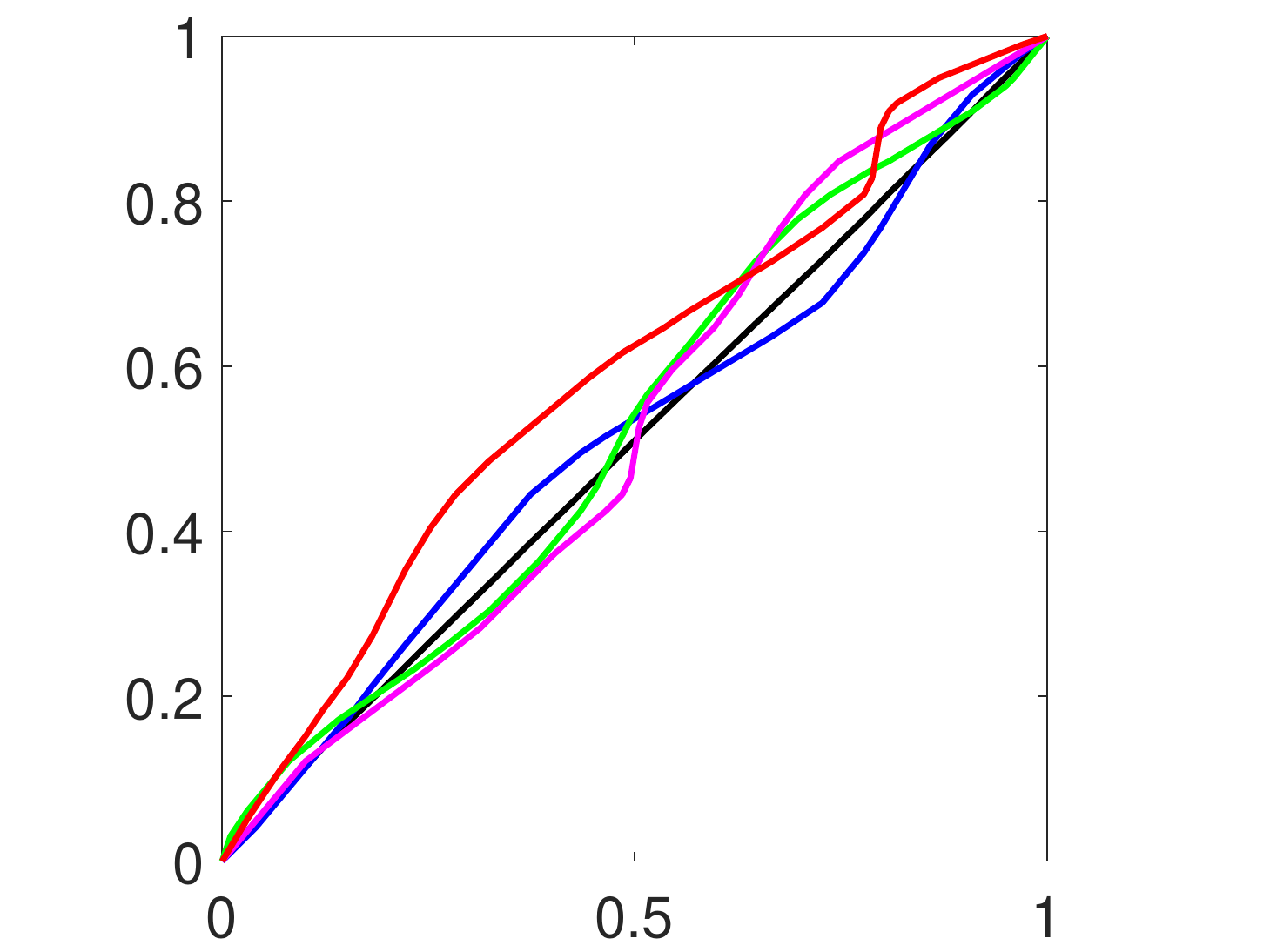}&\includegraphics[width=1.3in]{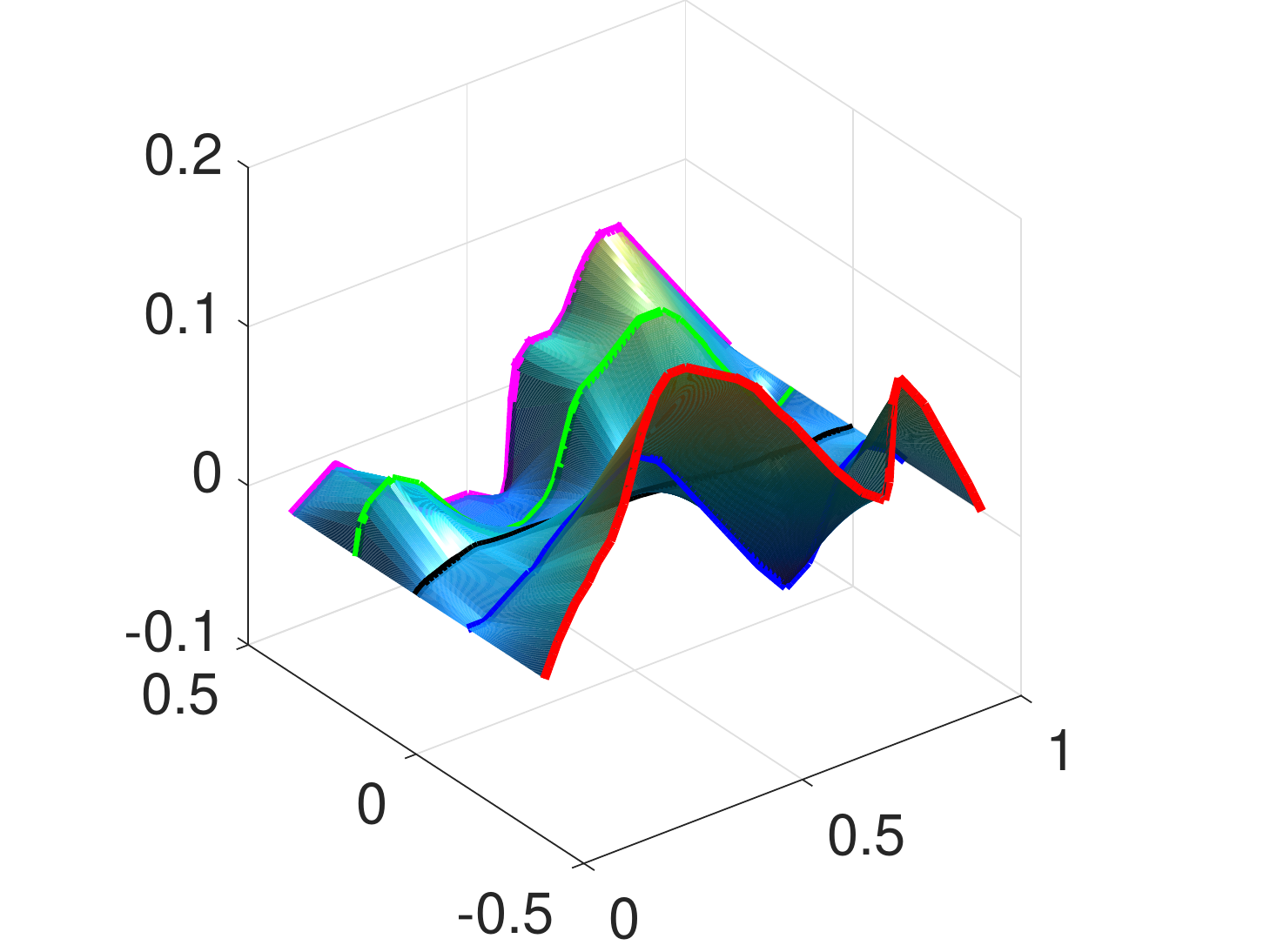}\\
    \hline
    \end{tabular}
    \caption{\small Amplitude and phase boxplot displays for sea surface temperature data. (a)\&(b) Amplitude boxplot and its surface display. (c)\&(d) Phase boxplot and its surface display.}\label{fig:ninoamplitude}
    \end{center}
    \end{small}
    \vspace{-5mm}
\end{figure}

We begin by displaying the separation of each of the variabilities in Figure \ref{fig:elnino}. First, note the crisp alignment of functions displayed as the amplitude component. It is clear from this panel that the only variability contained in this component is due to increasing or decreasing SST at each time point. Second, as stated earlier, the phase component is also significant in this dataset. Finally, the translation component displays the overall annual temperature variability. Figure \ref{fig:ninoamplitude}(a)\&(c) shows the amplitude and phase boxplots, respectively, with the two extremes in red and magenta, the two quartiles in blue and green, and the median in black. Additionally, in Figure \ref{fig:ninoamplitude}(b)\&(d), we show surface displays of the amplitude and phase boxplots. The amplitude surface boxplot is constructed by separating each of the boxplot functions according to the amplitude distances between them. The phase surface boxplot is constructed by first computing the deviation functions $h=\gamma_{id}-\gamma$ for each of the phase boxplot functions and then separating them according to the phase distances. We found that it is much more effective to display the phase surface boxplots using the deviation functions because the phase median is always very close to $\gamma_{id}$. In this case, a constant deviation function at 0 corresponds to this element of $\Gamma$.

%\begin{figure}[!t]
%\begin{center}
%\begin{small}
%    \begin{tabular}{|c|c|c|}
%    \hline
%    (a)&(b)&(c)\\
%    \hline
%    \includegraphics[width=1.4in]{elnino_warping_minmax1.pdf}&\includegraphics[width=1.4in]{elnino_warping_minmax_surface.pdf}&\includegraphics[width=1.4in]{elnino_dxdy.pdf}\\
%    \hline
%    \end{tabular}
%    \end{small}
%    \caption{\small Phase boxplot displays for sea surface temperature data. (a) Phase boxplot. (b) Surface display of phase boxplot. (c) Plot of phase ($x$-axis) vs. amplitude ($y$-axis) distances of each function in the data from the median.}\label{fig:ninophase}
%    \end{center}
%    \vspace{-5mm}
%\end{figure}

Figure \ref{fig:ninoamplitude}(a)\&(b) illustrates that as we traverse the amplitude boxplot from one extreme function to the median and then to the other extreme, the SST peak appearing toward the beginning of the year becomes steeper and the valley appearing toward the end of the year becomes deeper. Such amplitude variability is natural in this setting and corresponds to warmer vs. colder SST across years. The phase boxplot in Figure \ref{fig:ninoamplitude}(c)\&(d) also shows natural variability in the given data. The red phase function, representing one of the two extremes, is above the median phase (black) for the entire period of time; this indicates that during this year, the SST changes occured later than the median time. The magenta phase function represents the other extreme; it is below the median phase for the first half of the time period and above the median phase for the remainder of the time. This indicates that during that year, the SST changes came earlier than the median time for the initial half of the year, and later for the latter half.

\begin{figure}[!t]
\begin{center}
\begin{small}
    \begin{tabular}{|cc|cc|}
    \hline
    (a)&(b)&(c)&(d)\\
    \hline
    \includegraphics[width=1.3in]{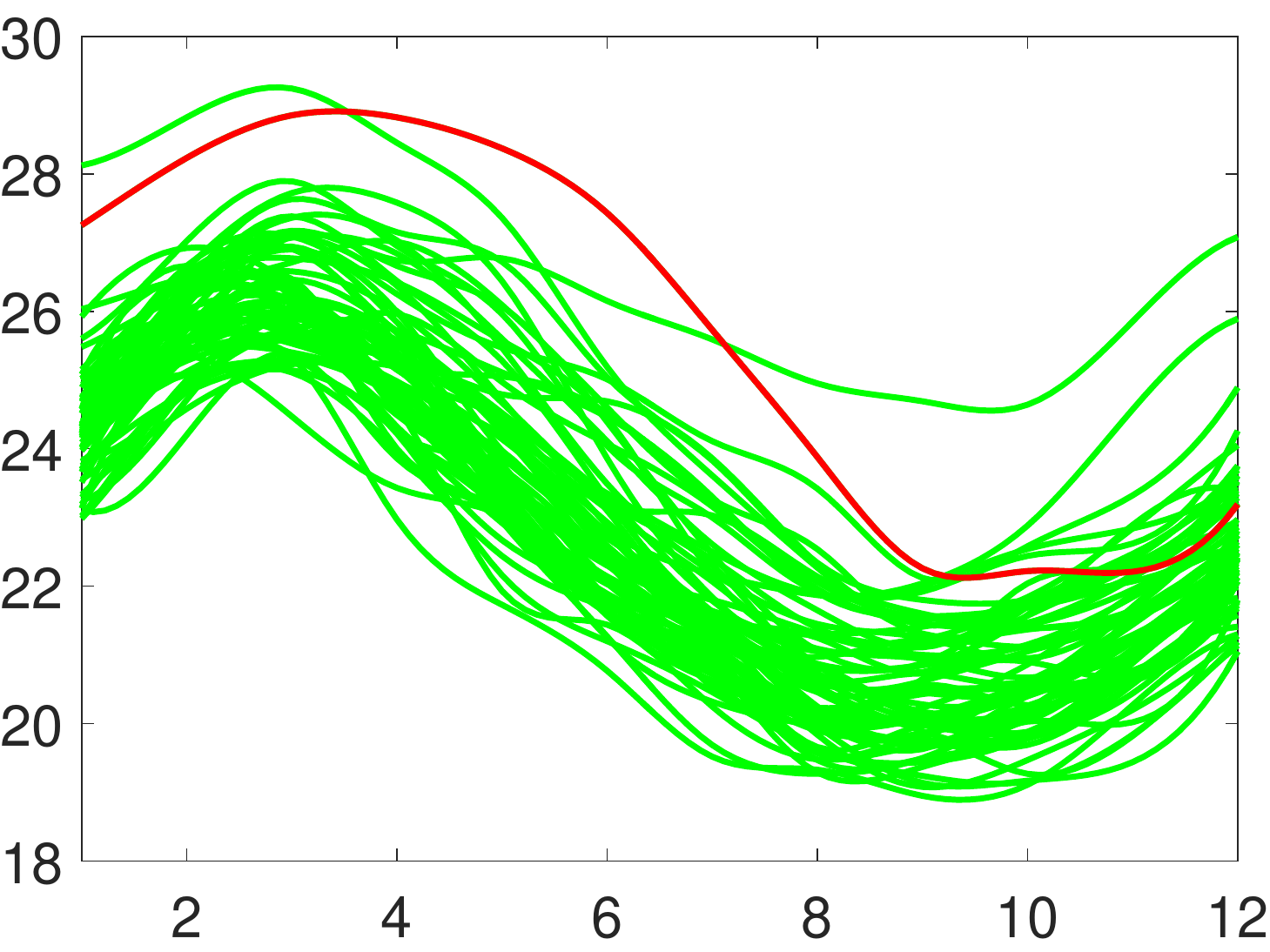}&\includegraphics[width=1.3in]{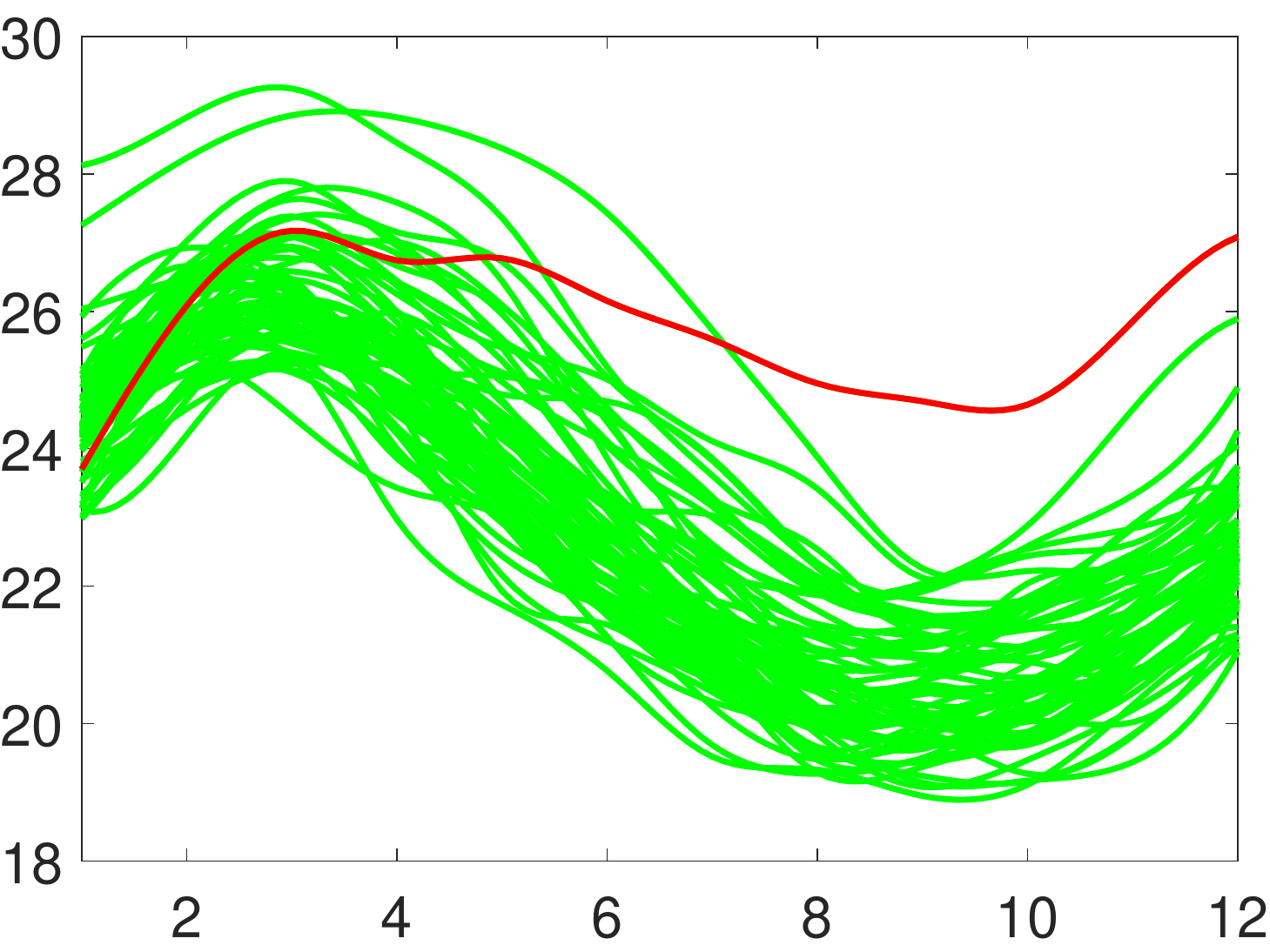}&\includegraphics[width=1.3in]{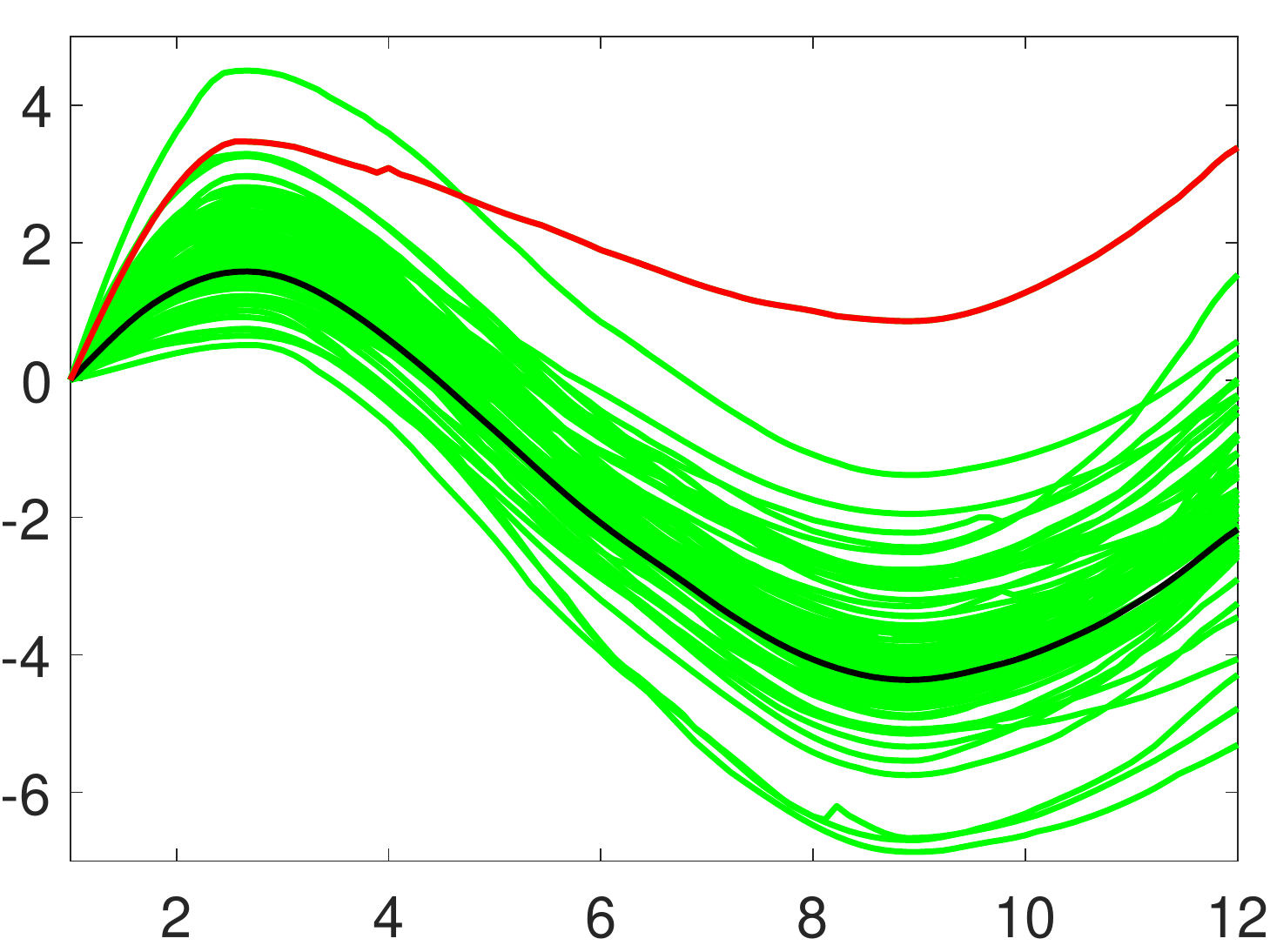}&\includegraphics[width=1.3in]{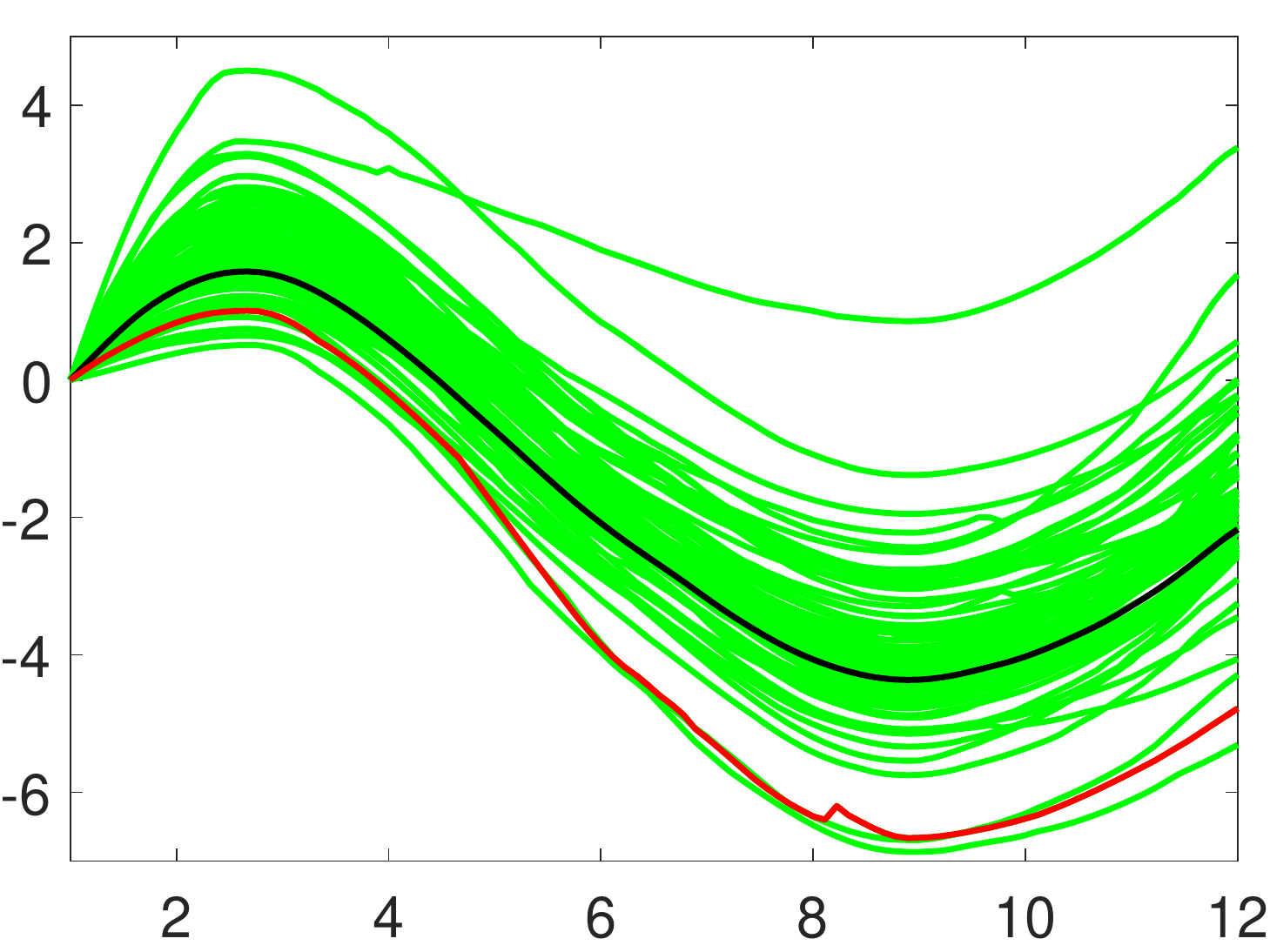}\\
    \hline
    \end{tabular}
    \begin{tabular}{|c|cc|c|}
    \hline
    (e)&(f)&(g)&(h)\\
    \hline
    \includegraphics[width=1.3in]{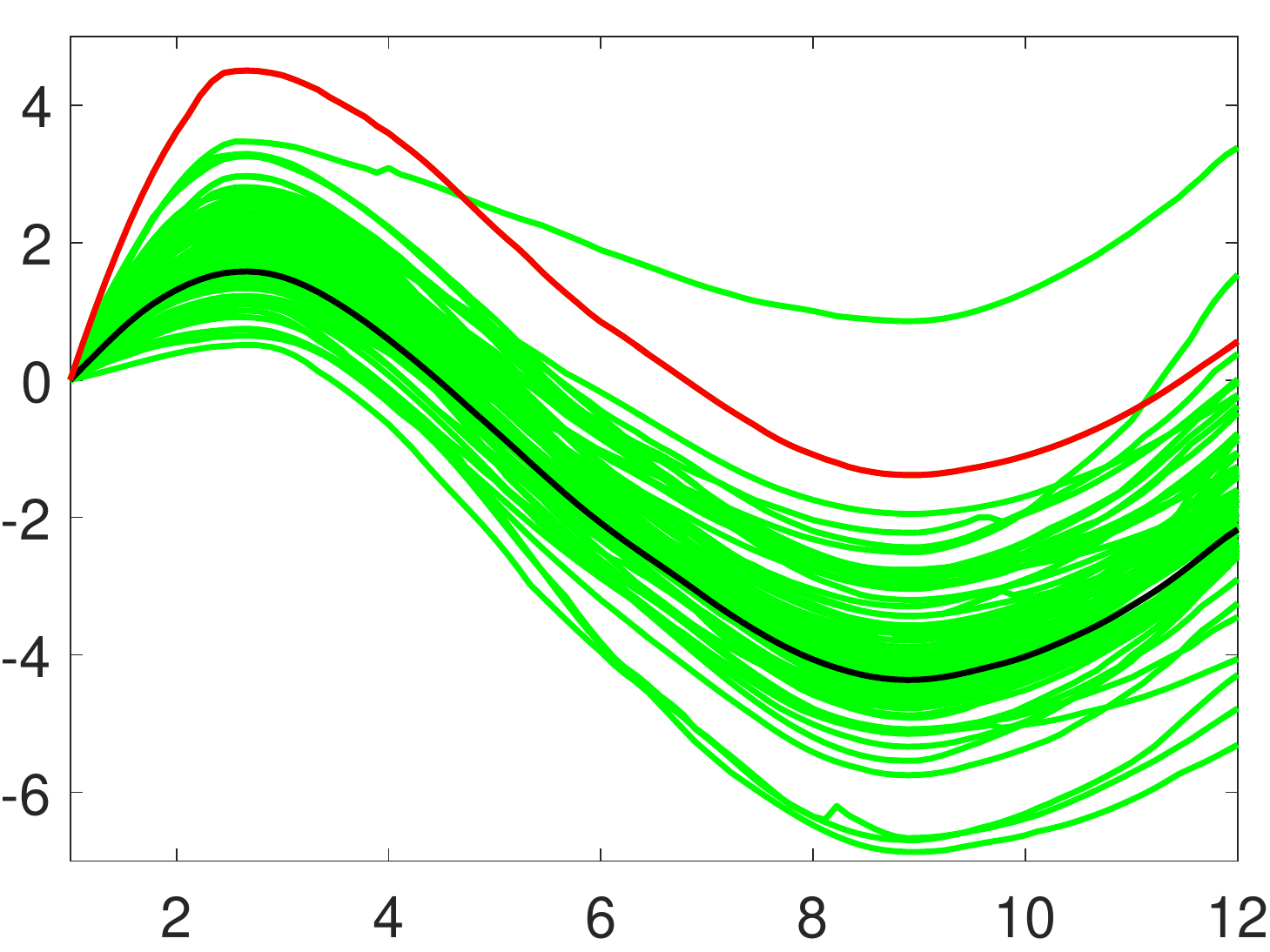}&\includegraphics[width=1.3in]{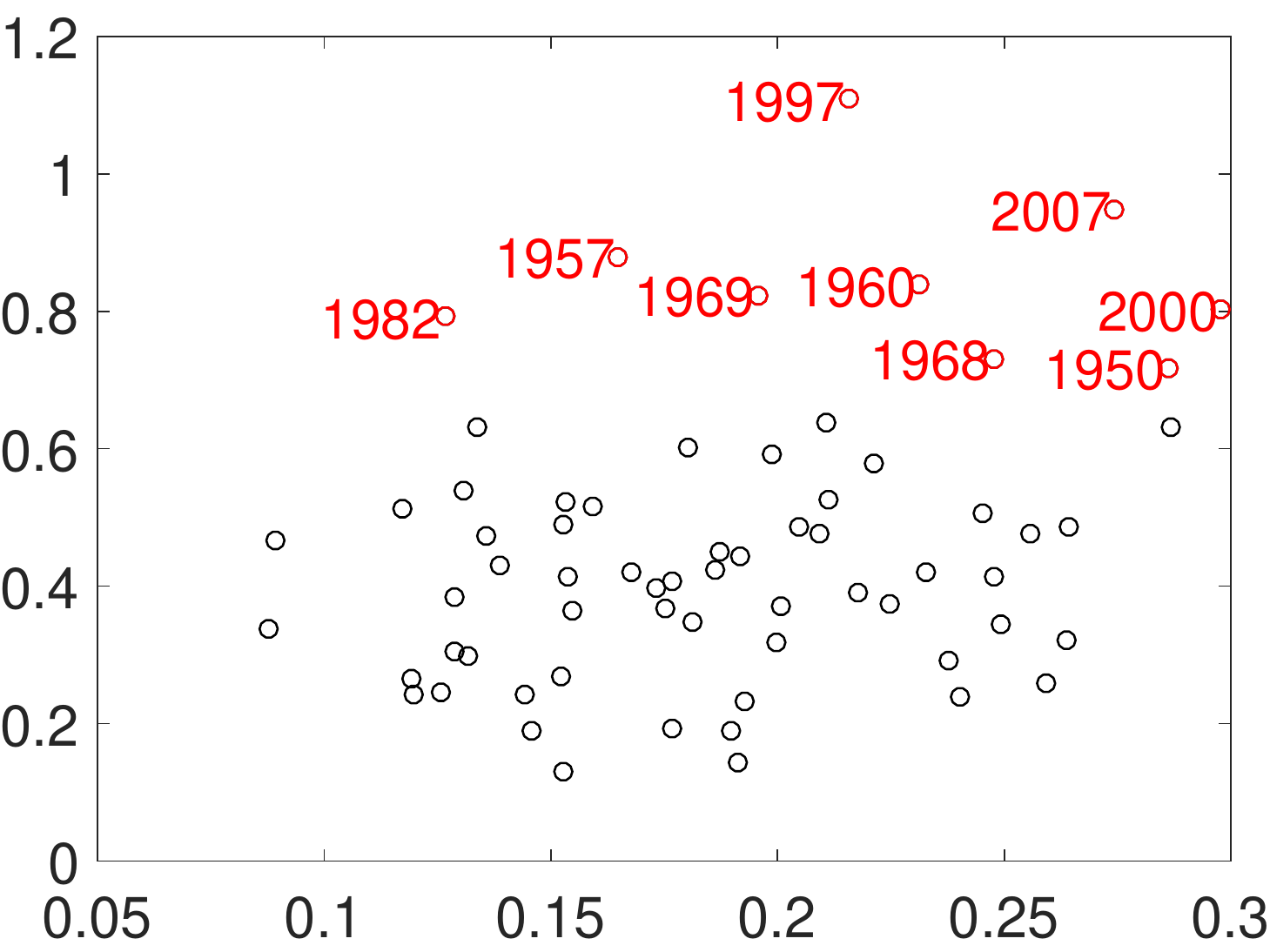}&\includegraphics[width=1.3in]{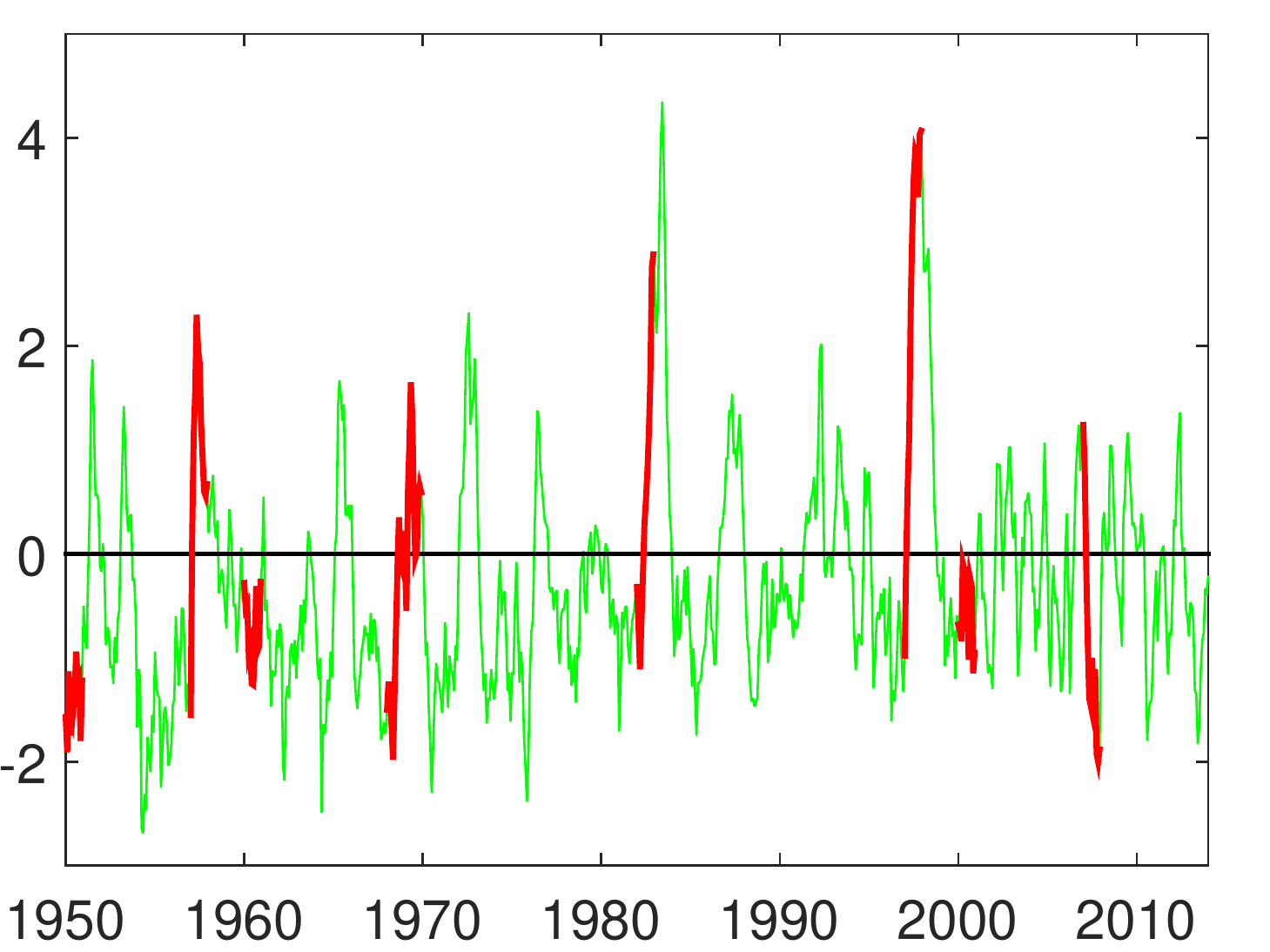}&\includegraphics[width=1.5in]{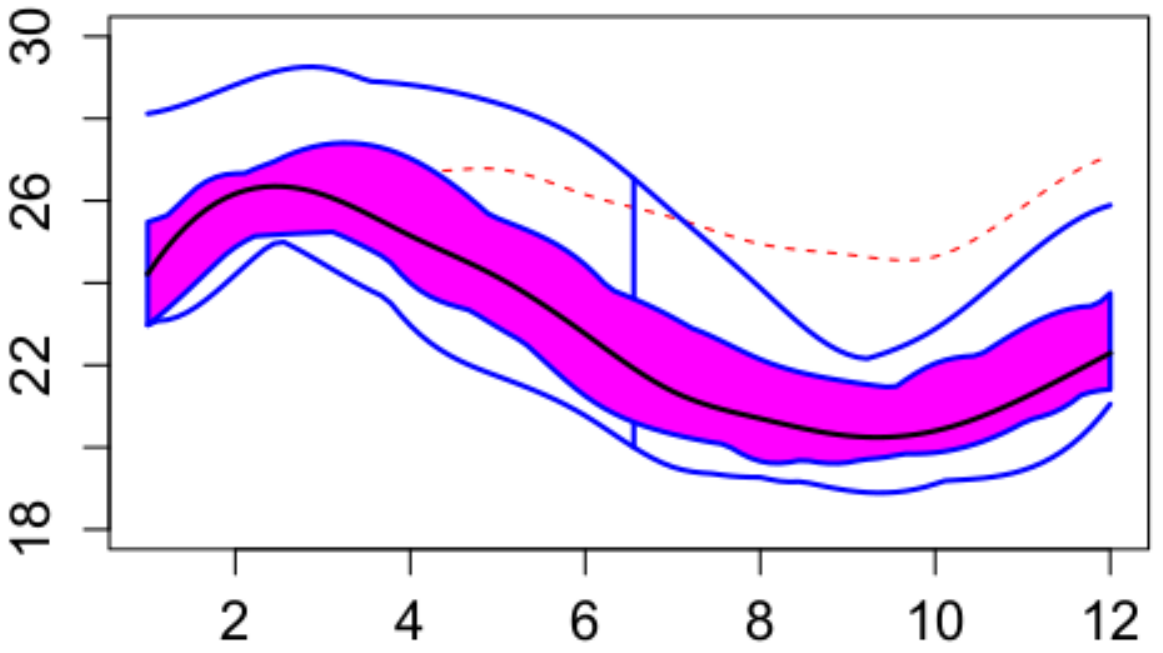}\\
    \hline
    \end{tabular}
    \end{small}
    \caption{\small (a) Translation outlier: Year 1983. (b) Translation outlier: Year 1997. (c) Amplitude outlier: Year 1997. (d) Mild amplitude outlier: Year 2007. (e) Mild amplitude outlier: Year 1957. (f) Plot of phase ($x$-axis) vs. amplitude ($y$-axis) distances of each function in the data from the median. (g) Plot of annual sea surface temperature anomalies used to identify El Ni\~{n}o and La Ni\~{n}a. (h) Functional boxplot generated by the method of \cite{citeulike:10107686}.}\label{fig:ninoampoutlier}
    \end{center}
    \vspace{-5mm}
\end{figure}

As a final set of results on this dataset, we study outlying SST years. First, in Figure \ref{fig:ninoampoutlier}(f), we display a scatterplot of the amplitude ($D_a$, $y$-axis) and phase ($D_p$, $x$-axis) distances of each annual SST function from the median. Large distances on either the $x$- or $y$-axis may imply the presence of phase or amplitude outliers. We label a few points of interest with large amplitude distances in the plot, because it is very effective in data exploration. We proceed by using the proposed formal outlier detection procedures to examine whether any amplitude or phase outliers are present in the data. The translation boxplot shows that the years 1997 and 1983 are detected as translation outliers. Our procedure also flagged 1997 as a regular amplitude outlier, and 2007 and 1957 as mild amplitude outliers. No phase outliers were detected in this data due to mild interannual variability, which indicates that across years, no single hot or cold month comes at a very exceptional time compared to other years. Figure \ref{fig:ninoampoutlier}(a)\&(b) displays the original SST functions for 1983 and 1997 in red with all other SST functions plotted in green in the background. These translation outliers had average annual SSTs that were significantly higher than those of other years in this dataset. Figure \ref{fig:ninoampoutlier}(c)-(e) displays the amplitude components of all functions in green and the outliers in red. As amplitude outliers, the SST functions of 1997, 2007 and 1957 had significantly different `shapes' from the other years. It is important to note that the 1997 amplitude outlier has a very small downward slope during the middle of the year resulting in abnormally high SST in the winter months, behavior that is typical for El Ni\~{n}o events. On the other hand, the 2007 mild amplitude outlier has a very steep downward slope during the middle of the year, resulting in abnormally low SST in the winter months; this behavior is more consistent with a La Ni\~{n}a event. Figure \ref{fig:ninoampoutlier}(h) shows the functional boxplot of \cite{citeulike:10107686} applied to this dataset.

We can confirm the validity of the reported results based on what is currently known about strong El Ni\~{n}o and La Ni\~{n}a events. To do this, we use the sea surface temperature anomaly plot shown in Figure \ref{fig:ninoampoutlier}(g), where we have highlighted the years identified in Figure \ref{fig:ninoampoutlier}(f) in red. Years with abnormally high sea surface temperatures, which correspond to El Ni\~{n}o events, have high positive peaks in this plot (e.g., 1983 and 1997). Years with abnormally low sea surface temperatures, which correspond to La Ni\~{n}a events, have low negative peaks (e.g., 1954, 1964, 1971, 1976, and 2007). According to a National Climatic Data Center report, the winter of 1997-1998 was the second warmest and seventh wettest winter since 1895, which also corresponds to a record breaking El Ni\~{n}o event\footnote{http://www1.ncdc.noaa.gov/pub/data/techrpts/tr9802/tr9802.pdf}. Furthermore, the 1982-1983 El Ni\~{n}o is considered one of the strongest El Ni\~{n}o events since the collection of records began\footnote{http://www.fcst-office.com/HardRock/Meteo241/El\%20Nino\%201982-1983/ProjectThree.html}. While we are not able to flag all significant El Ni\~{n}o or La Ni\~{n}a years as translation or amplitude outliers, we are able to detect some of the strongest events. We hypothesize that this is due to the events stretching over multiple years, which is not captured in our annual SST data. In the future, we propose to extend this study using a multiscale approach, where multiyear SST functions are used for visualization and detection of amplitude and phase outliers.

\subsection{Real Data Study 2: Berkeley Growth Data}

\begin{figure}[!t]
\begin{center}
\begin{small}
    \begin{tabular}{|c|c|c|c|}
    \hline
    (a)&(b)&(c)&(d)\\
    \hline
    \includegraphics[width=1.3in]{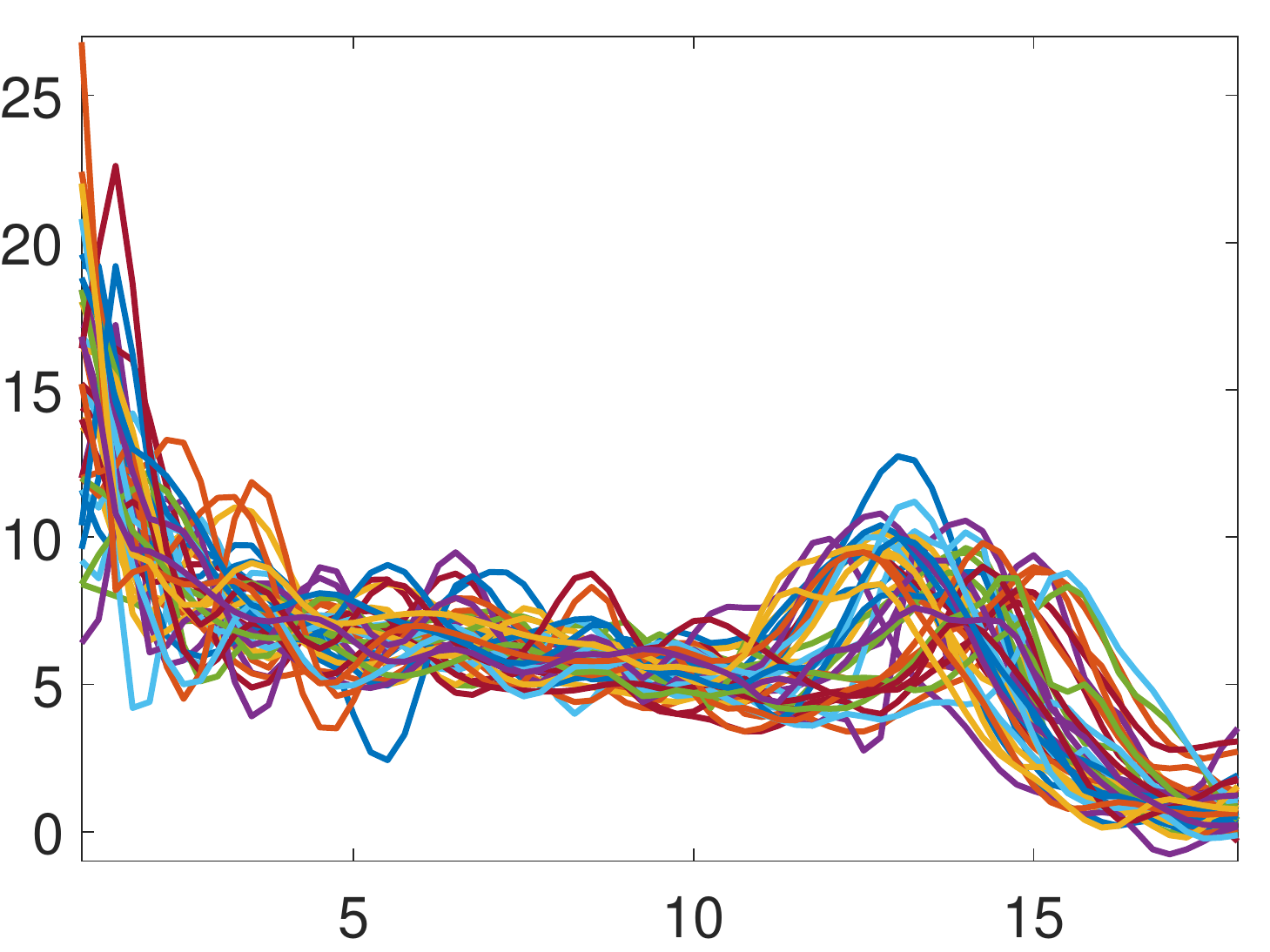}&\includegraphics[width=1.3in]{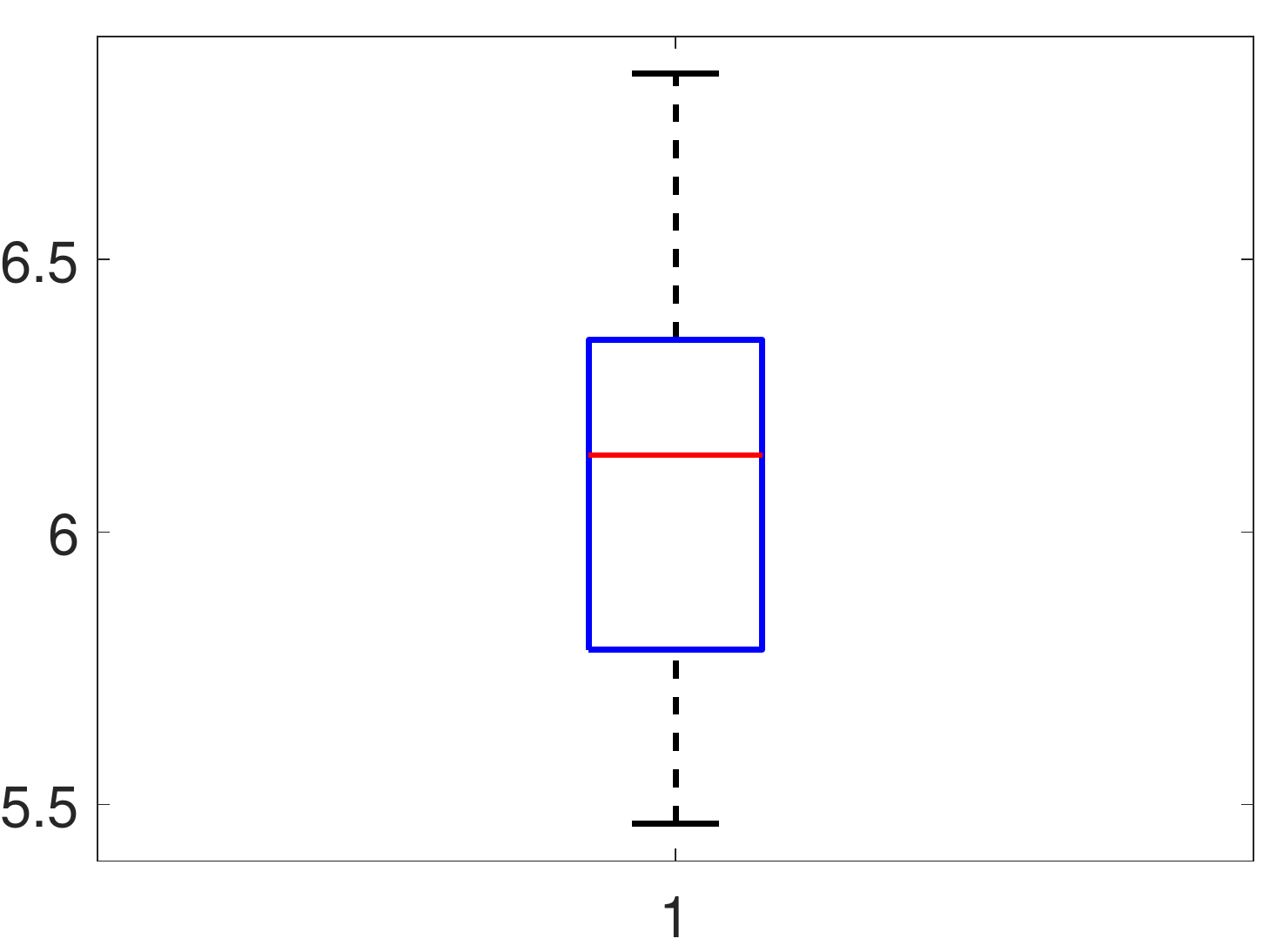}&\includegraphics[width=1.3in]{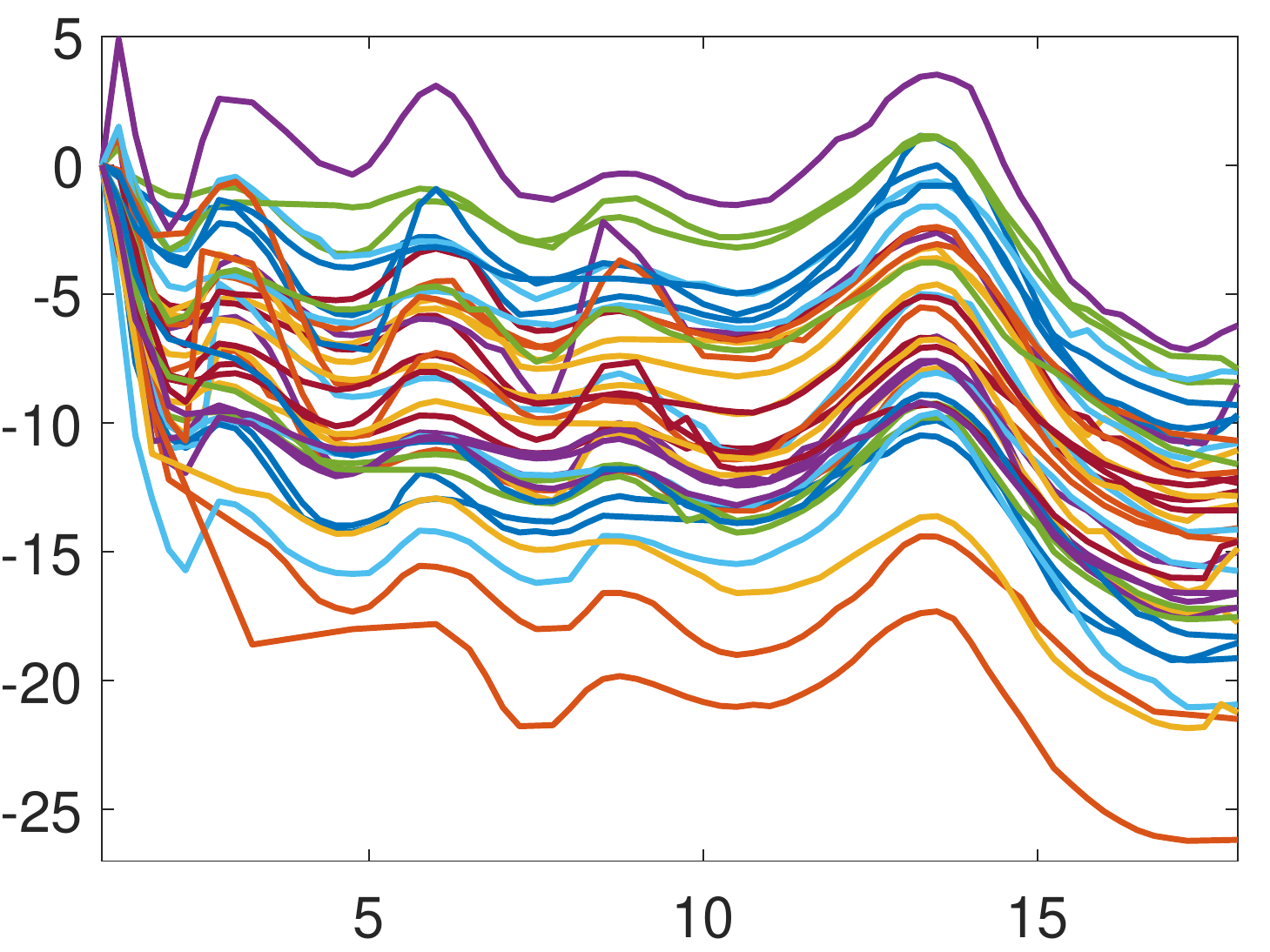}&\includegraphics[width=1.3in]{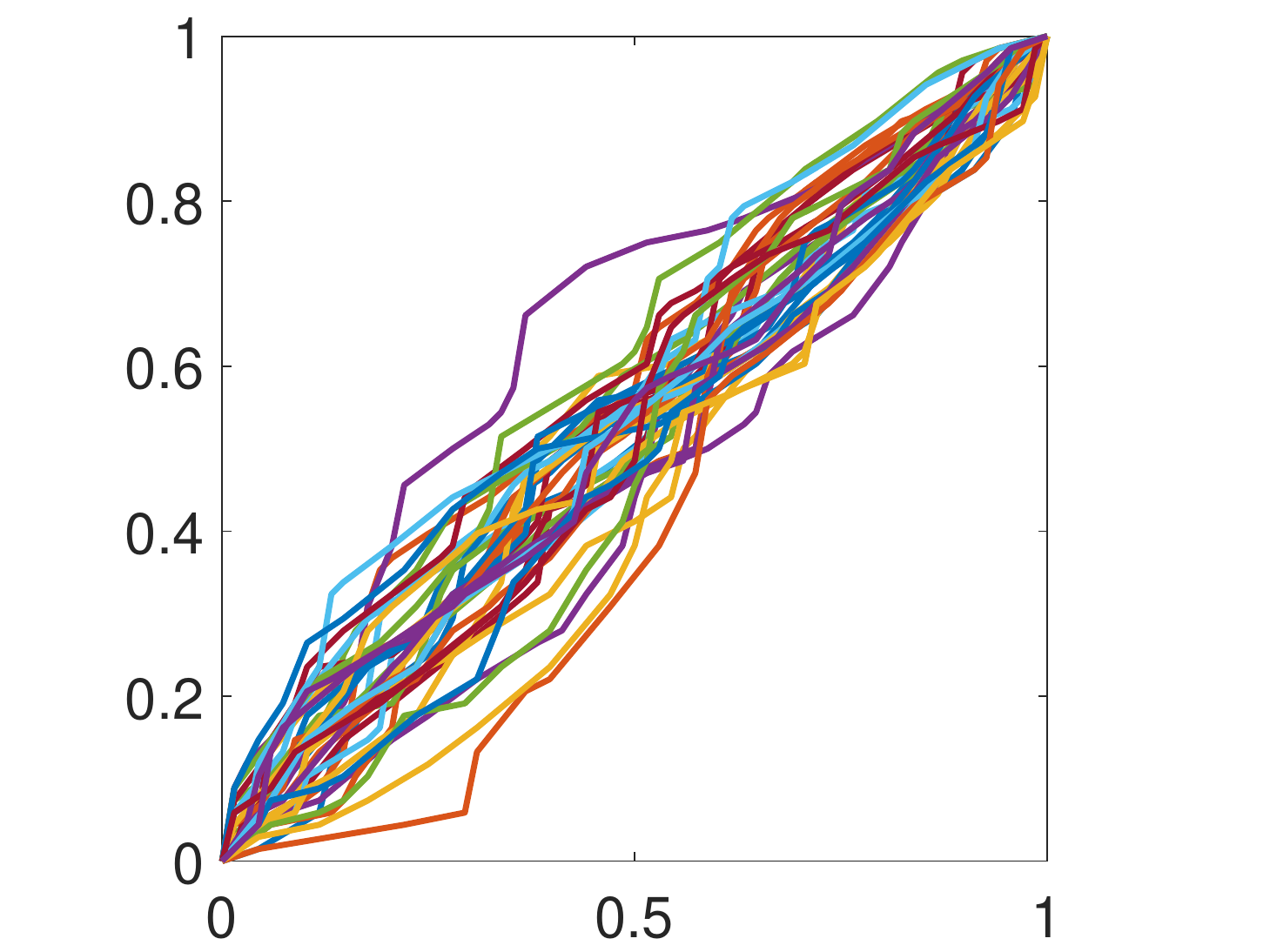}\\
    \hline
    \hline
    \includegraphics[width=1.3in]{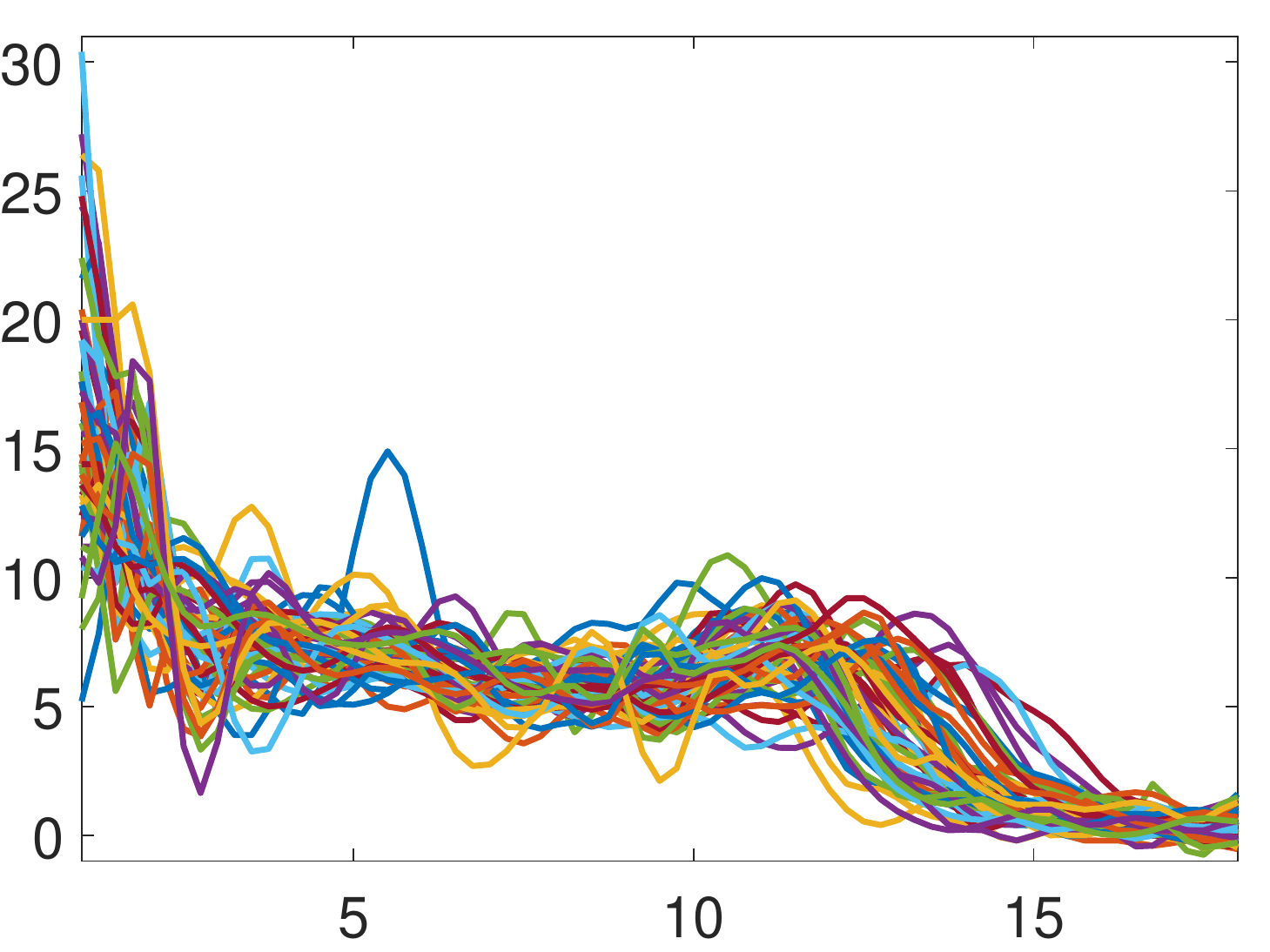}&\includegraphics[width=1.3in]{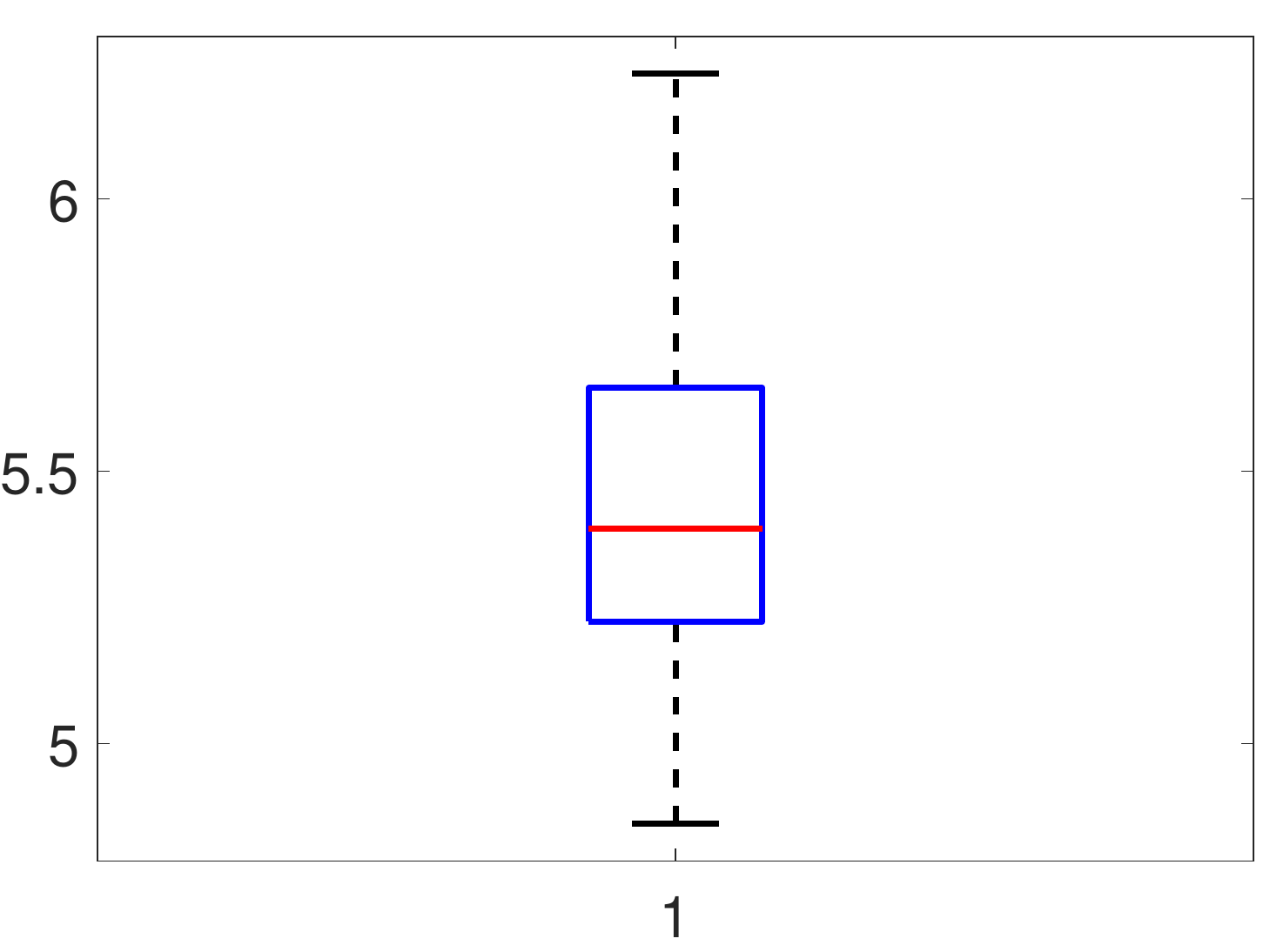}&\includegraphics[width=1.3in]{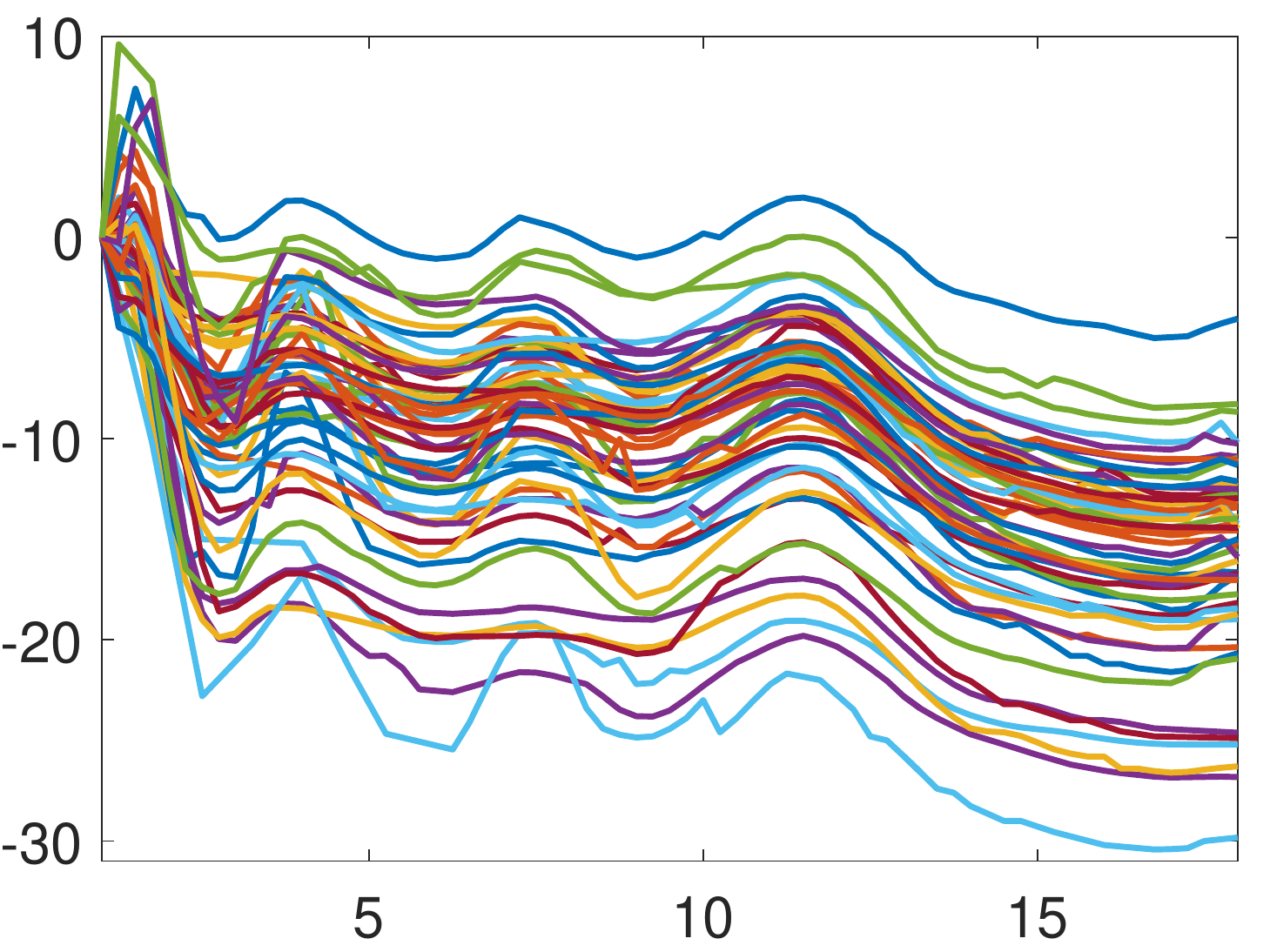}&\includegraphics[width=1.3in]{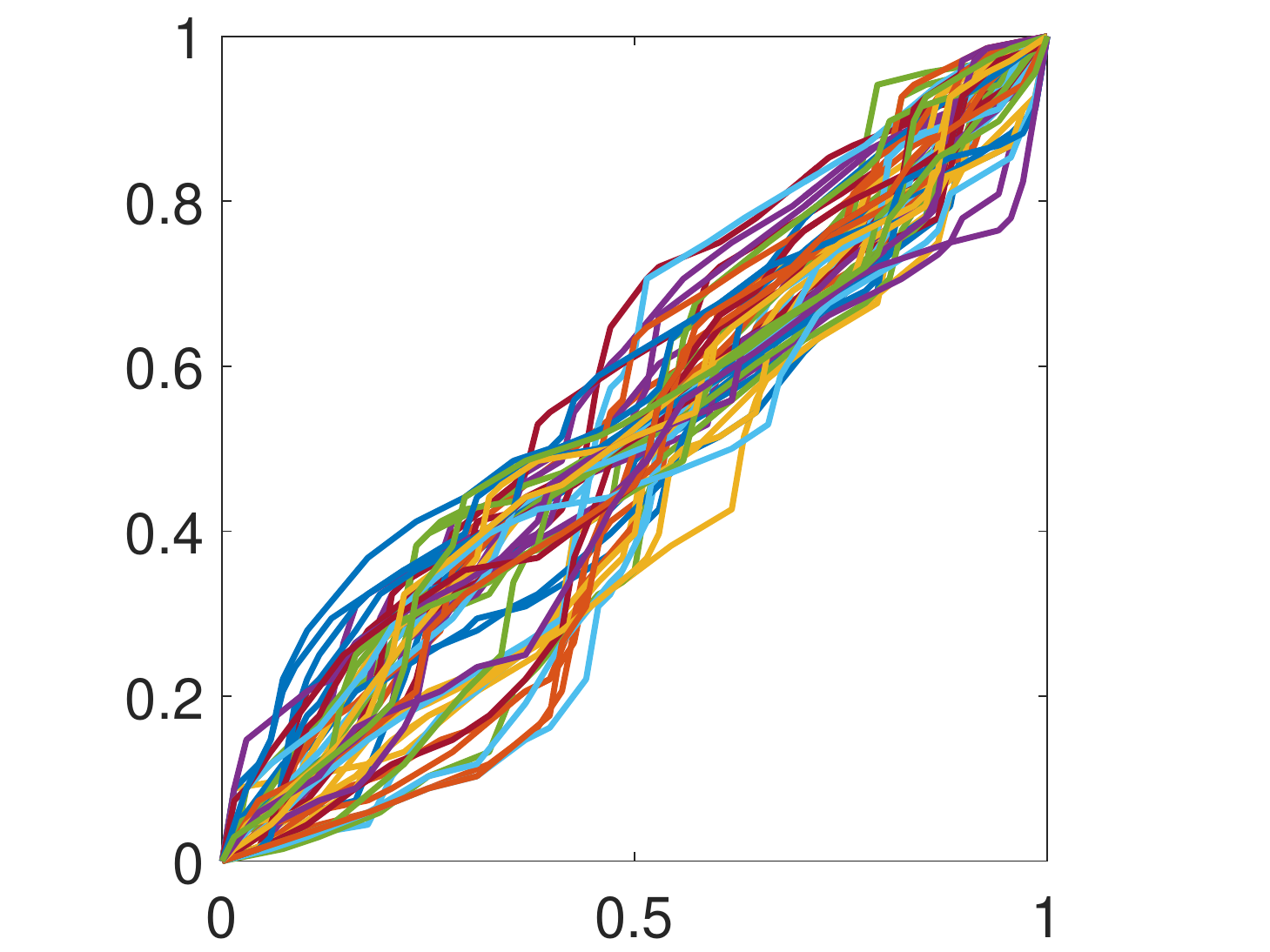}\\
    \hline
    \end{tabular}
    \end{small}
    \caption{\small Separation of translation, amplitude, and phase in the Berkeley growth data (top=boys; bottom=girls). (a) Original functions. (b) Translation. (c) Amplitude. (d) Phase.}\label{fig:bgd}
    \end{center}
    \vspace{-5mm}
\end{figure}

The Berkeley growth data is a collection of height growth functions for 39 boys and 54 girls from birth up to 18 years old (\cite{BGD}). We first transfer these original growth functions to growth rate functions by taking a derivative. Figure \ref{fig:bgd}(a) shows the original growth velocity functions for boys (top) and girls (bottom). This is another example where clear warping variability is present in the data; that is, individual boys and girls have growth spurts (peaks in the velocity functions) at different times. If this warping variability is not taken into account, the overall structure of the data can be destroyed when computing summary statistics such as the mean or median. Figure \ref{fig:bgd} shows the decomposition of the growth velocity functions into translation, amplitude, and phase components. We find significant phase variability in both boy and girl groups. %Figure \ref{fig:bgdrainbow} shows all rainbow plots for amplitude and phase. Again the amplitude and phase functions are colored according to their distance from the median (displayed in black) with blue corresponding to small distances and red corresponding to large distances.

%\begin{figure}[!t]
%\begin{center}
%\begin{small}
%    \begin{tabular}{|c|c|c|c|}
%    \hline
%    (a)&(b)&(c)&(d)\\
%    \hline
%    \includegraphics[width=1.3in]{berkeley_male_amplitude_rainbow.pdf}&\includegraphics[width=1.3in]{berkeley_female_amplitude_rainbow.pdf}&\includegraphics[width=1.3in]{berkeley_male_warping_rainbow.pdf}&\includegraphics[width=1.3in]{berkeley_female_warping_rainbow.pdf}\\
%    \hline
%    \end{tabular}
%    \end{small}
%    \caption{\small (a) Boy amplitude rainbow plot. (b) Girl amplitude rainbow plot. (c) Boy phase rainbow plot. (d) Girl phase rainbow plot.}\label{fig:bgdrainbow}
%    \end{center}
%    \vspace{-5mm}
%\end{figure}

\begin{figure}[!t]
\begin{small}
\begin{center}
    \begin{tabular}{|cc|cc|}
    \hline
    (a)&(b)&(c)&(d)\\
    \hline
    \includegraphics[width=1.3in]{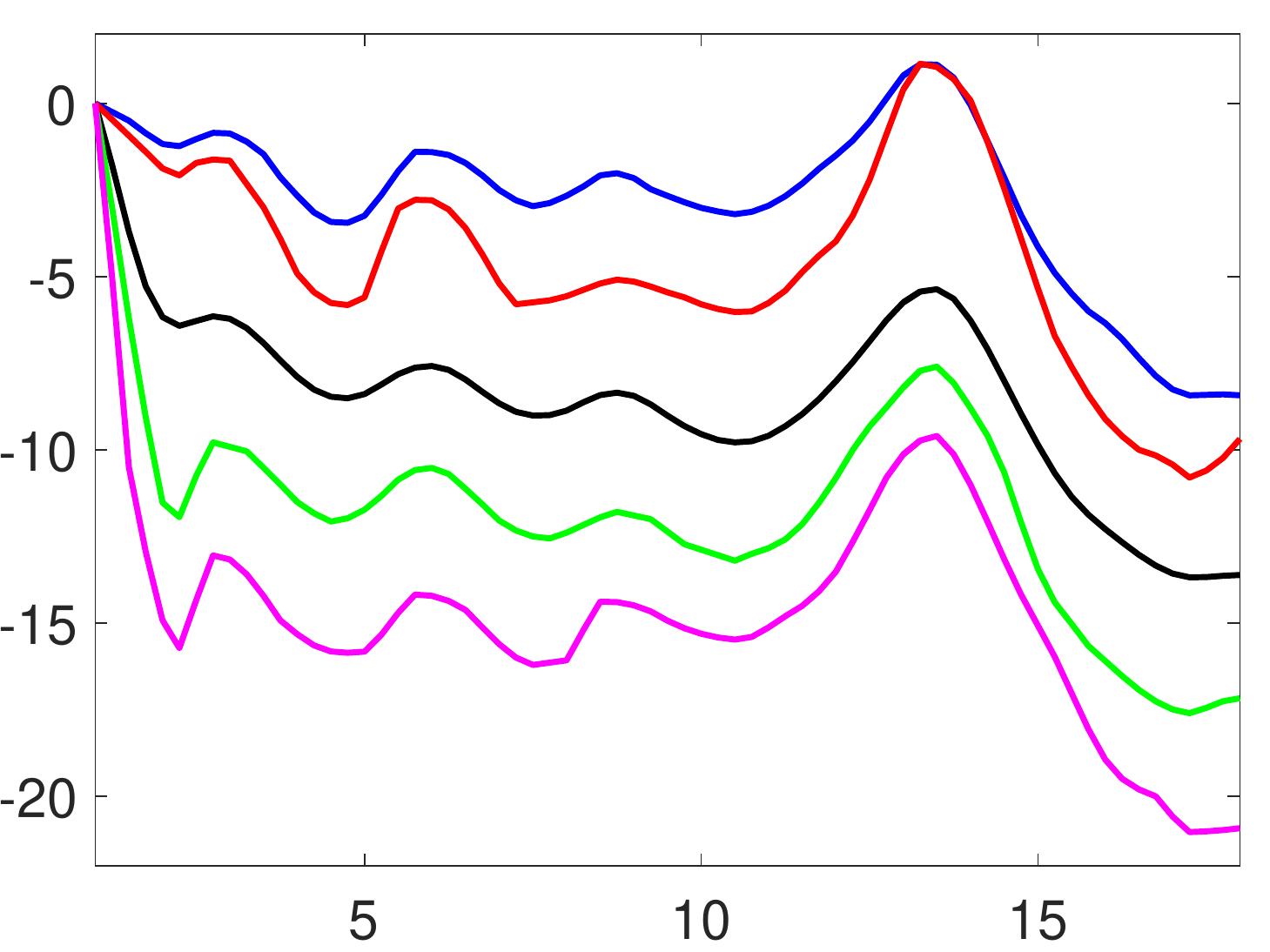}&\includegraphics[width=1.3in]{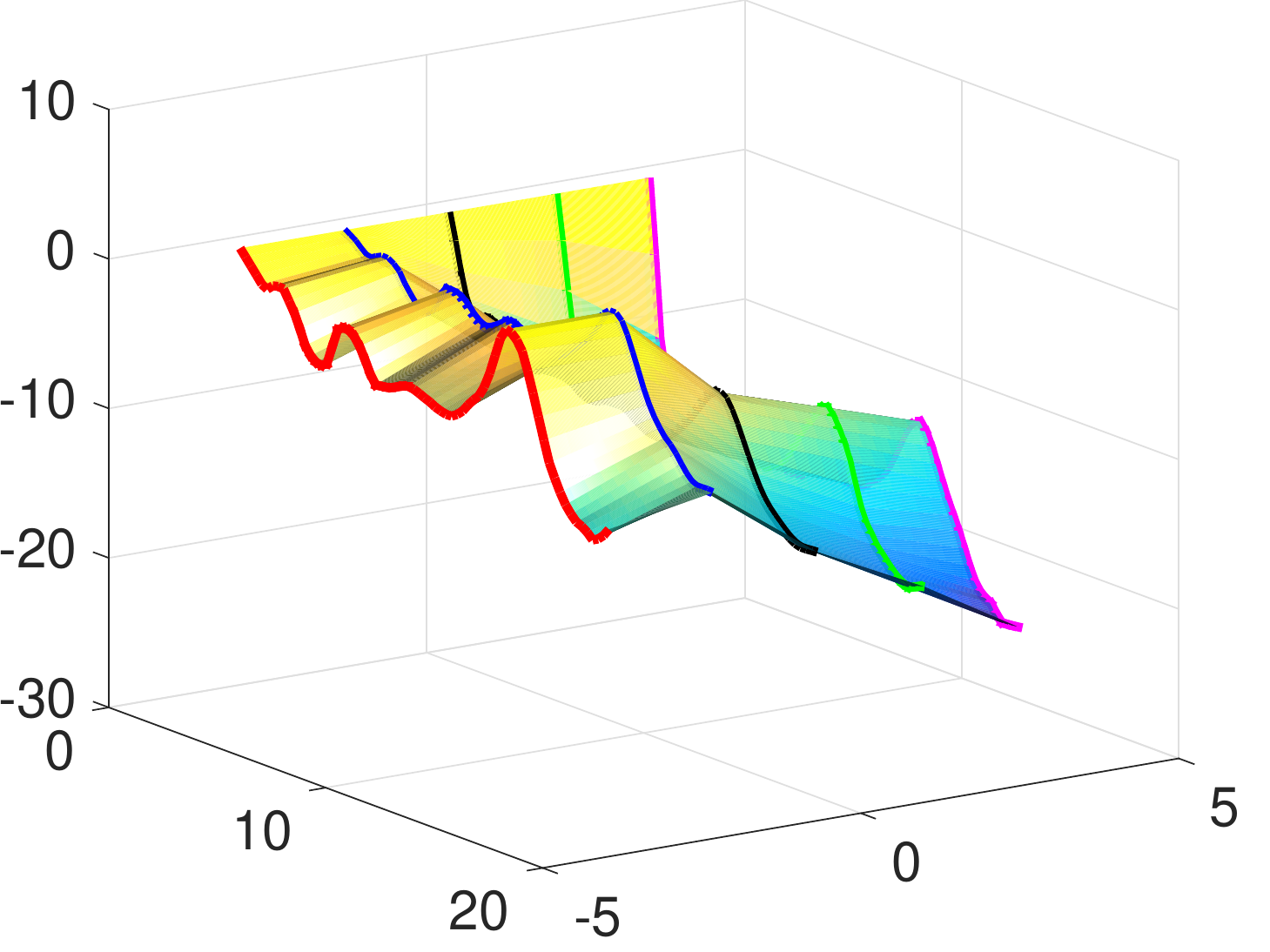}&\includegraphics[width=1.3in]{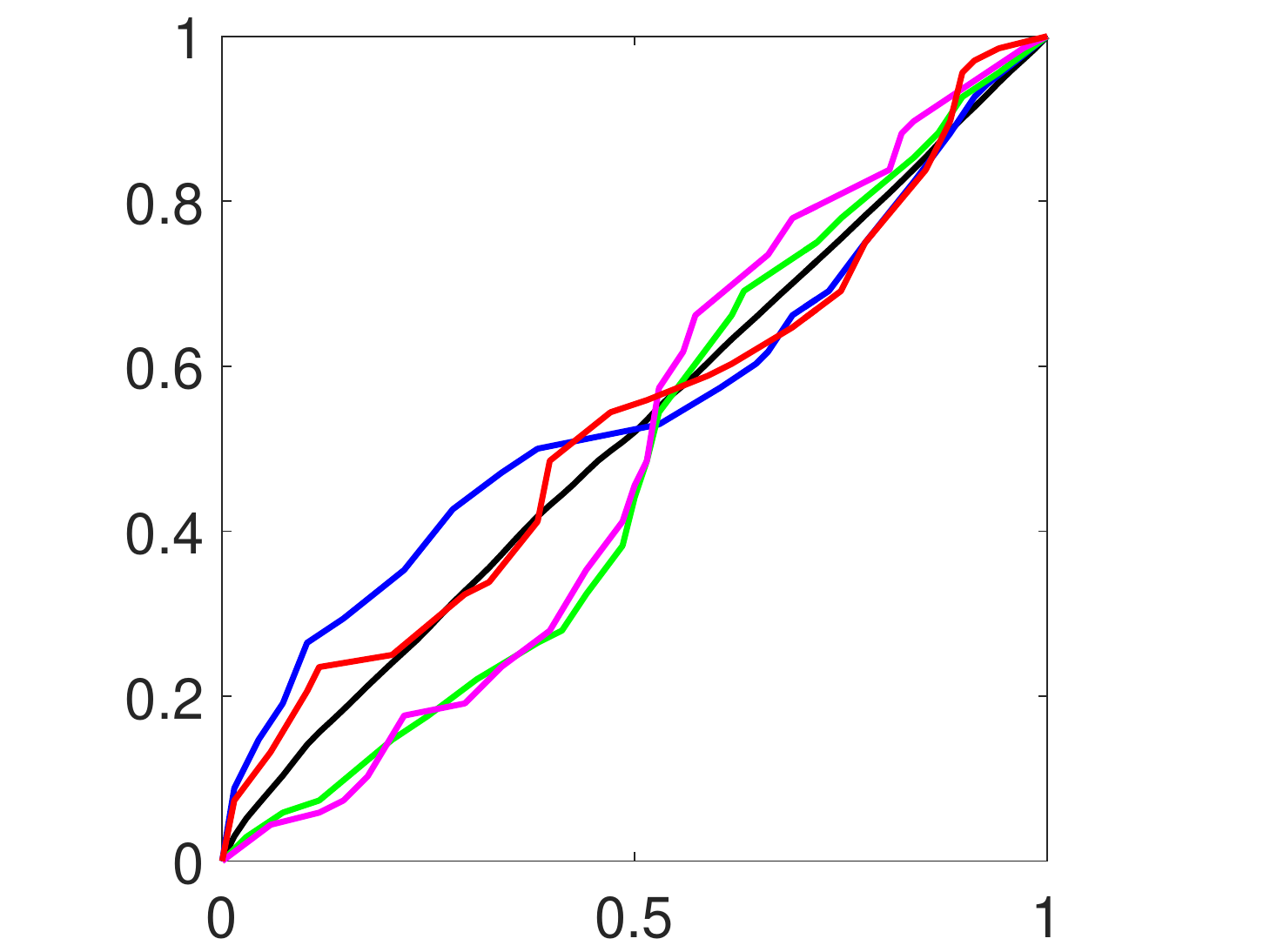}&\includegraphics[width=1.3in]{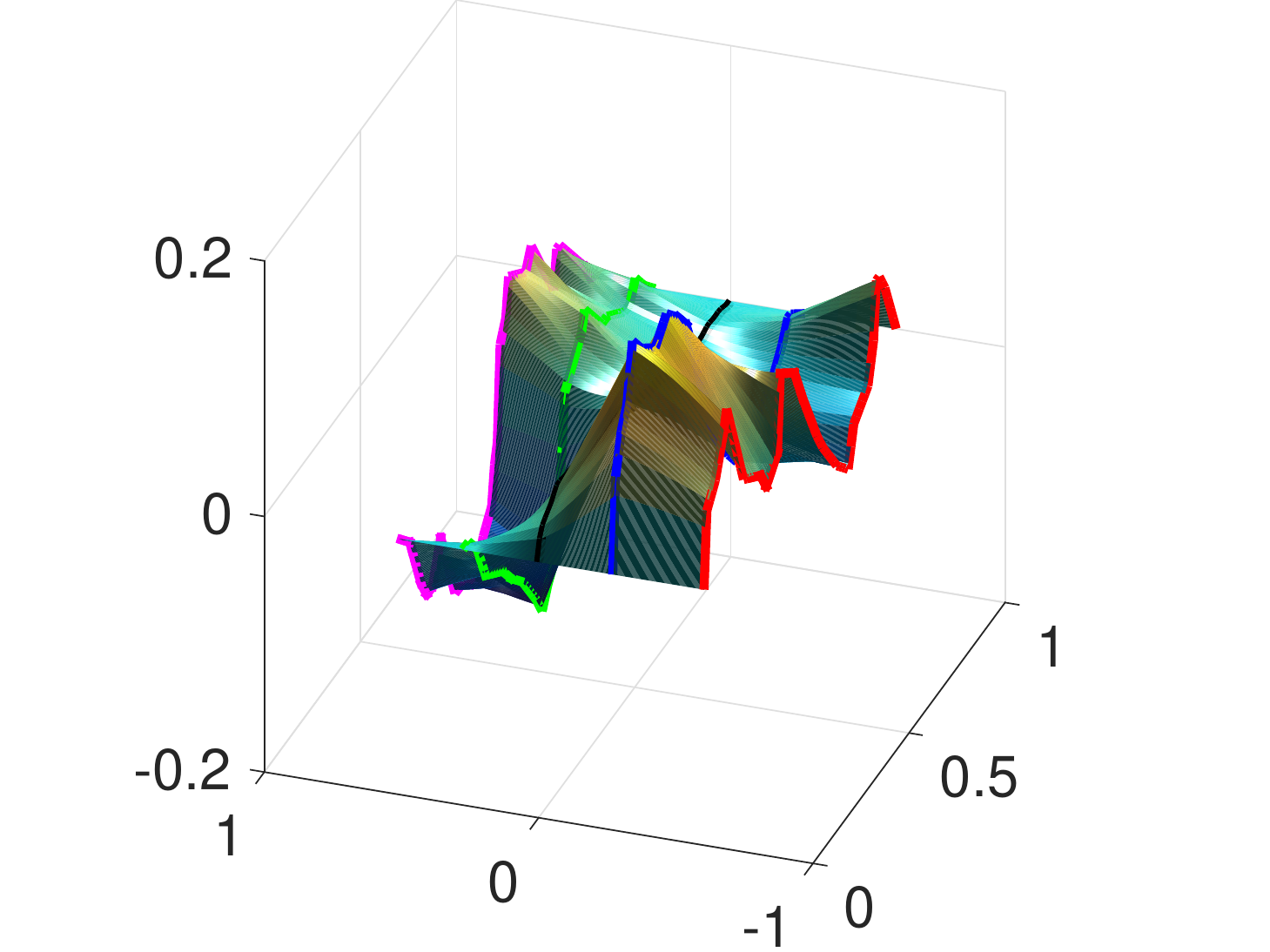}\\
    \hline
    \includegraphics[width=1.3in]{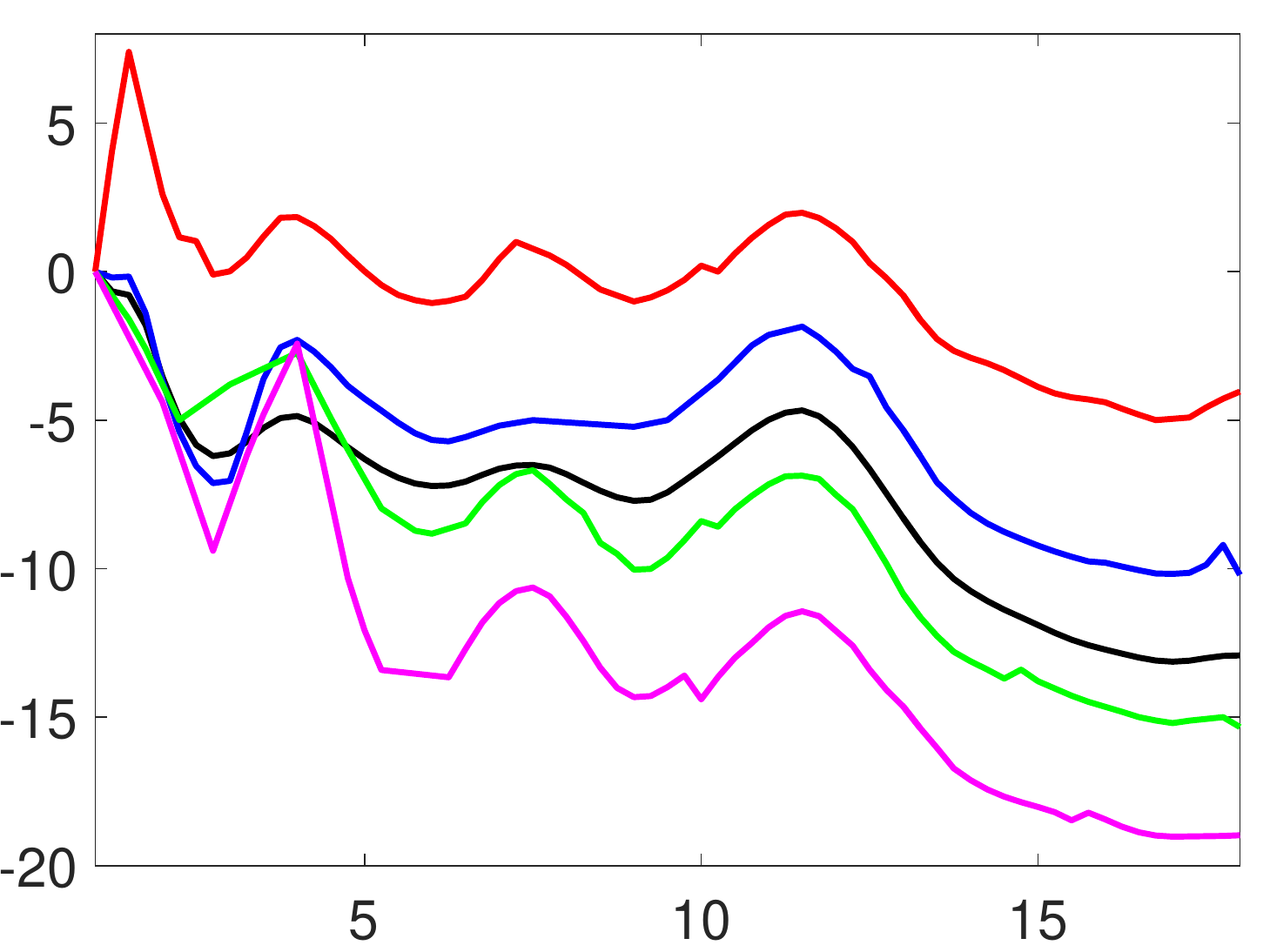}&\includegraphics[width=1.3in]{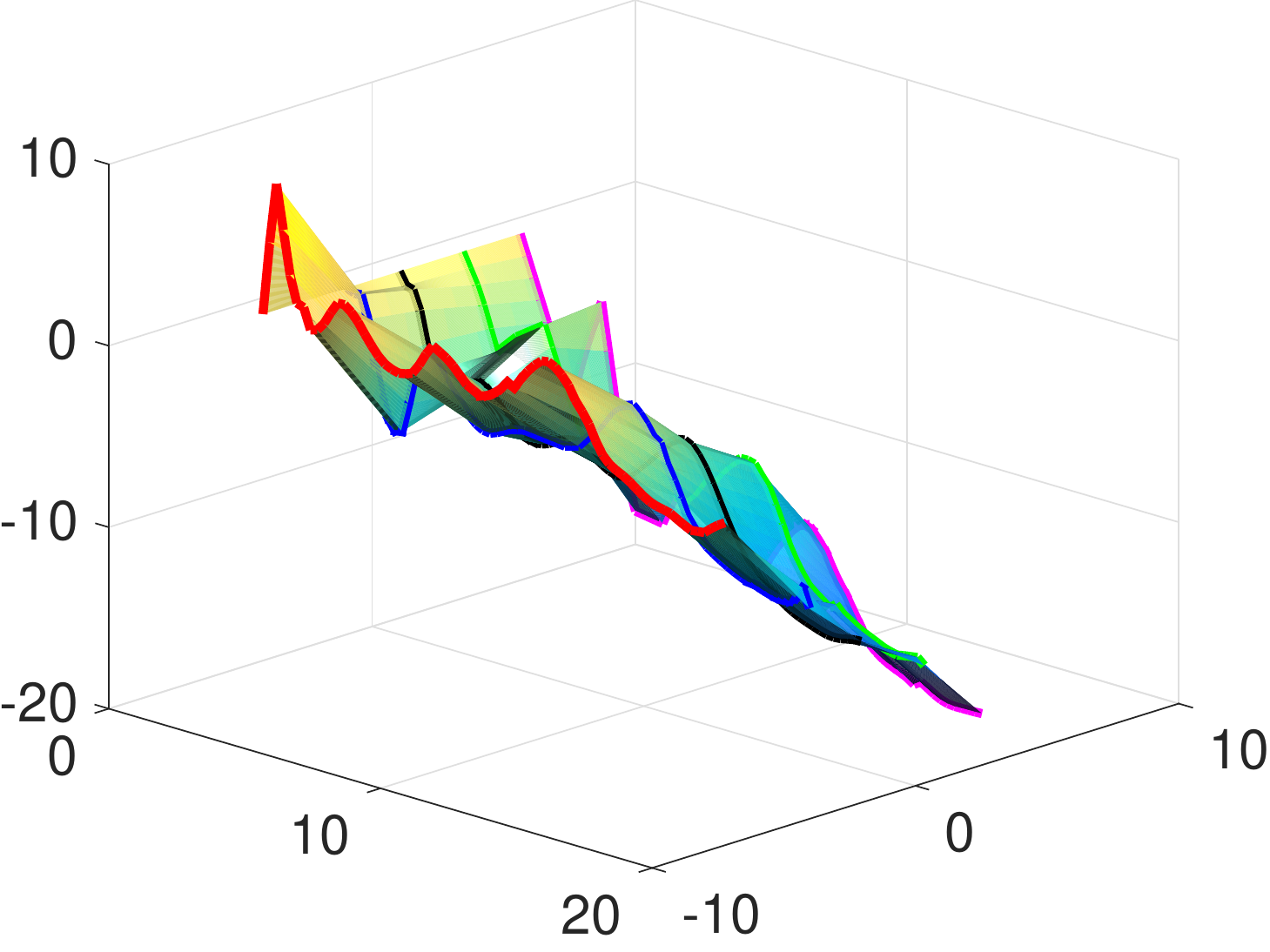}&\includegraphics[width=1.3in]{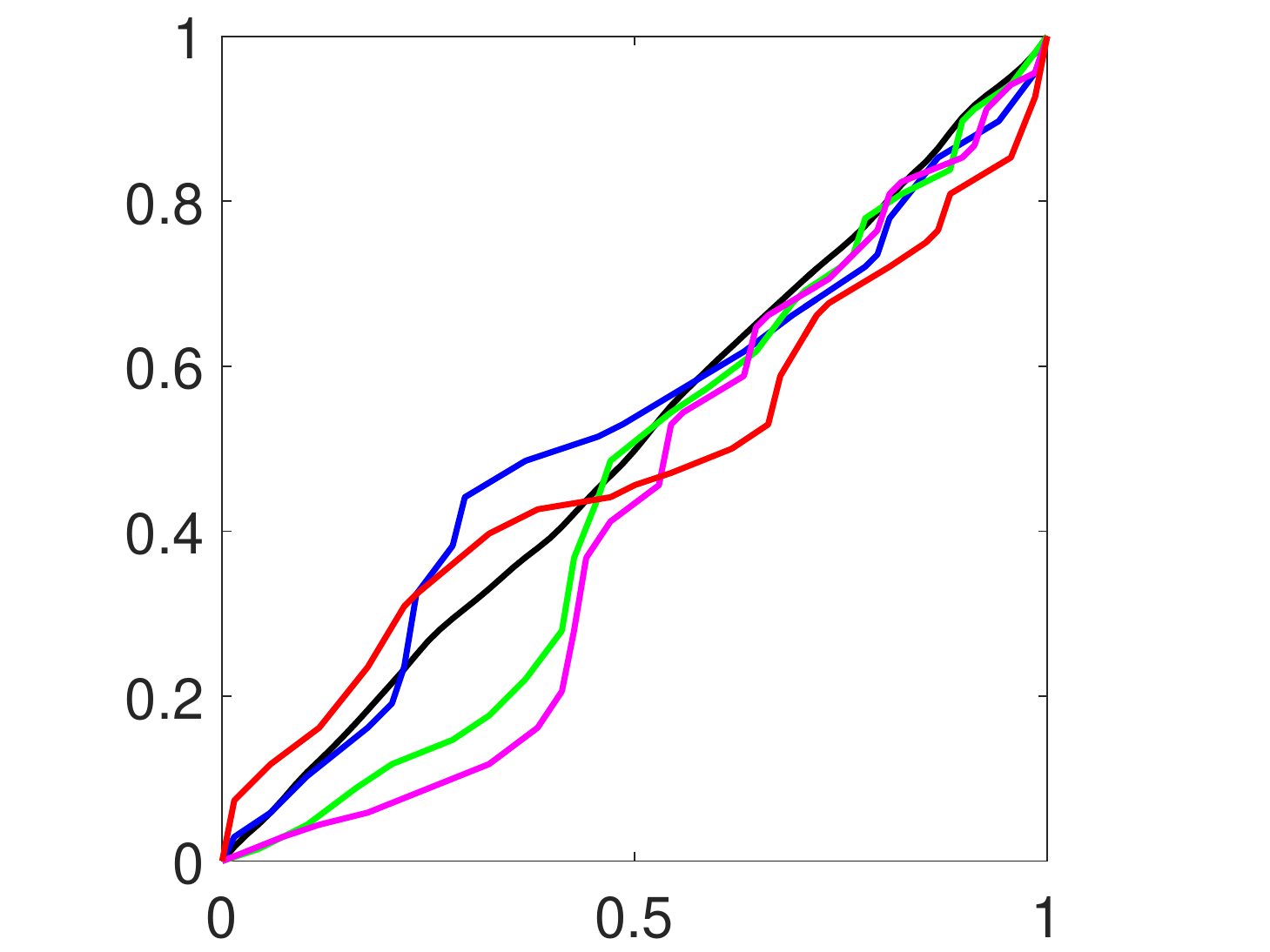}&\includegraphics[width=1.3in]{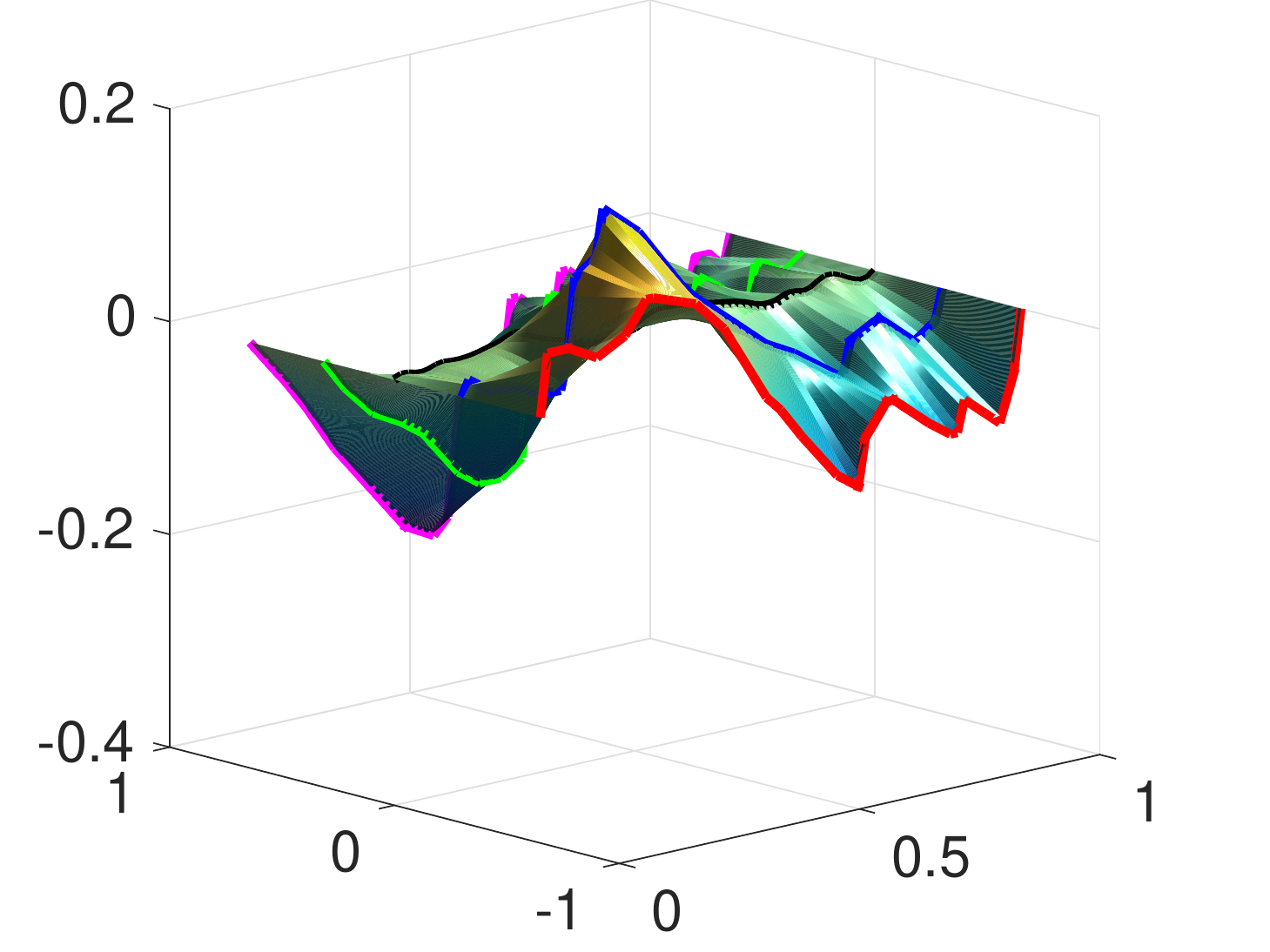}\\
    \hline
%    (a)&(b)&(c)\\
%    \hline
%    \includegraphics[width=1.3in]{berkeley_female_amplitude_minmax_1.pdf}&\includegraphics[width=1.3in]{berkeley_female_amplitude_minmax_2.pdf}&\includegraphics[width=1.3in]{berkeley_female_amplitude_minmax_3.pdf}\\
%    \hline
%    (d)&(e)&(f)\\
%    \hline
%    \includegraphics[width=1.3in]{berkeley_female_amplitude_minmax_4.pdf}&\includegraphics[width=1.3in]{berkeley_female_amplitude_minmax_5.pdf}&\includegraphics[width=1.3in]{berkeley_female_amplitude_minmax_surface.pdf}\\
%    \hline
    \end{tabular}
    \caption{\small Amplitude and phase boxplot displays for velocity growth curves in the Berkeley growth data (top=boys; bottom=girls). (a)\&(b) Amplitude boxplot and its surface display. (c)\&(d) Phase boxplot and its surface display.}\label{fig:bgdamplitude}
    \end{center}
    \end{small}
    \vspace{-5mm}
\end{figure}

Figure \ref{fig:bgdamplitude} shows the amplitude and phase boxplots for boys and girls. In Figure \ref{fig:bgdamplitude}(a)\&(b), the median amplitude (black) captures four growth spurts for boys (the first three being fairly small) and three growth spurts for girls (the first two being fairly small). The amplitude boxplot for girls (bottom) captures a very interesting source of variability where growth velocity functions mainly differ in the number and sizes of growth spurts. For boys (top), the main variability is in the initial growth velocity and the sizes of the growth spurts. In Figure \ref{fig:bgdamplitude}(c)\&(d), the phase boxplots show a lot of initial phase variability in both boys and girls, which stabilizes after approximately nine years of age. This initial variability shows natural timing variation in growth; for example, the magenta phase extreme for boys (top) implies that the corresponding boy's initial growth spurts occurred earlier than the median, while the last growth spurt occurred at a very similar time to the median.

\begin{figure}[!t]
\begin{small}
\begin{center}
    \begin{tabular}{|cc|ccc|}
    \hline
    \multicolumn{2}{|c|}{(a)}&\multicolumn{3}{|c|}{(b)}\\
    \hline
    \includegraphics[width=1.0in]{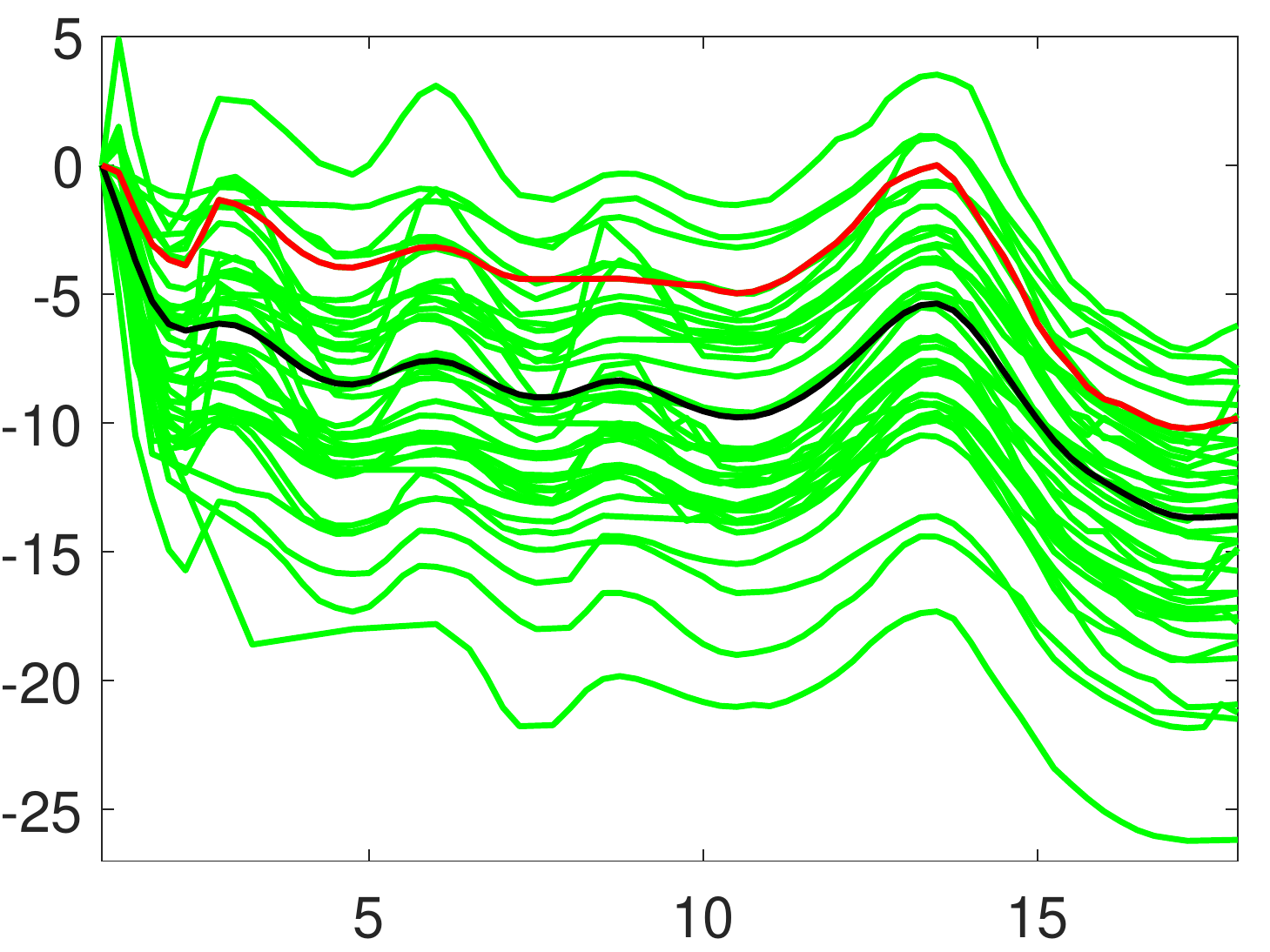}&\includegraphics[width=1.0in]{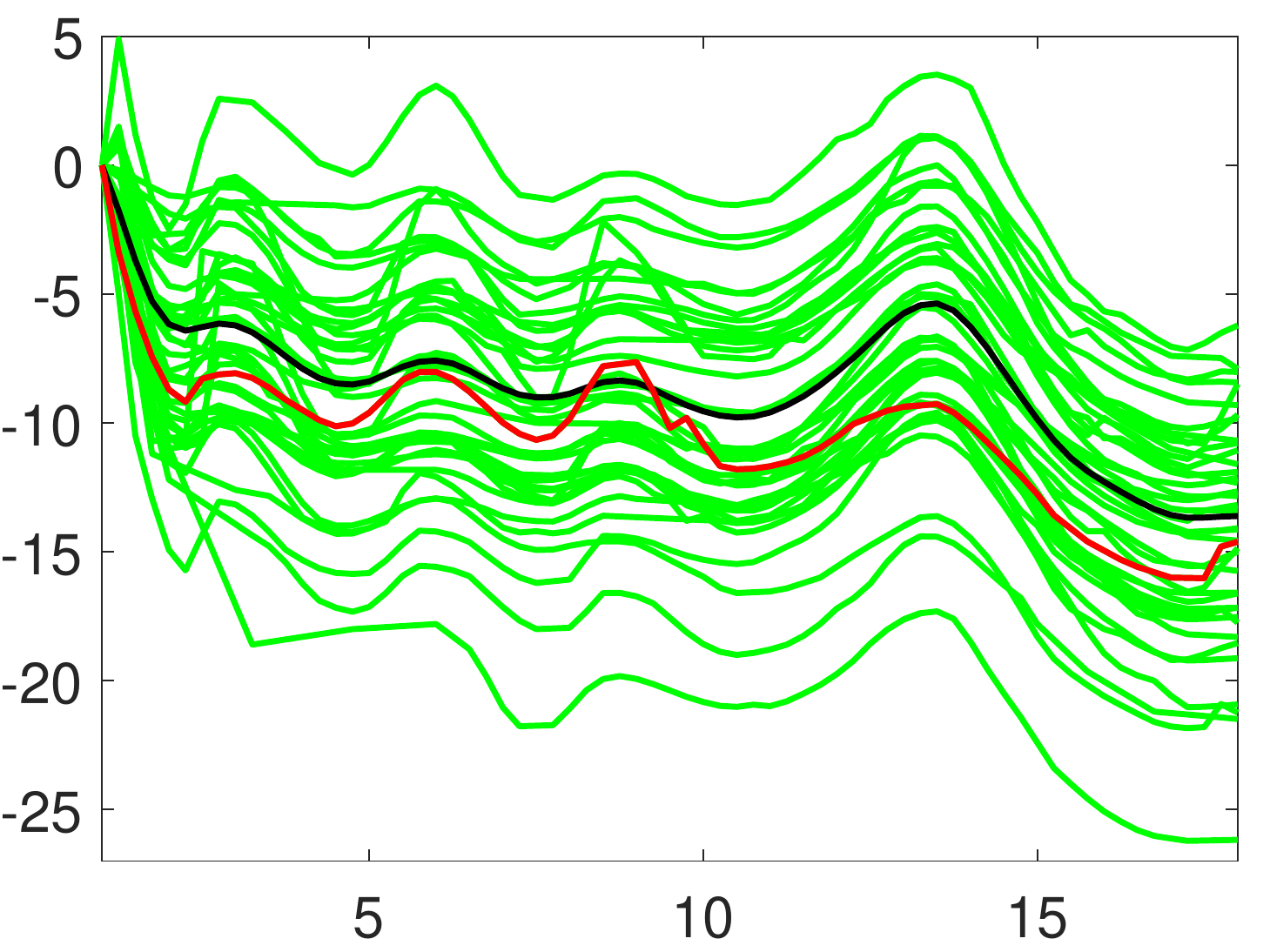}&\includegraphics[width=1.0in]{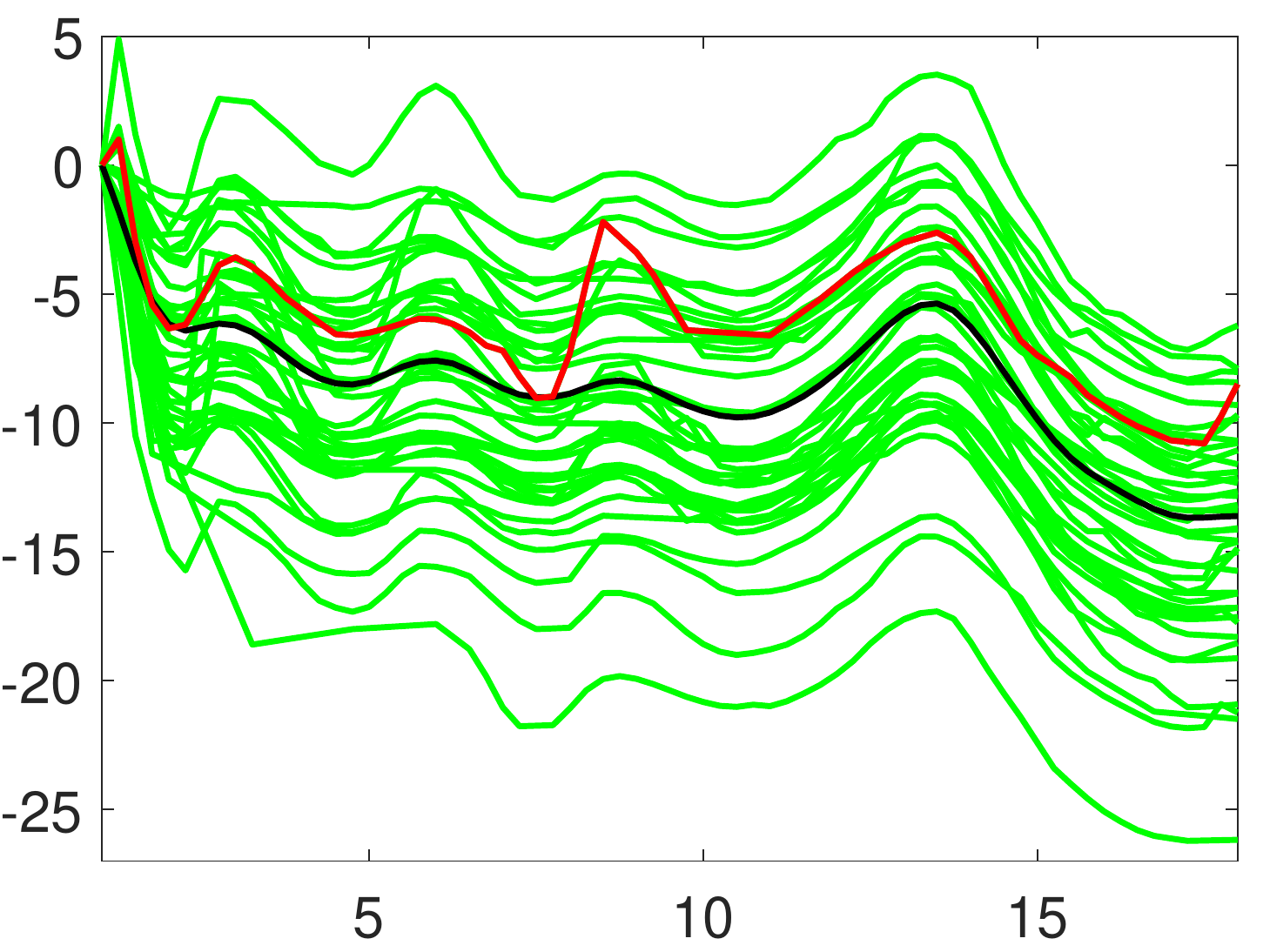}&\includegraphics[width=1.0in]{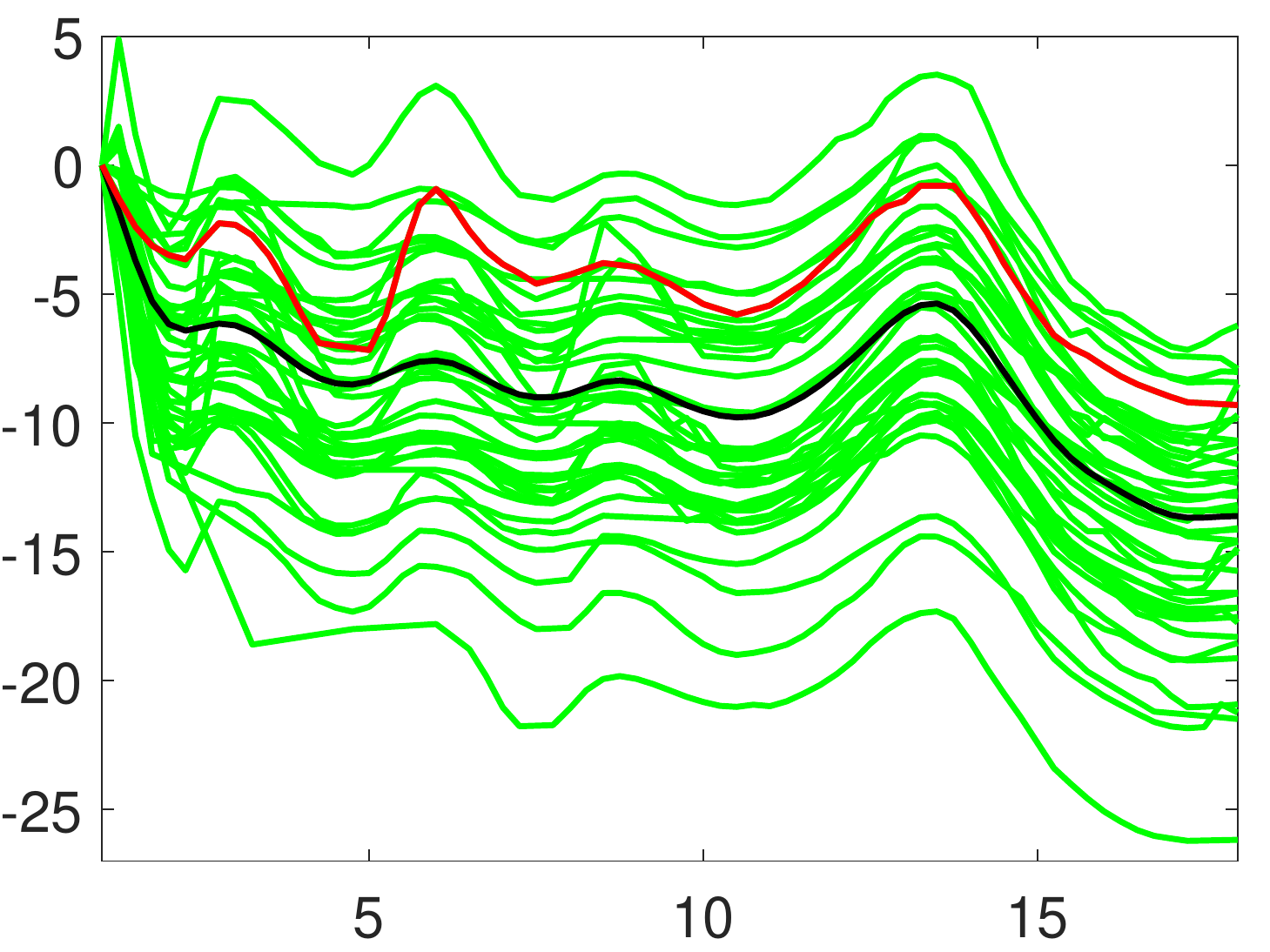}&\includegraphics[width=1.0in]{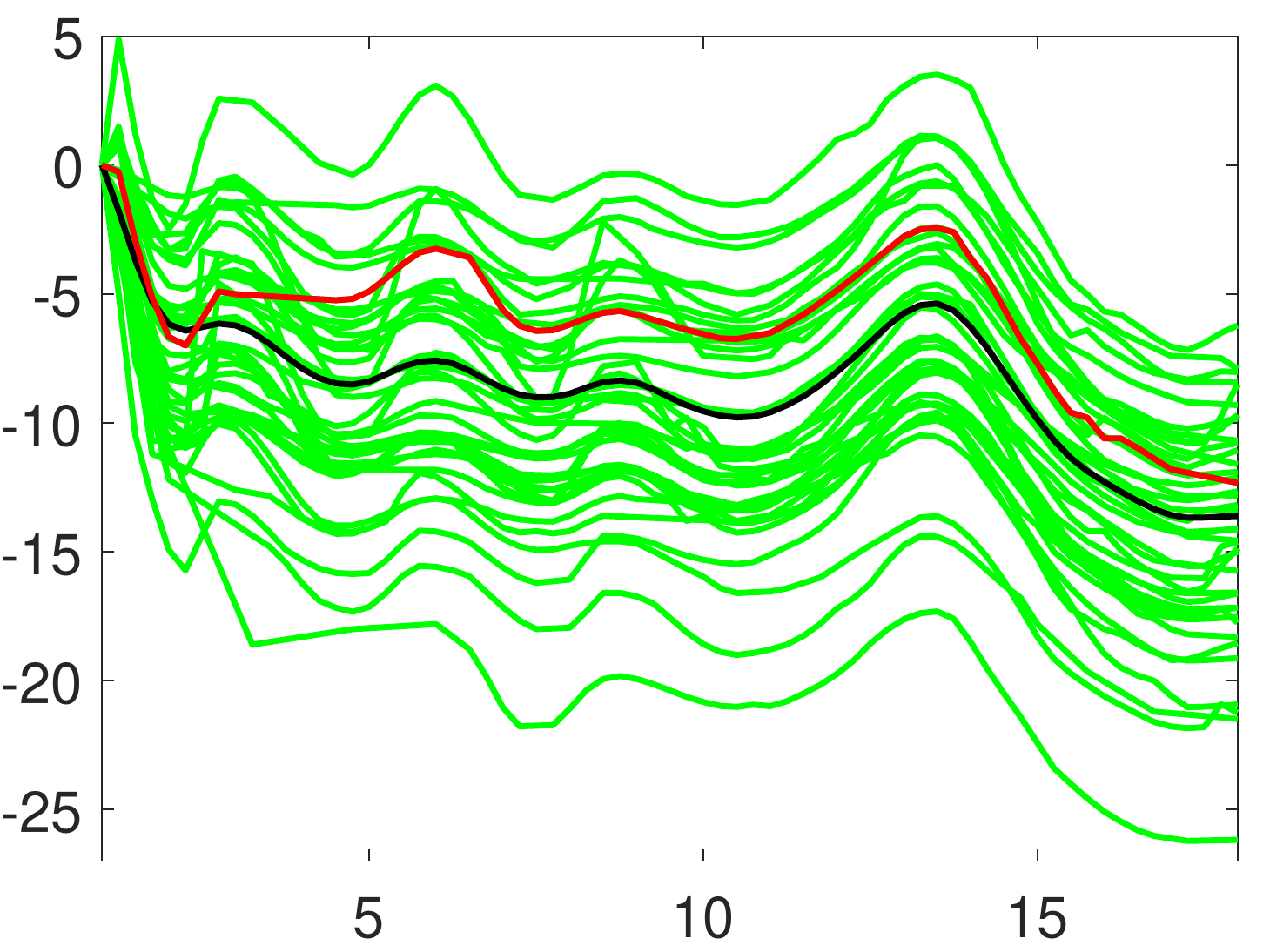}\\
    \hline
    \end{tabular}
    \begin{tabular}{|cc|cc|}
    \hline
    \multicolumn{2}{|c|}{(c)}&\multicolumn{2}{|c|}{(d)}\\
    \hline
    \includegraphics[width=1.3in]{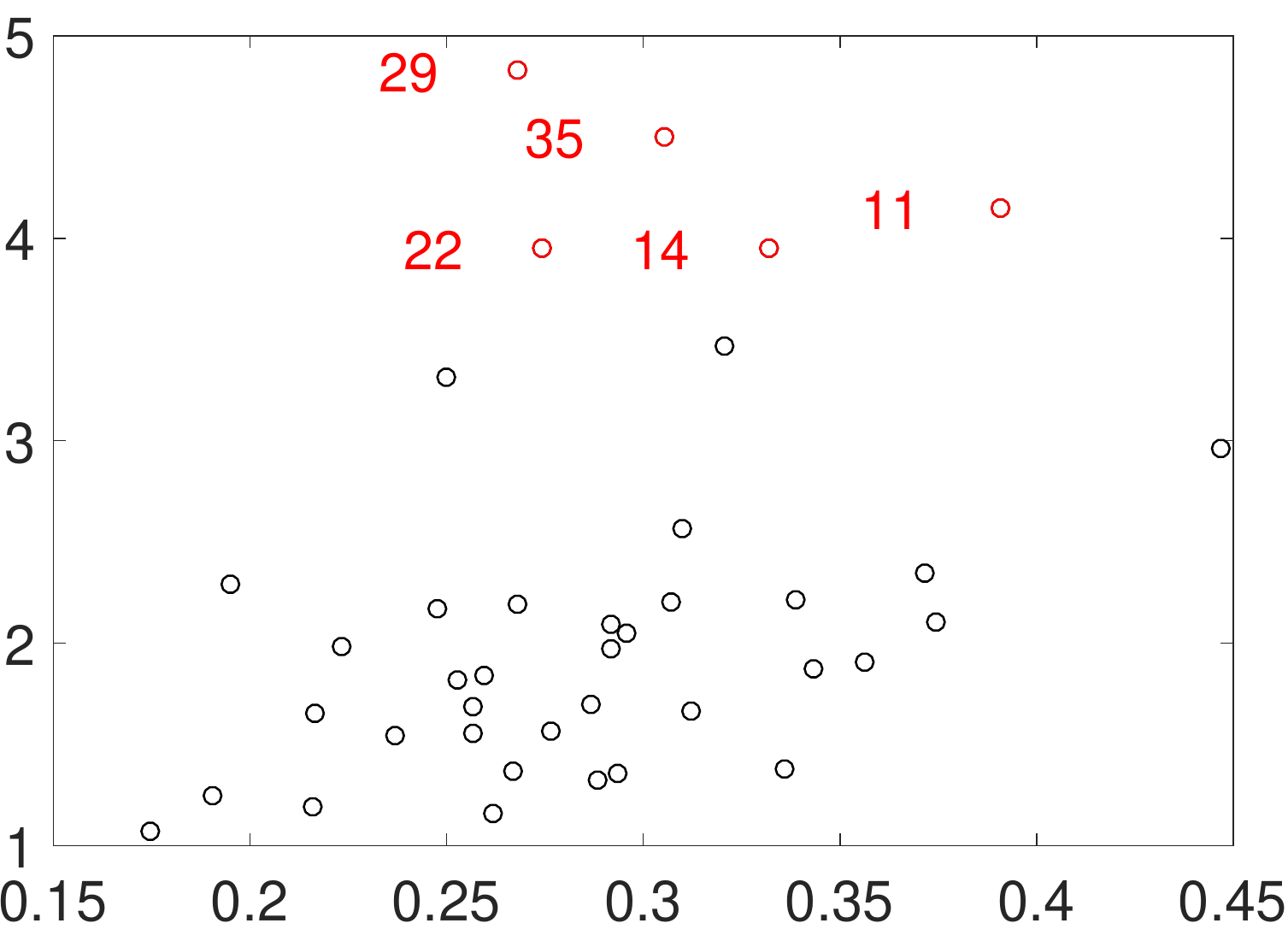}&\includegraphics[width=1.3in]{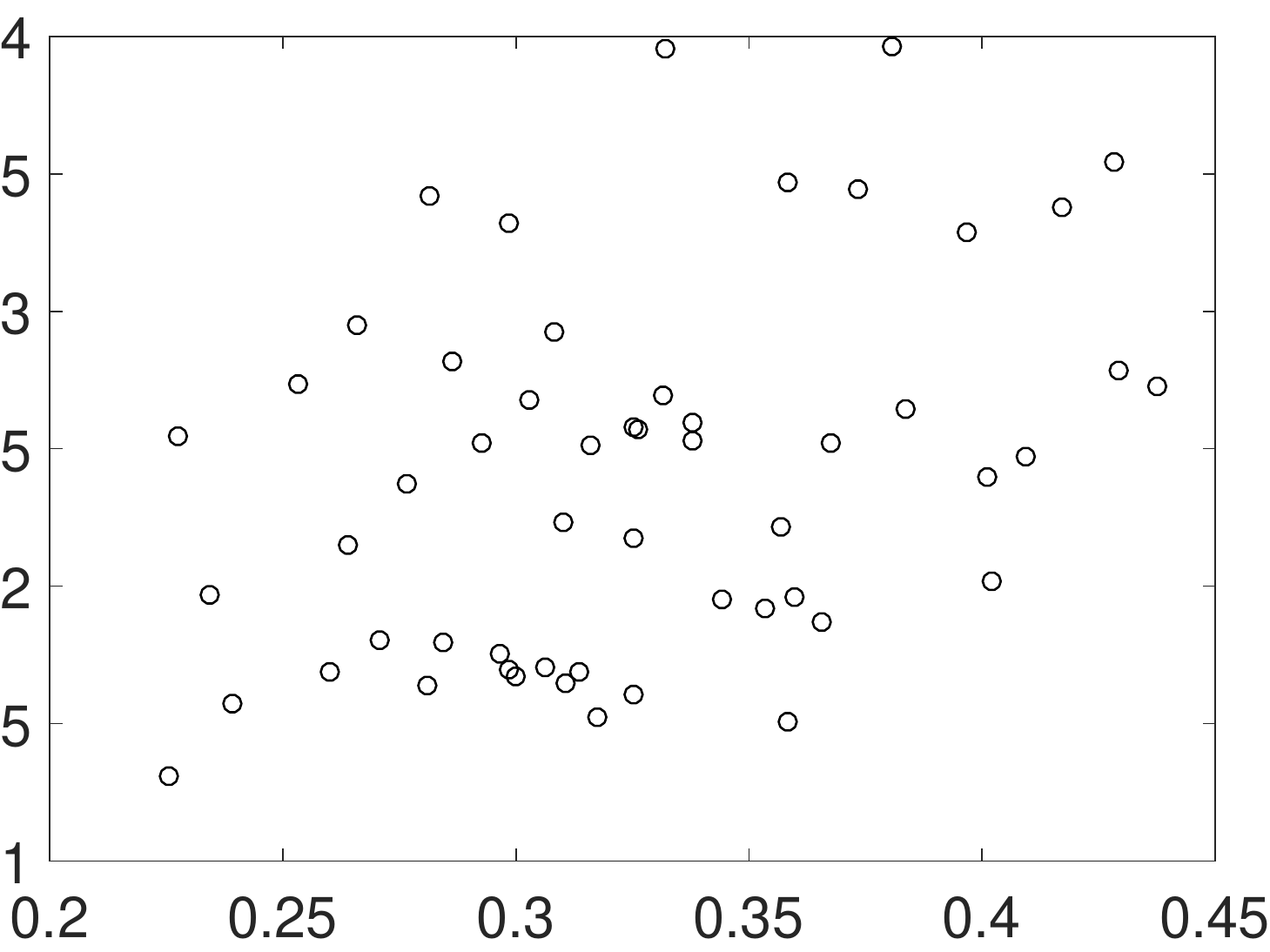}&\includegraphics[width=1.4in]{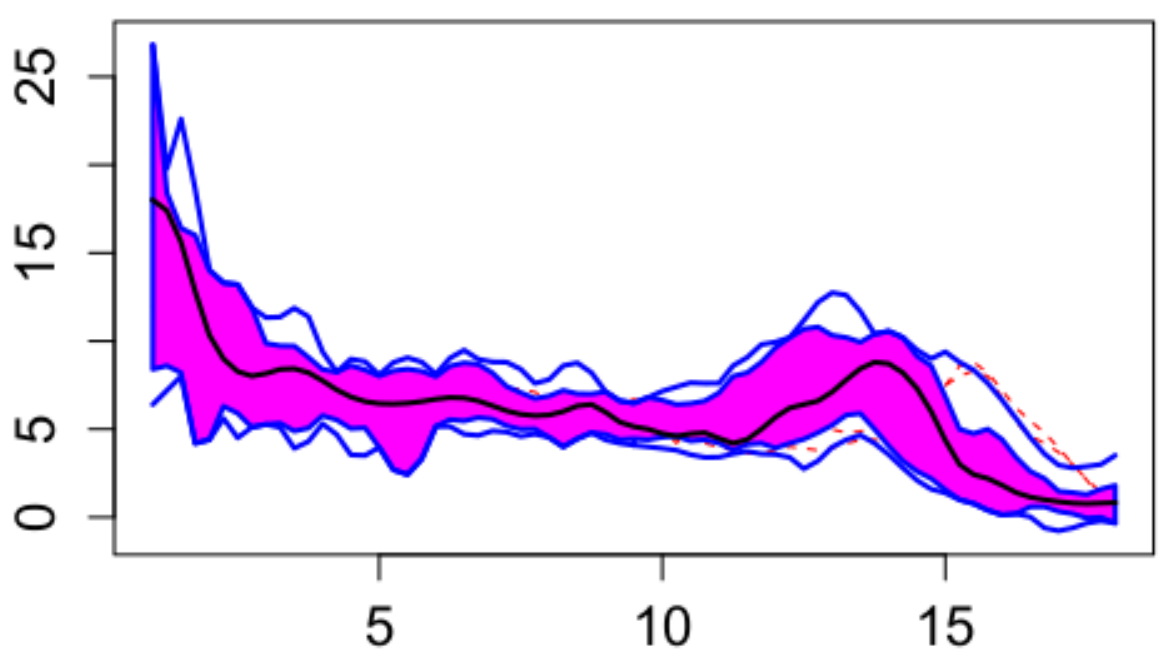}&\includegraphics[width=1.4in]{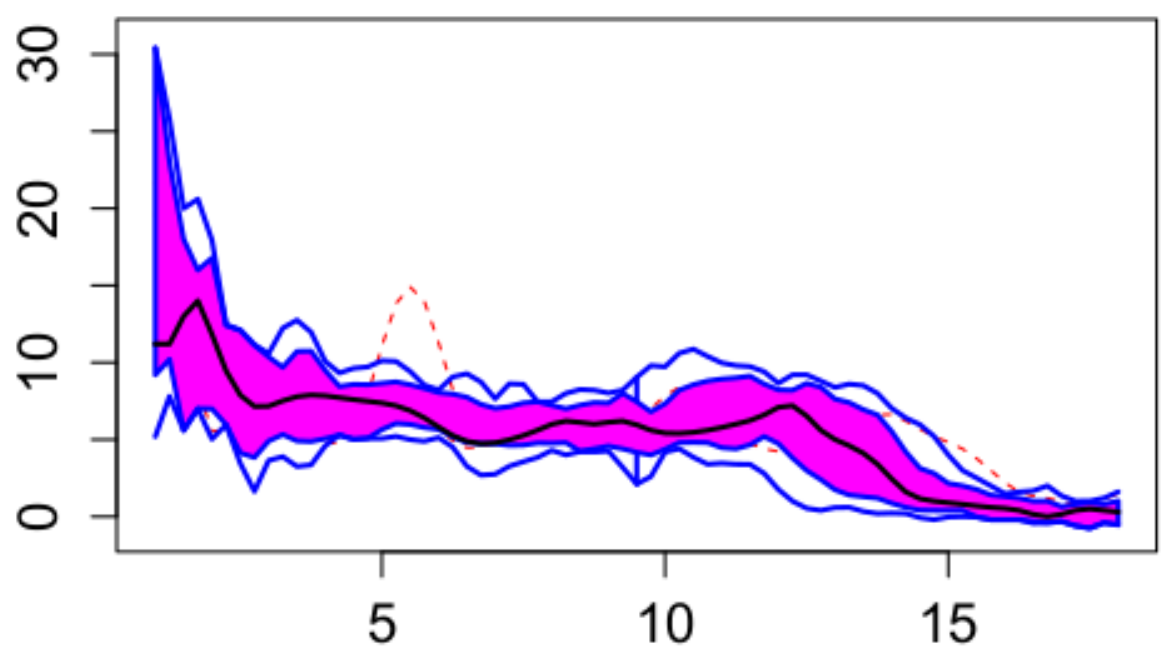}\\
    \hline
    \end{tabular}
    \caption{\small (a) Amplitude outliers: boys 29 and 35. (b) Mild amplitude outliers: boys 11, 22 and 14. (c) Plot of phase ($x$-axis) vs. amplitude ($y$-axis) distances of each function in the data from the median for boys (left) and girls (right). (d) Functional boxplots generated by the method of \cite{citeulike:10107686} for boys (left) and girls (right).}\label{fig:bgdampoutlier}
    \end{center}
    \end{small}
    \vspace{-5mm}
\end{figure}

We close with a study of amplitude and phase outliers for the boys' and girls' growth velocity functions. Using the proposed framework, boys 11, 22 and 14 were identified as mild amplitude outliers, and boys 29 and 35 were identified as regular amplitude outliers; no phase or translation outliers were detected. Figure \ref{fig:bgdampoutlier} displays all of the aligned outlying functions in red with the rest of the data in green. From these plots, we notice that the outlying observations are significantly different in `shape' from the growth velocity functions of all other boys. The main difference comes in emphasis and/or deemphasis of one or more of the growth spurts. No outliers were detected in the girl velocity growth curve dataset. In Figure \ref{fig:bgdampoutlier}(c), we display the phase vs. amplitude distance plots for boys and girls and highlight the outlying observations. Figure \ref{fig:bgdampoutlier}(d) gives a comparison to the functional boxplots of \cite{citeulike:10107686} on the same dataset. Because of the lack of separation of amplitude and phase, the given boxplots are harder to interpret than the proposed method. In particular, it is difficult to separate the variability due to the number and magnitude of growth spurts (amplitude) and their timing (phase) in the display.

\subsection{Real Data Study 3: PQRST Complexes from ECG Signals}

The electrocardiogram (ECG) is a diagnostic tool that is routinely used to assess electrical and muscular functions of the heart and is very popular for diagnosing and monitoring various heart diseases; thus, studying ECG patterns is integral to cardiac medicine. In this work, we construct amplitude and phase boxplots for 80 PQRST complexes segmented from a long ECG signal using the method in \cite{biosignals}. The original data came from the PTB Diagnostic ECG Database (\cite{PTB1}), which was obtained from PhysioNet (\cite{goldberger}).

\begin{figure}[!t]
\begin{center}
\begin{small}
    \begin{tabular}{|c|c|c|c|}
    \hline
    (a)&(b)&(c)&(d)\\
    \hline
    \includegraphics[width=1.3in]{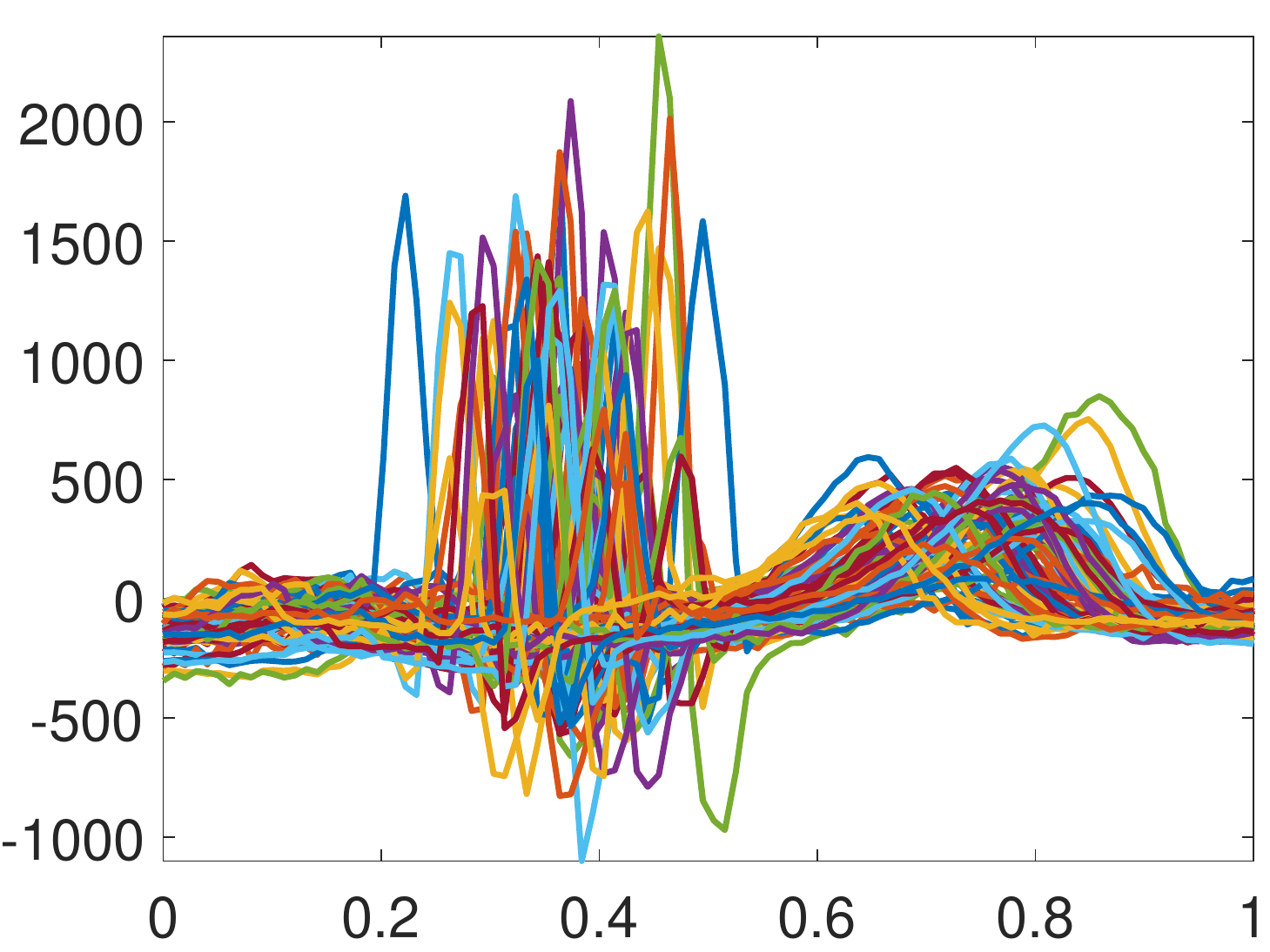}&\includegraphics[width=1.3in]{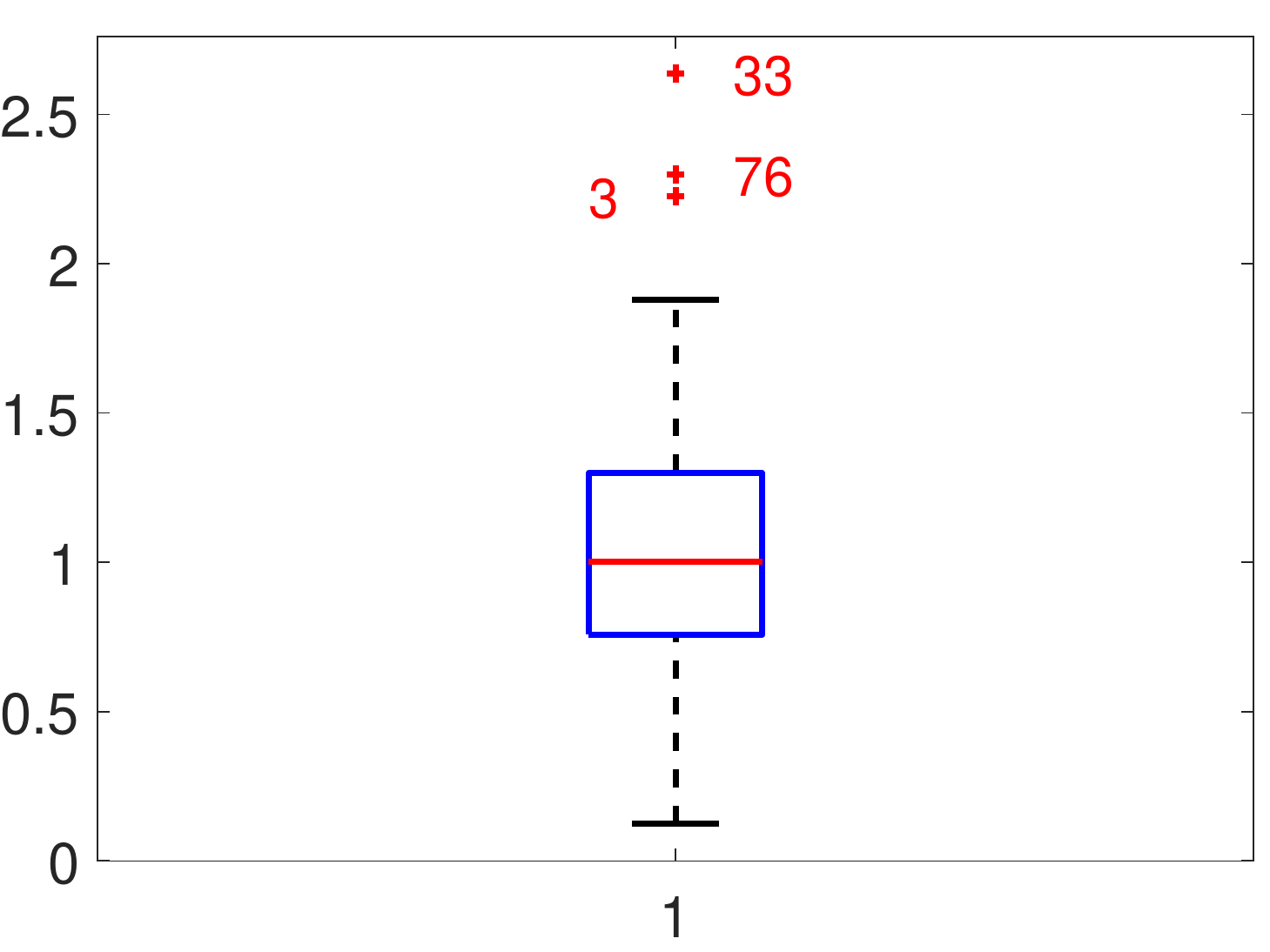}&\includegraphics[width=1.3in]{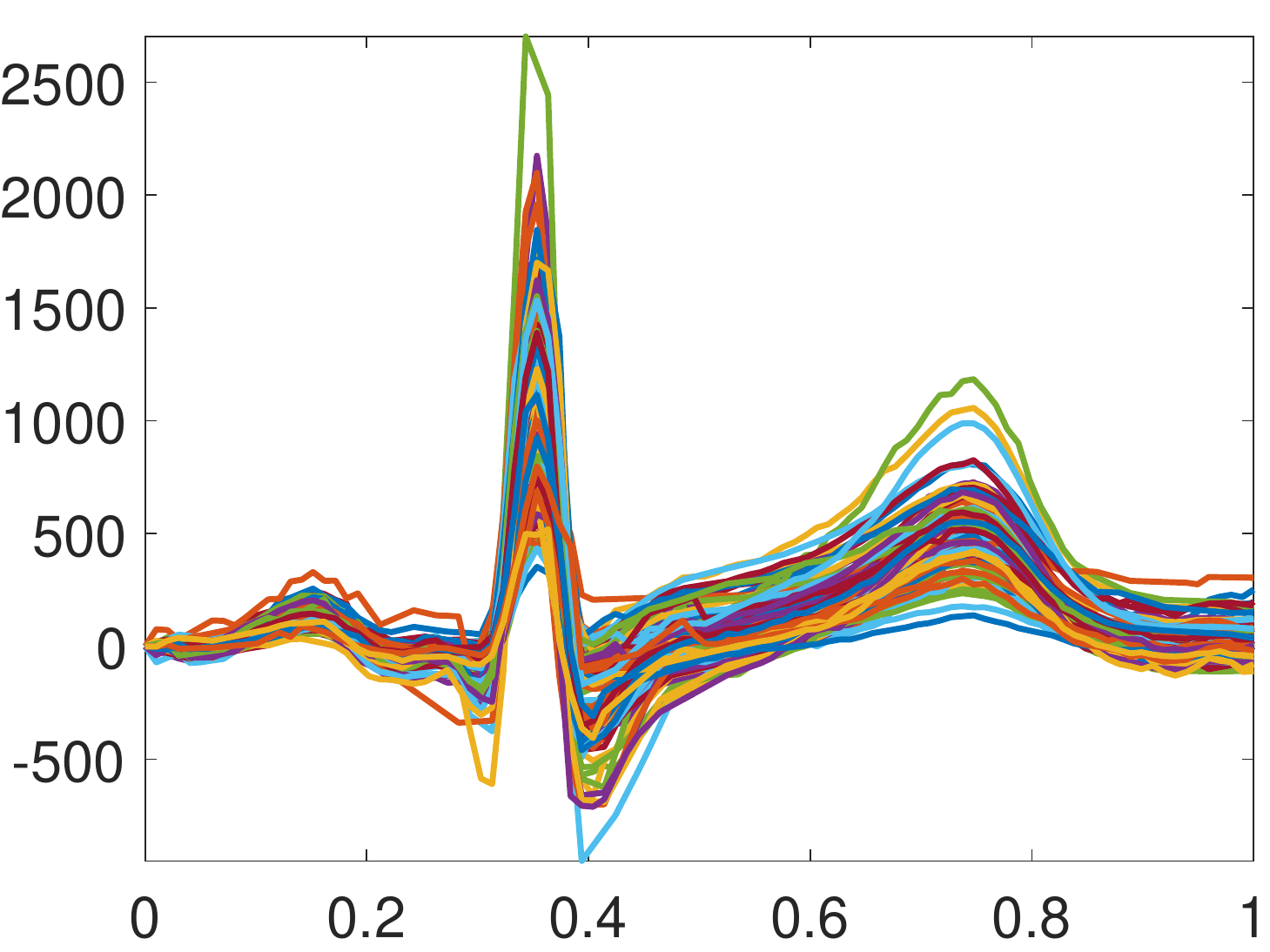}&\includegraphics[width=1.3in]{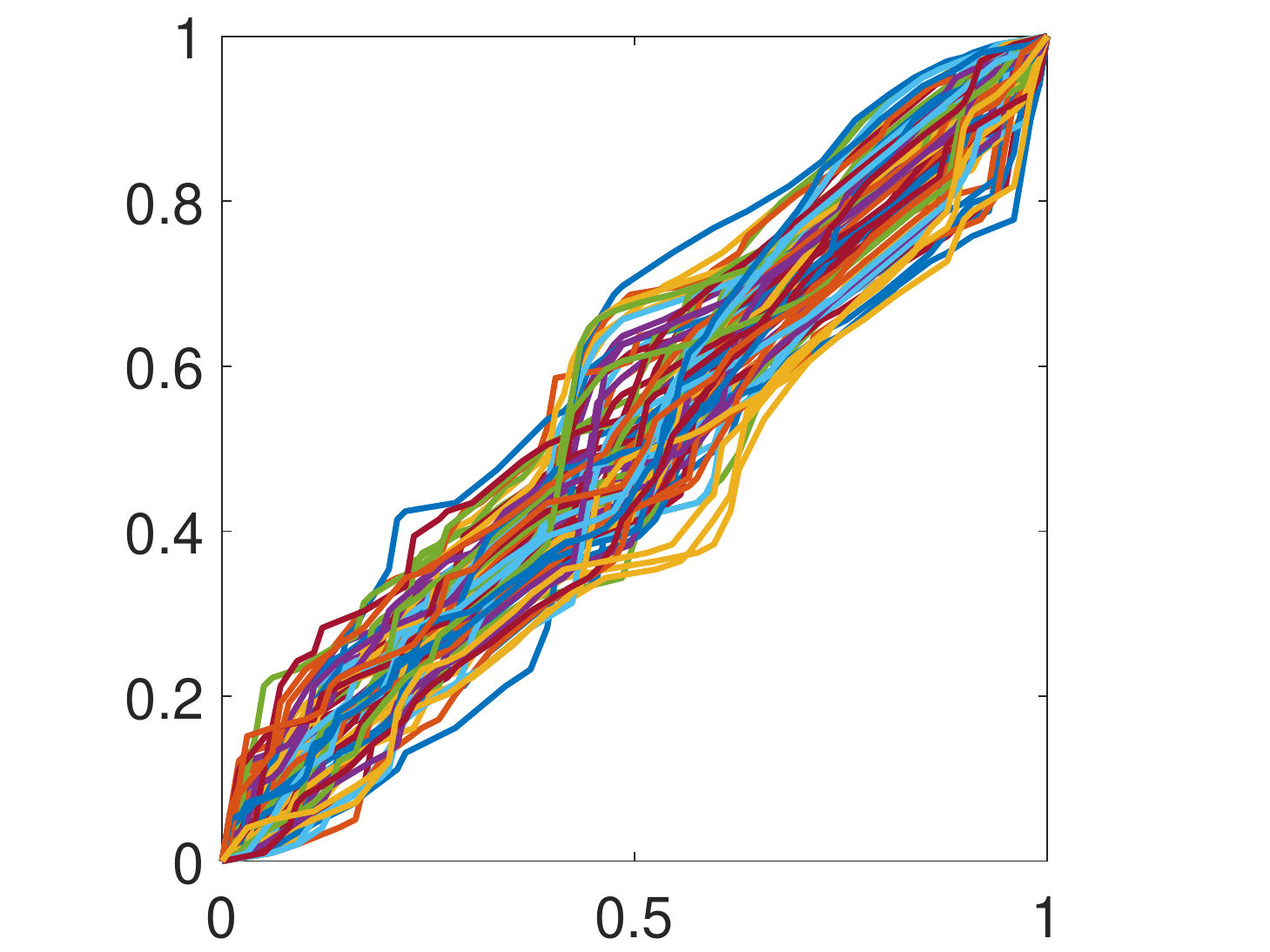}\\
    \hline
    \end{tabular}
    \end{small}
    \caption{\small Separation of translation, amplitude, and phase in PQRST complexes. (a) Original complexes. (b) Translation. (c) Amplitude. (d) Phase.}\label{fig:ecg}
    \end{center}
    \vspace{-5mm}
\end{figure}

\begin{figure}[!t]
\begin{small}
\begin{center}
    \begin{tabular}{|cc|cc|}
    \hline
    (a)&(b)&(c)&(d)\\
    \hline
    \includegraphics[width=1.3in]{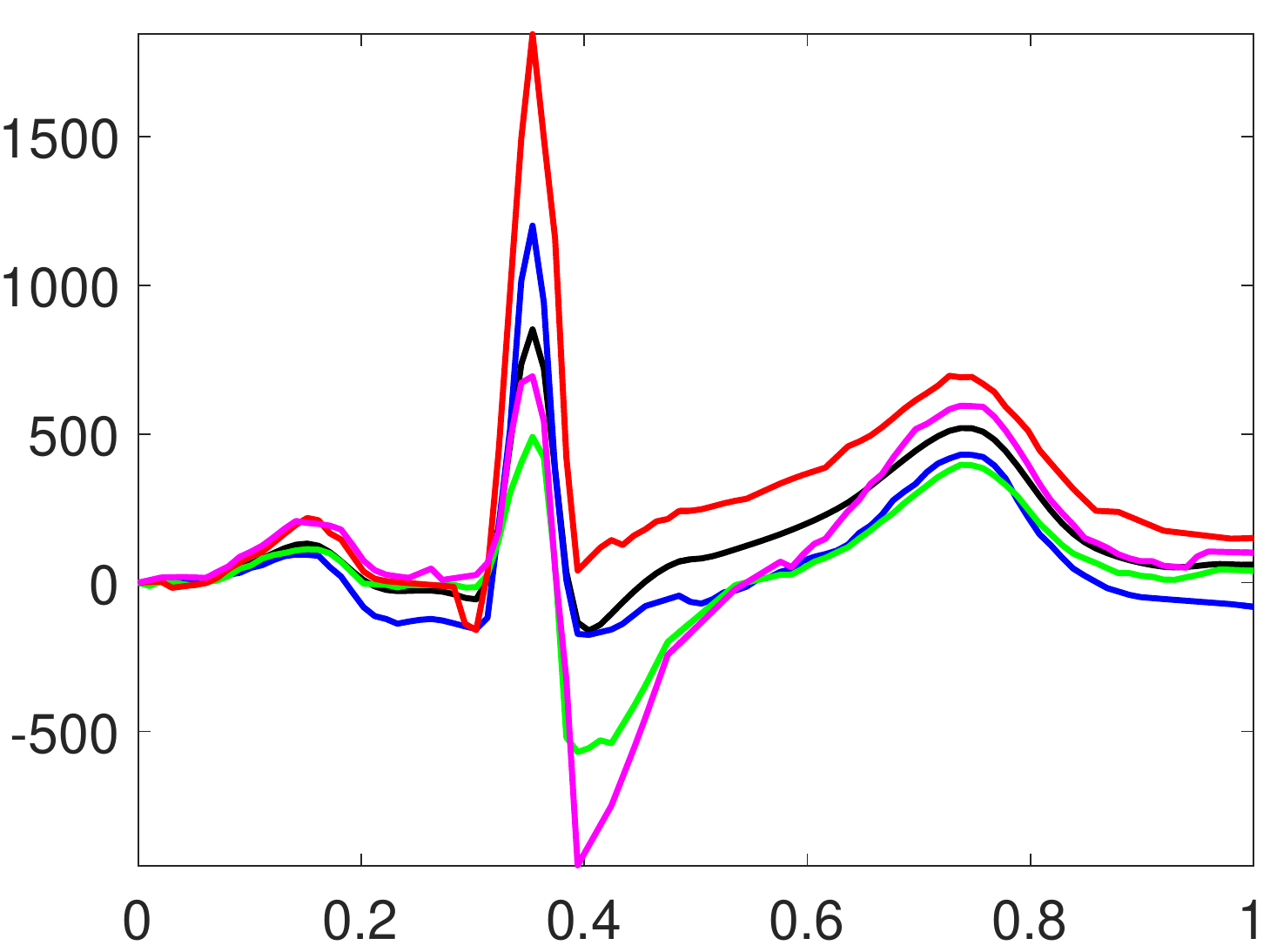}&\includegraphics[width=1.3in]{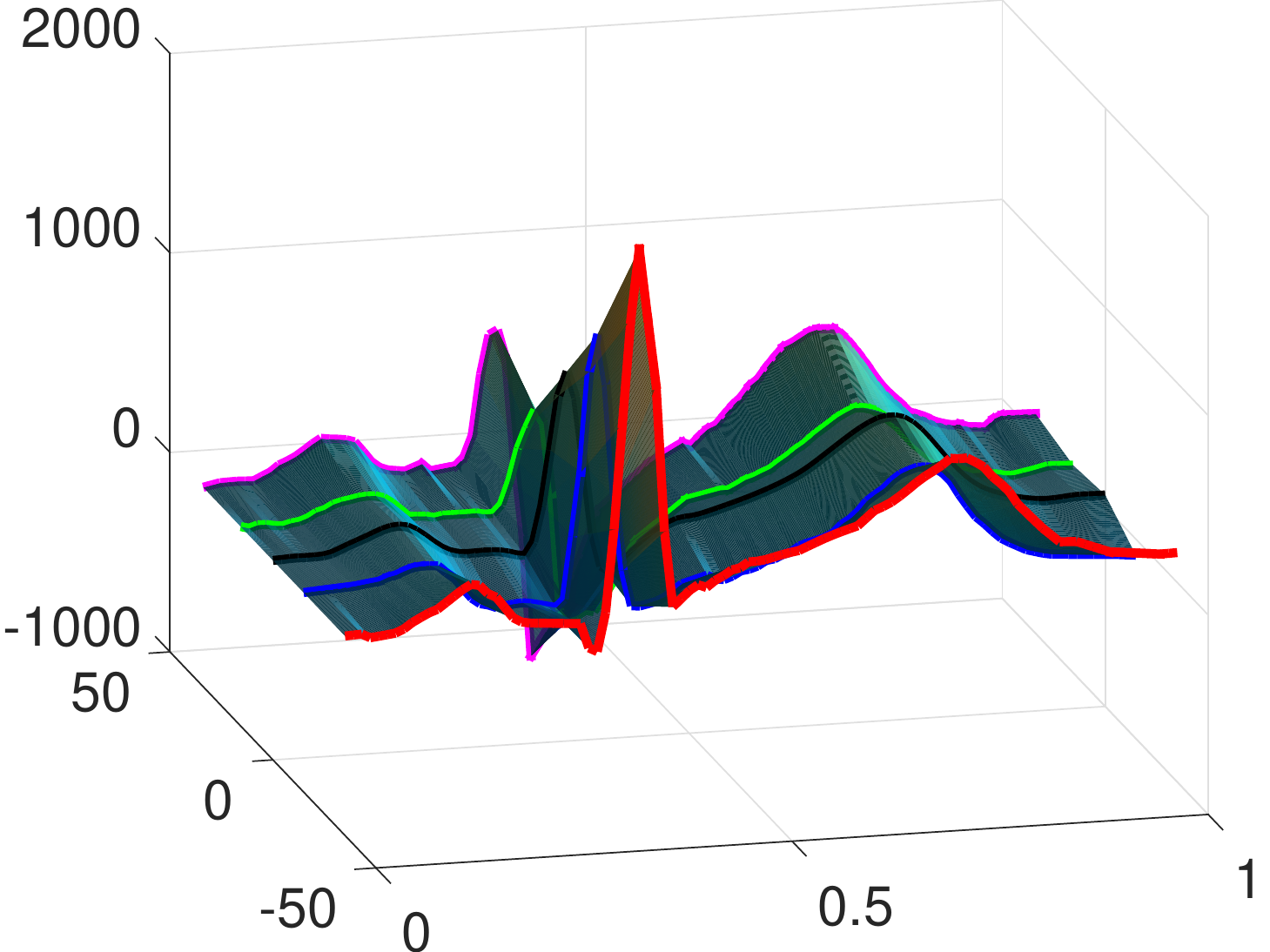}&\includegraphics[width=1.3in]{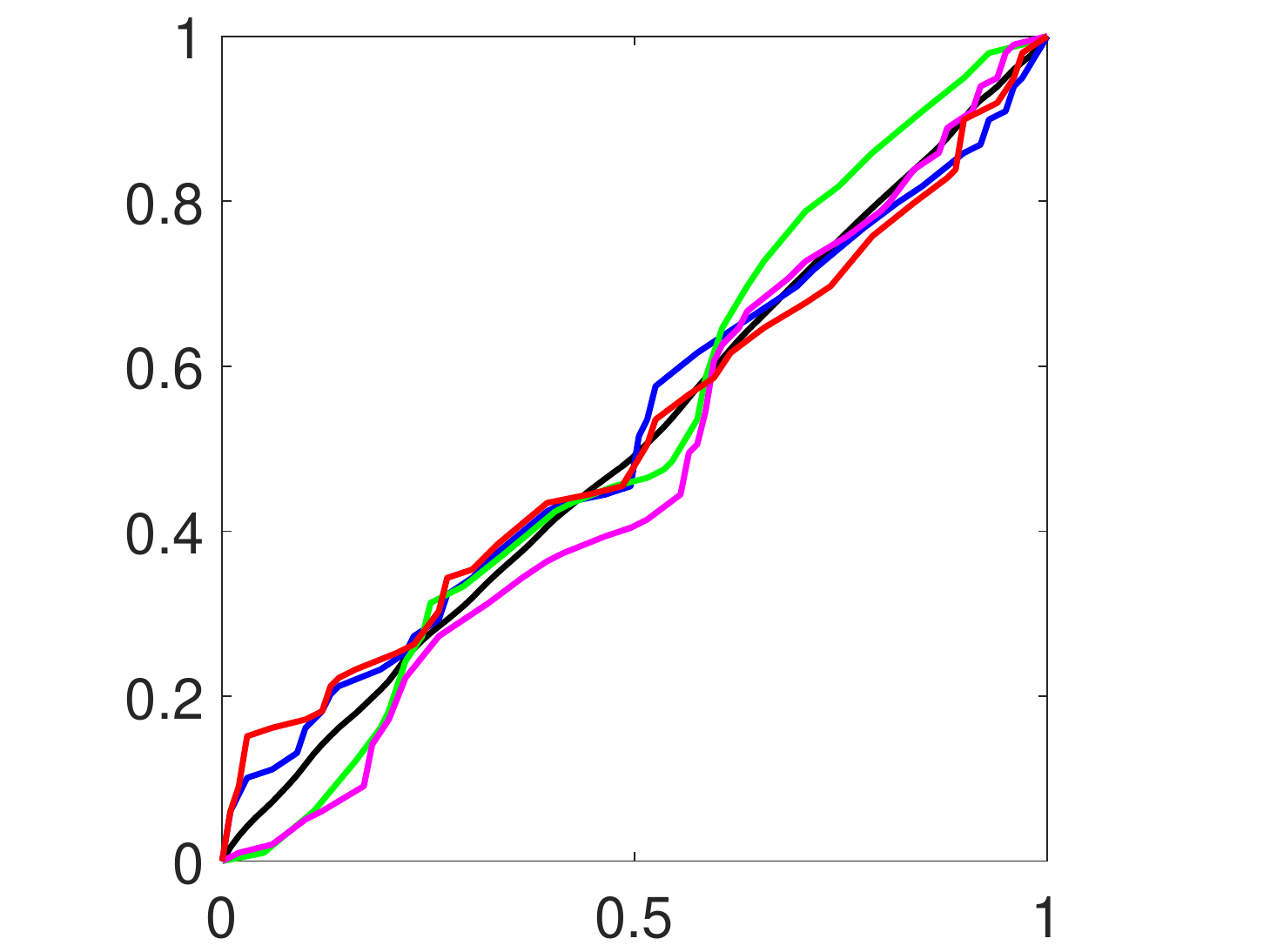}&\includegraphics[width=1.3in]{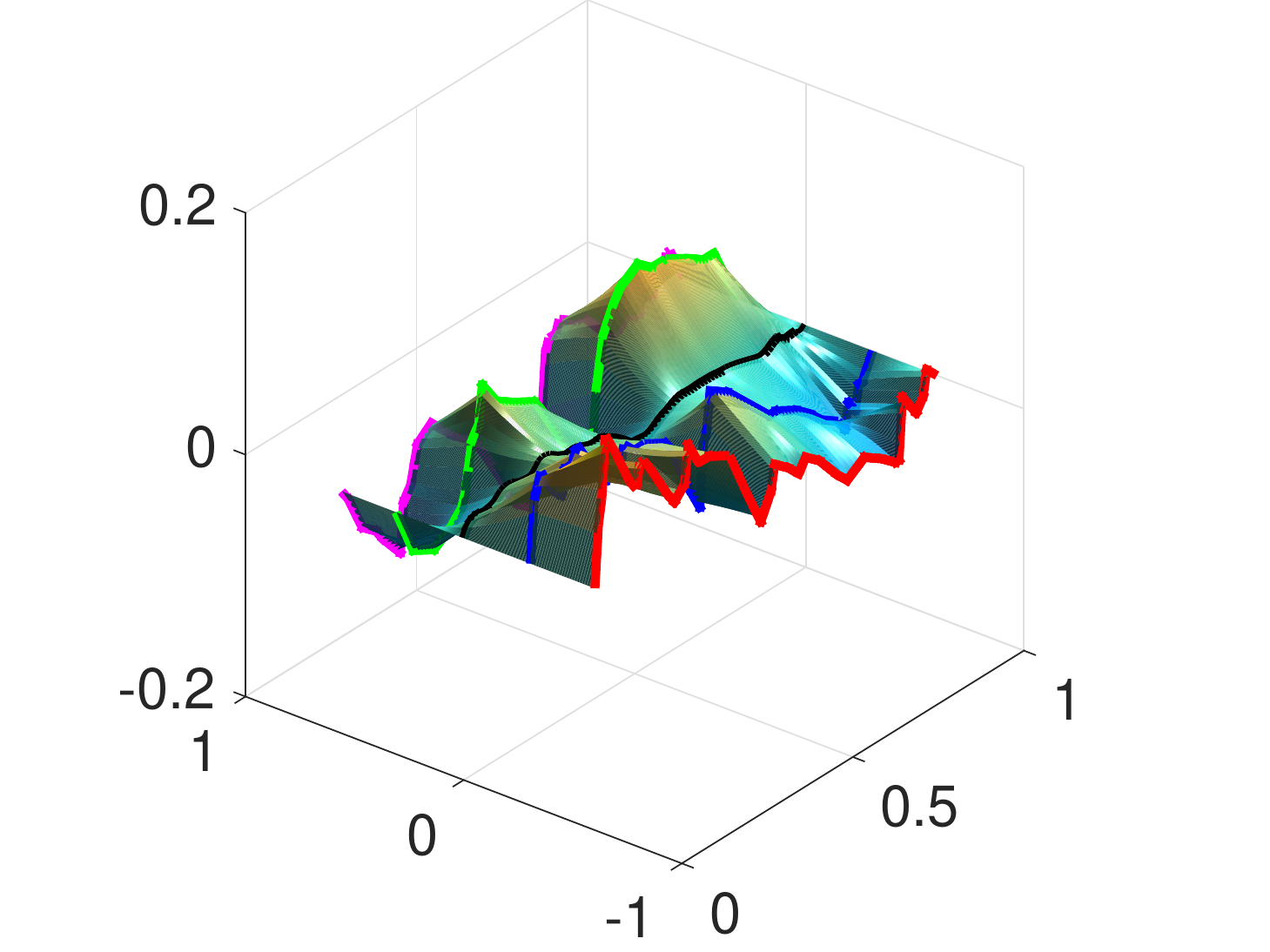}\\
    \hline
    \end{tabular}
    \caption{\small Amplitude and phase boxplot displays for ECG data. (a)\&(b) Amplitude boxplot and its surface display. (c)\&(d) Phase boxplot and its surface display.}\label{fig:ecgamplitude}
    \end{center}
    \end{small}
    \vspace{-5mm}
\end{figure}

Figure \ref{fig:ecg} shows the decomposition of the PQRST complexes into the translation, amplitude and phase components. It is well known in cardiology that there can be variation among intervals between the PQRST waves, which corresponds to phase variability of the PQRST complexes. In addition, the heights of the waves constitute the amplitude component. It is clear from Figure \ref{fig:ecg}(c) that the PQRST features are very well aligned in the amplitude component; this manifests itself in significant phase variability shown in Figure \ref{fig:ecg}(d). Next, in Figure \ref{fig:ecgamplitude}, we show the amplitude and phase boxplots. The amplitude boxplot (panels (a)\&(b)) is especially effective at displaying the variability in the heights of the R and S waves. The phase boxplot (panels (c)\&(d)) is harder to interpret in this case; one of the dominant features is the variability in the initial timing, corresponding to the P wave of the PQRST complexes.

%\begin{figure}[!t]
%\begin{center}
%\begin{small}
%    \begin{tabular}{|c|c|c|}
%    \hline
%    (a)&(b)&(c)\\
%    \hline
%    \includegraphics[width=1.3in]{ECG_amplitude_minmax_1.pdf}&\includegraphics[width=1.3in]{ECG_amplitude_minmax_2.pdf}&\includegraphics[width=1.3in]{ECG_amplitude_minmax_3.pdf}\\
%    \hline
%    (d)&(e)&(f)\\
%    \hline
%    \includegraphics[width=1.3in]{ECG_amplitude_minmax_4.pdf}&\includegraphics[width=1.3in]{ECG_amplitude_minmax_5.pdf}&\includegraphics[width=1.3in]{ECG_amplitude_minmax_surface.pdf}\\
%    \hline
%    \end{tabular}
%    \end{small}
%    \caption{\small Amplitude boxplot displays for ECG data. (a)\&(e) Amplitude boxplot extremes. (b)\&(d) Quartiles. (c) Median. (f) Surface display of amplitude boxplot.}\label{fig:ecgamplitude}
%    \end{center}
%    \vspace{-5mm}
%\end{figure}

%\begin{figure}[!t]
%\begin{center}
%\begin{small}
%    \begin{tabular}{|c|c|c|}
%    \hline
%    (a)&(b)&(c)\\
%    \hline
%    \includegraphics[width=1.4in]{ECG_warping_minmax.pdf}&\includegraphics[width=1.4in]{ECG_warping_minmax_surface.pdf}&\includegraphics[width=1.4in]{ECG_dxdy.pdf}\\
%    \hline
%    \end{tabular}
%    \end{small}
%    \caption{\small Phase boxplot displays for ECG data. (a) Phase boxplot. (b) Surface display of phase boxplot. (c) Plot of phase ($x$-axis) vs. amplitude ($y$-axis) distances of each function in the data from the median.}\label{fig:ecgphase}
%    \end{center}
%    \vspace{-5mm}
%\end{figure}

\begin{figure}[!t]
\begin{center}
\begin{small}
    \begin{tabular}{|ccc|c|}
    \hline
    \multicolumn{3}{|c|}{(a)}&(b)\\
    \hline
    \includegraphics[width=1.3in]{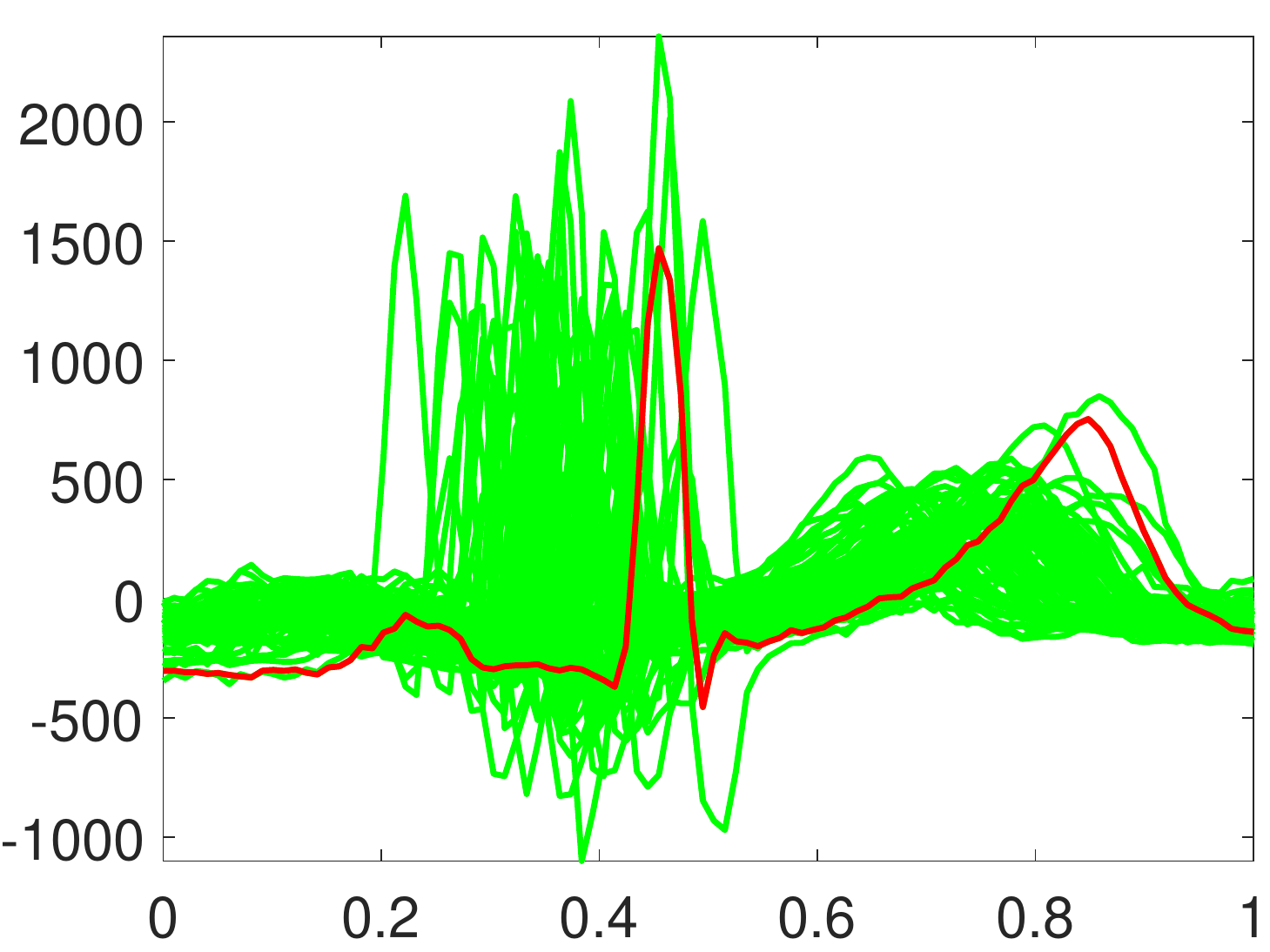}&\includegraphics[width=1.3in]{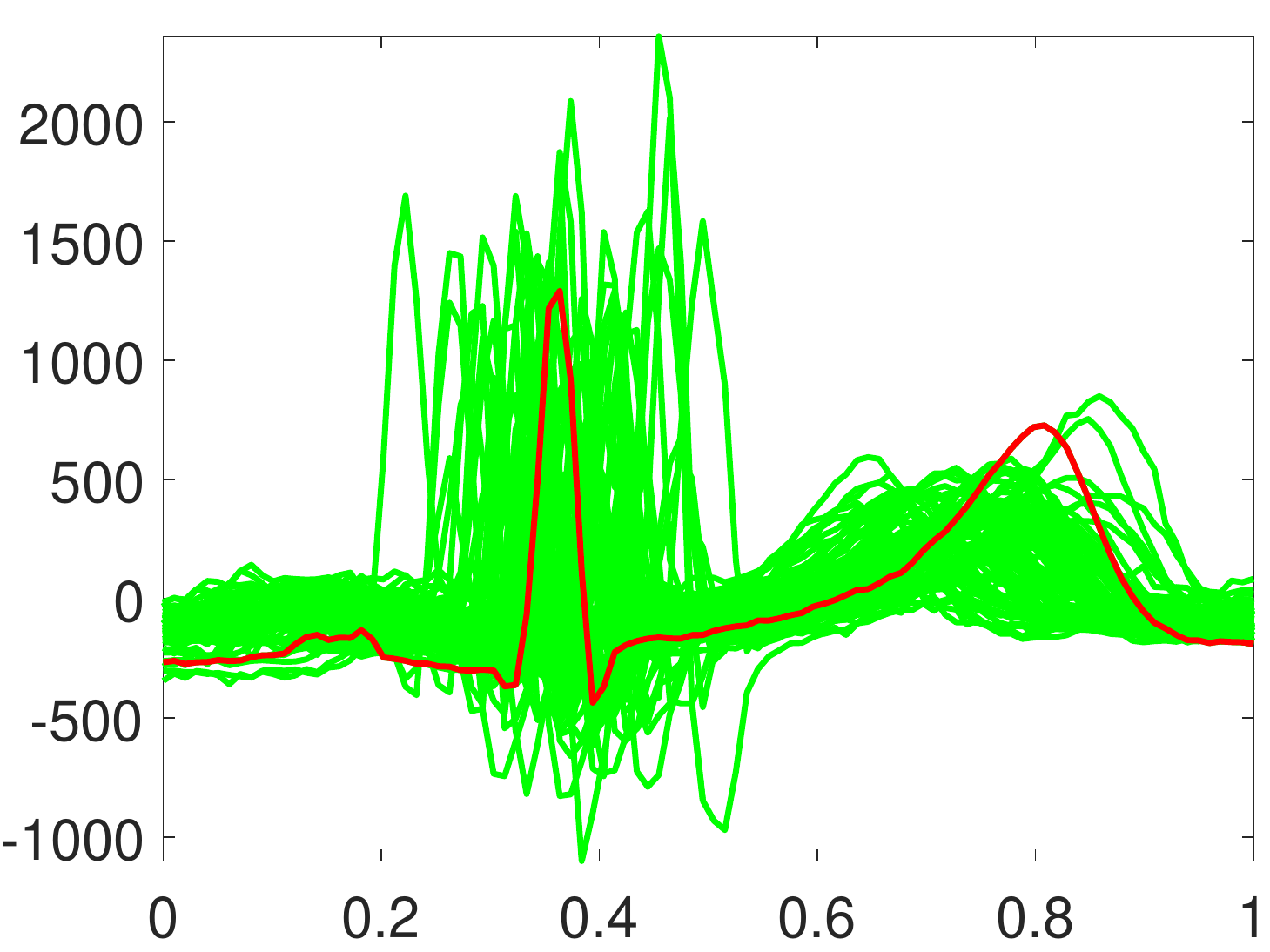}&\includegraphics[width=1.3in]{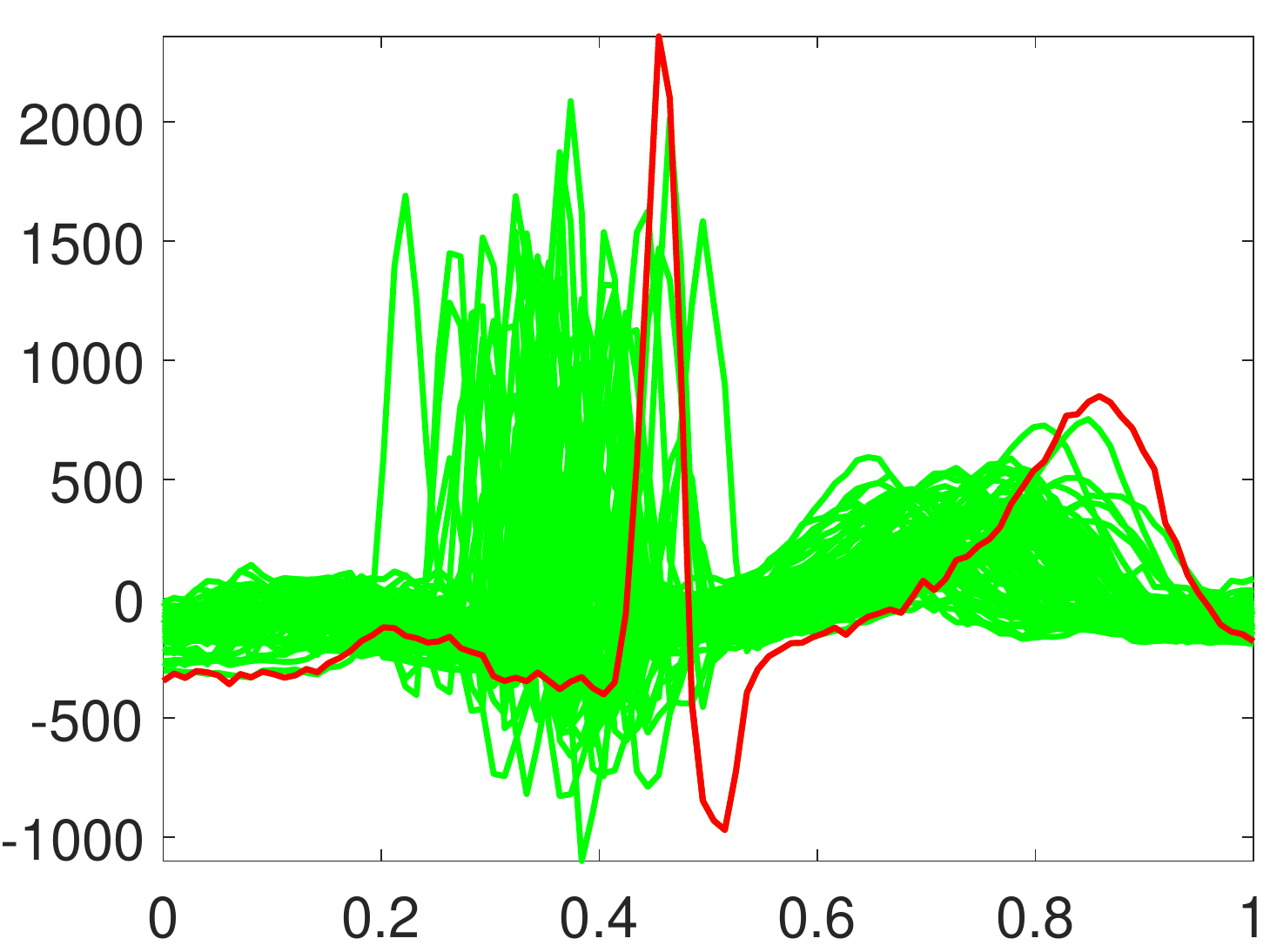}&\includegraphics[width=1.3in]{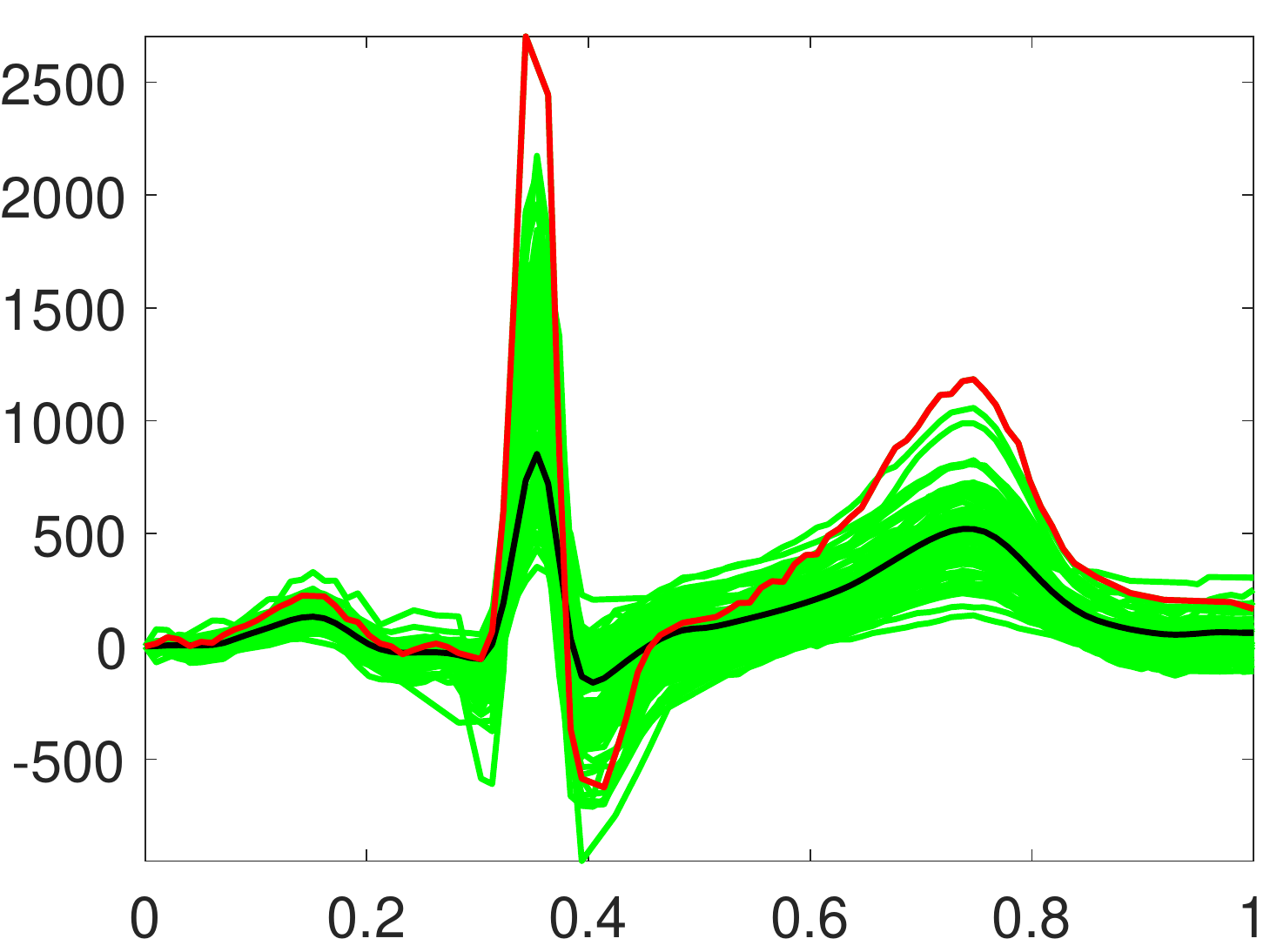}\\
    \hline
    \end{tabular}
    \begin{tabular}{|c|c|}
    \hline
    (c)&(d)\\
    \hline
    \includegraphics[width=1.5in]{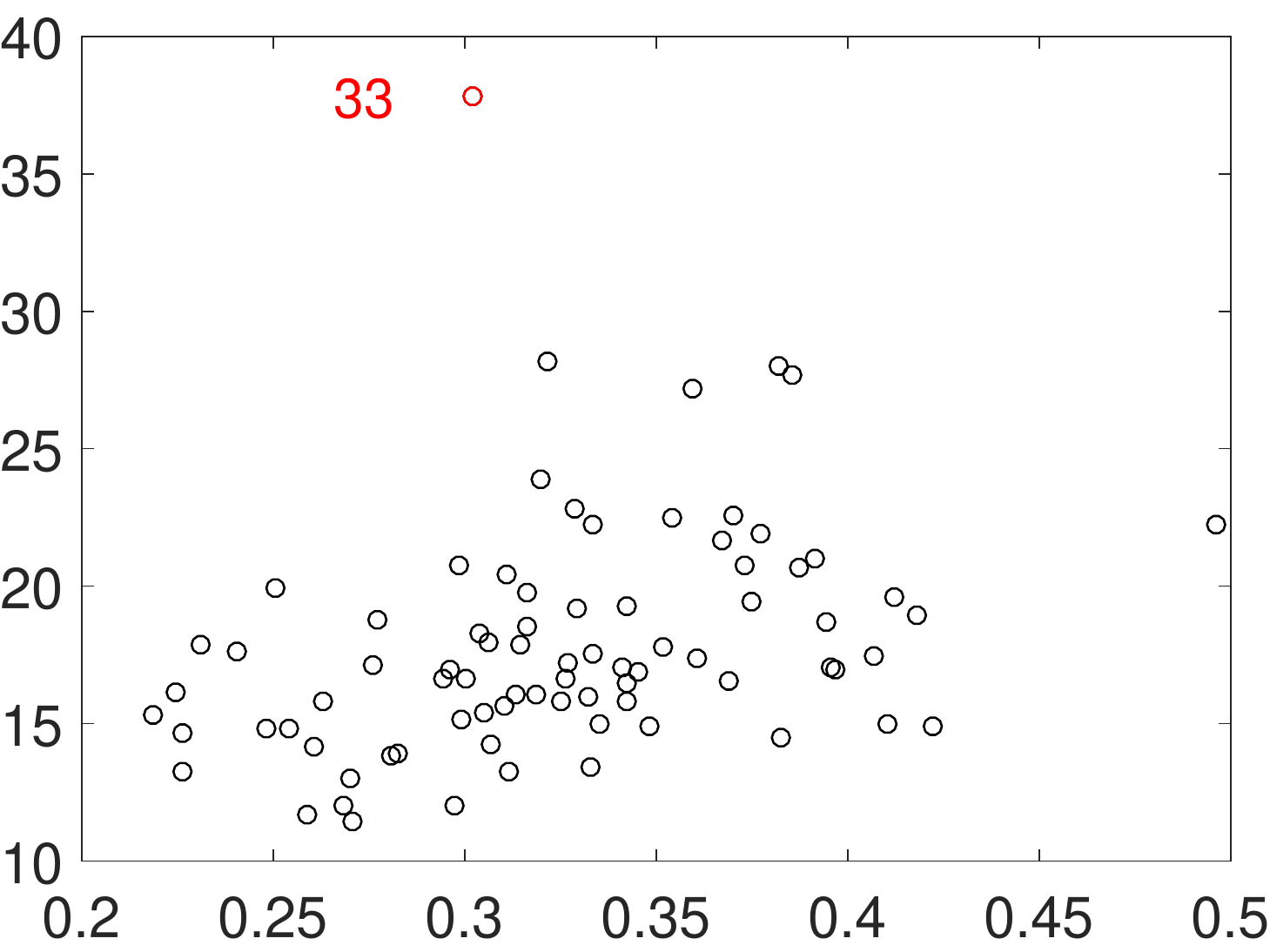}&\includegraphics[width=1.7in]{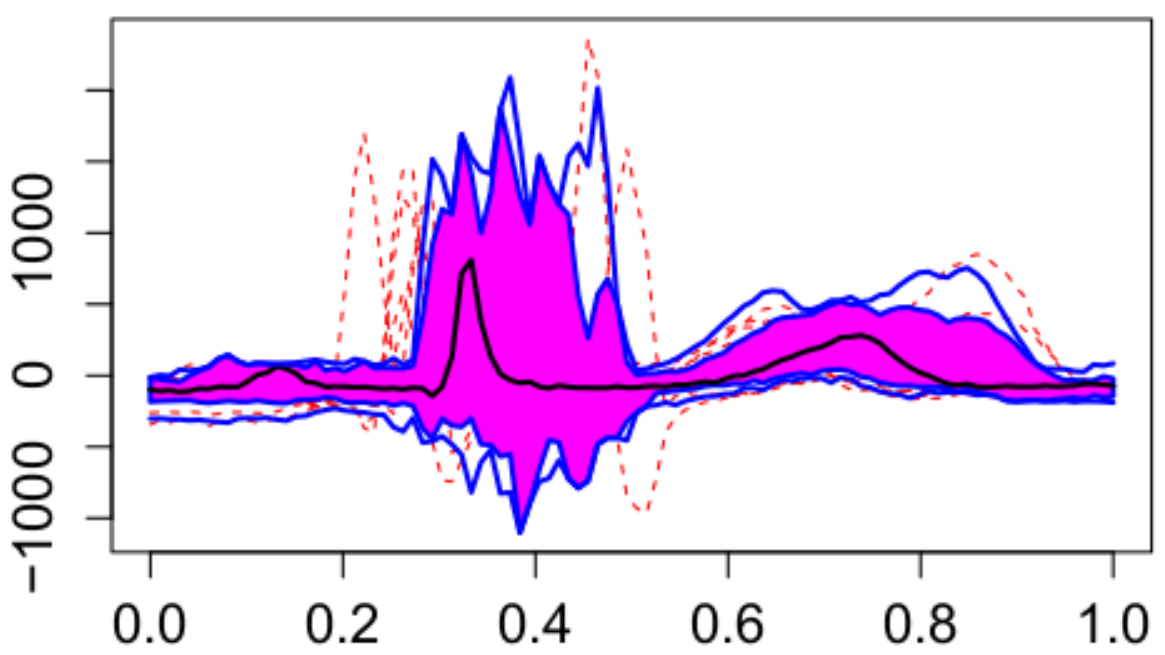}\\
    \hline
    \end{tabular}
    \end{small}
    \caption{\small (a) Translation outliers: complexes 3, 76 and 33. (b) Mild amplitude outlier: complex 33. (c) Plot of phase ($x$-axis) vs. amplitude ($y$-axis) distances of each function in the data from the median. (d) Functional boxplot generated by the method of \cite{citeulike:10107686}.}\label{fig:ecgampoutlier}
    \end{center}
    \vspace{-5mm}
\end{figure}

Finally, we consider outlier detection in this ECG dataset and display the outlying functions in Figure \ref{fig:ecgampoutlier} in red with the original or aligned data in green. We find that PQRST complexes 3, 76, and 33 are translation outliers, and PQRST complex 33 is a mild amplitude outlier. This mild amplitude outlier is confirmed by the very high amplitude distance shown in the phase vs. amplitude distance plot in Figure \ref{fig:ecgampoutlier}(c). Figure \ref{fig:ecgampoutlier}(d) provides a comparison to the functional boxplot of \cite{citeulike:10107686}. Their method identifies eight outliers in this case, which is mostly due to misaligned features of the given functions. For additional real data studies and the computational complexity of the proposed methods please see the Supplementary Material.

\section{Summary and Future Work}
\label{sec:conc}

In this work, we introduced the concept of boxplot-type visualizations for the translation, amplitude, and phase components of elastic functional data, allowing for independent analysis and outlier detection for each component. The proposed method is metric-based and relies on the geometry of underlying representation spaces. We provided a number of simulation results to compare this method to that of \cite{citeulike:10107686}. Finally, we show the versatility of the plots in visualizing and detecting outliers in real complex datasets including annual sea surface temperature, Berkeley growth data, and electrocardiogram PQRST complexes.

We have identified multiple future directions of research. First, we will formally study the robustness of the proposed procedure (especially the median amplitude and phase computation) to various types of outliers. Second, we will theoretically motivate the values of $k_a$ and $k_p$, which define amplitude and phase outlier cutoffs. Third, we will develop functional boxplots, which are able to display the covariation of amplitude and phase in functional data. This will require a single ranking procedure for the two components. Finally, we will focus on defining similar boxplot displays for more complex functional data including images and shapes of curves and surfaces. In the case of curves, the additional rotation and scale variabilities in the given data add a layer of difficulty; that is, they also require a visualization component. In the case of images and surfaces, the Riemannian geometry of the phase component is much more complicated and requires the development of novel, computationally efficient tools.

\noindent\textbf{Acknowledgments:} We would like to thank the reviewers for their valuable comments, which greatly improved the quality of this manuscript. This research was partially supported by NSF DMS 1613054 (to Sebastian Kurtek and Karthik Bharath), and the KAUST Office of Sponsored Research under award OSR-2015-CRG4-2582 (to Ying Sun).

\begin{small}
\bibliographystyle{Chicago}
\bibliography{biblio}
\end{small}
\end{document}